\documentclass[manuscript]{acmart}

\AtBeginDocument{%
  }

\setcopyright{acmlicensed}
\copyrightyear{2024}
\acmYear{2024}

\acmVolume{99}
\acmNumber{99}
\acmArticle{99}
\acmMonth{9}

\usepackage{algorithmic}
\usepackage{algorithm}
\usepackage{amsmath}
\usepackage{amsfonts}
\usepackage{multirow}
\usepackage{hyperref}
\usepackage{subfigure}
\usepackage{shadowtext}
\usepackage{soul} 
\usepackage{xcolor} 
\definecolor{mycolor}{RGB}{230, 230, 230} 
\sethlcolor{mycolor}

\usepackage{array} 

\newcolumntype{L}[1]{>{\raggedright\arraybackslash}p{#1}} %
\newcolumntype{C}[1]{>{\centering\arraybackslash}p{#1}} %
\newcolumntype{R}[1]{>{\raggedleft\arraybackslash}p{#1}} %
\usepackage{amsmath} %
\usepackage{enumitem} %

\begin{document}

\title{Lossless and Privacy-Preserving Graph Convolution Network for Federated Item Recommendation}

\author{Guowei Wu}
\email{wuguowei2023@email.szu.edu.cn}

\affiliation{%
	\institution{College of Computer Science and Software Engineering, Shenzhen University}
	\streetaddress{3688\# Nanhai Avenue, Nanshan District}
	\city{Shenzhen}
	\postcode{518060}
	\country{China}
}

\author{Weike Pan}
\authornote{Corresponding author.}
\email{panweike@szu.edu.cn}
\affiliation{%
	\institution{College of Computer Science and Software Engineering, Shenzhen University}
	\streetaddress{3688\# Nanhai Avenue, Nanshan District}
	\city{Shenzhen}
	\postcode{518060}
	\country{China}
}

\author{Qiang Yang}
\email{qyang@cse.ust.hk}
\affiliation{%
	\institution{WeBank AI Lab, WeBank, China and Hong Kong University of Science and Technology}
	\city{Hong Kong}
	\country{China}
}

\author{Zhong Ming}
\email{mingz@szu.edu.cn}
\affiliation{%
	\institution{Shenzhen Technology University, Guangdong Laboratory of 
Artificial Intelligence and Digital Economy (SZ), Shenzhen}
	\city{Shenzhen}
	\country{China}
}

\renewcommand{\shortauthors}{Wu et al.}

\begin{abstract}
Graph neural network (GNN) has emerged as a state-of-the-art solution for item recommendation. However, existing GNN-based recommendation methods rely on a centralized storage of fragmented user-item interaction sub-graphs and training on an aggregated global graph, which will lead to privacy concerns.
As a response, some recent works develop GNN-based federated recommendation methods by exploiting decentralized and fragmented user-item sub-graphs in order to preserve user privacy. However, due to privacy constraints, the graph convolution process in existing federated recommendation methods is incomplete compared with the centralized counterpart, causing a degradation of the recommendation performance. In this paper, we propose a novel lossless and privacy-preserving graph convolution network (LP-GCN), which fully completes the graph convolution process with decentralized user-item interaction sub-graphs while ensuring privacy. It is worth mentioning that its performance is equivalent to that of the non-federated (i.e., centralized) counterpart. Moreover, we validate its effectiveness through both theoretical analysis and empirical studies. Extensive experiments on three real-world datasets show that our LP-GCN outperforms the existing federated recommendation methods. The code will be publicly available once the paper is accepted.
\end{abstract}




\maketitle

\section{Introduction}
With the proliferation of the Internet and mobile devices, finding the preferred information from a huge amount of online data can be challenging in fields like e-commerce~\cite{cen2019trust} and fintech~\cite{cao2021data}. Fortunately, recommender systems can help alleviate this situation by filtering out some irrelevant information and suggesting items that match user interests. Traditional recommendation models rely on the server to collect and manage users' historical data, such as browsing, clicks, and purchases, to train a centralized model. However, this behavior data often contains users' private information, which they might not want to share, even with the server of a shopping site.

In 2016, the European Union introduced the General Data Protection Regulation (GDPR)\footnote{\url{https://gdpr-info.eu}}, and other countries have also implemented various privacy protection regulations~\cite{CCPA, ISL}. As a result, many traditional recommendation models may not be able to offer their services in the future if they do not comply with these regulations. Therefore, we need a new recommendation framework that can provide services while protecting user privacy.

Federated learning~\cite{GoogleFed,yang2019federated} is a distributed machine learning framework that provides an effective solution for protecting user privacy. Federated learning distributes a model to multiple participants, who collaboratively train the model. Each user's original data remains local and is never shared with other participants. Instead, the model parameters, such as gradients, are transmitted to facilitate training. Typically, federated models can achieve performance comparable to their corresponding un-federated counterparts. Therefore, federating existing recommendation models allows recommendation algorithms to function while protecting user privacy. Recently, many federated recommendation algorithms, such as FCF~\cite{FCF} and FedPerGNN~\cite{wu2022federated}, have been proposed, which apply federated learning to the existing recommendation algorithms.

Federated recommendation can be divided into cross-user federated recommendation (CUFR) and cross-organization federated recommendation (COFR) based on the participants~\cite{lin2022generic}. In CUFR, many individual users participate in model construction, typically using mobile devices with limited computational and communication resources. The server coordinates training and aggregates parameters, with sufficient capabilities to handle these tasks. In contrast, COFR involves fewer participants, like banks or hospitals, which have more robust and high-performance resources. In this paper, we focus on cross-user federated recommendation. It is worth noting that in cross-user federated learning, "client" and "user" refer to the same concept, so we will use them interchangeably from here on.

\begin{figure*}[ht]
	\subfigure[GNN-based centralized recommendation]{
		\label{fig:GNN-based centralized recommendations}
		\centering
		\includegraphics[width=0.42\linewidth]{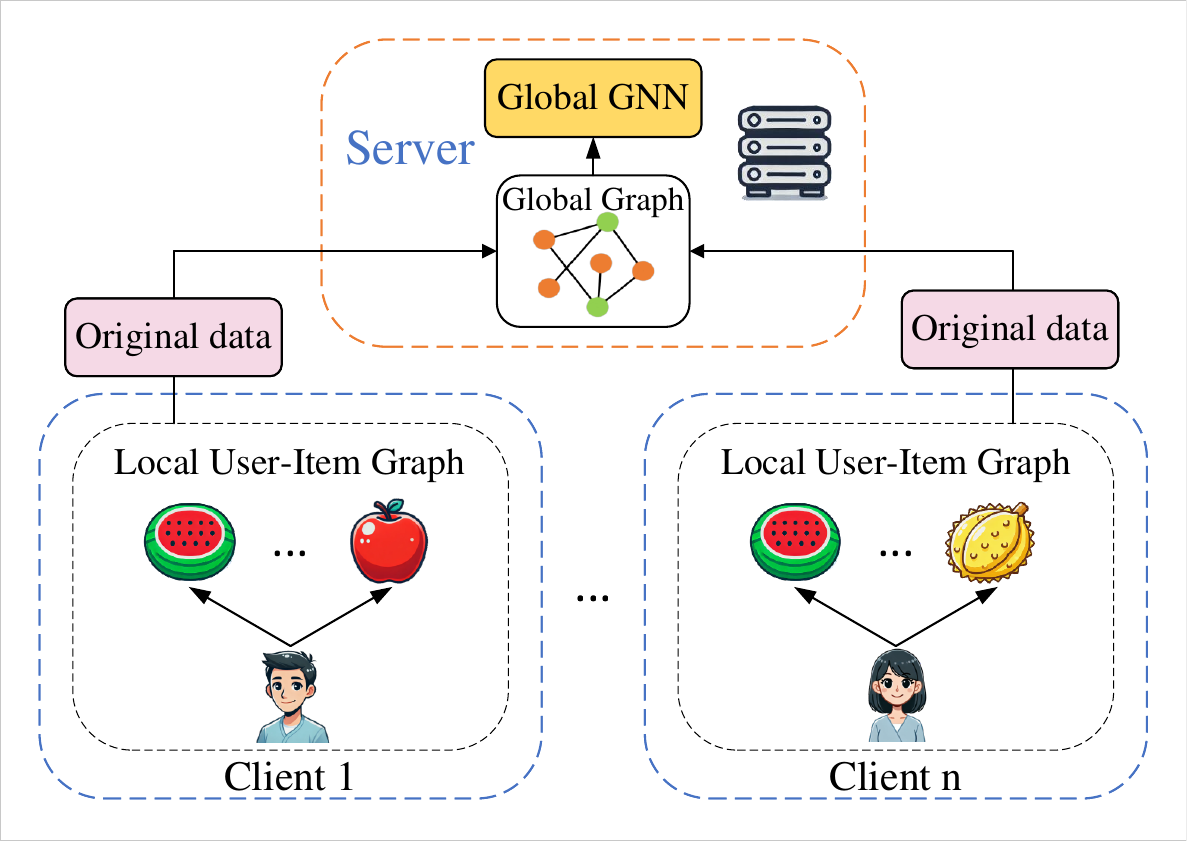}
	}%
	\subfigure[GNN-based federated recommendation]{
		\label{fig:GNN-based federated recommendation mode}
		\centering
		\includegraphics[width=0.548\linewidth]{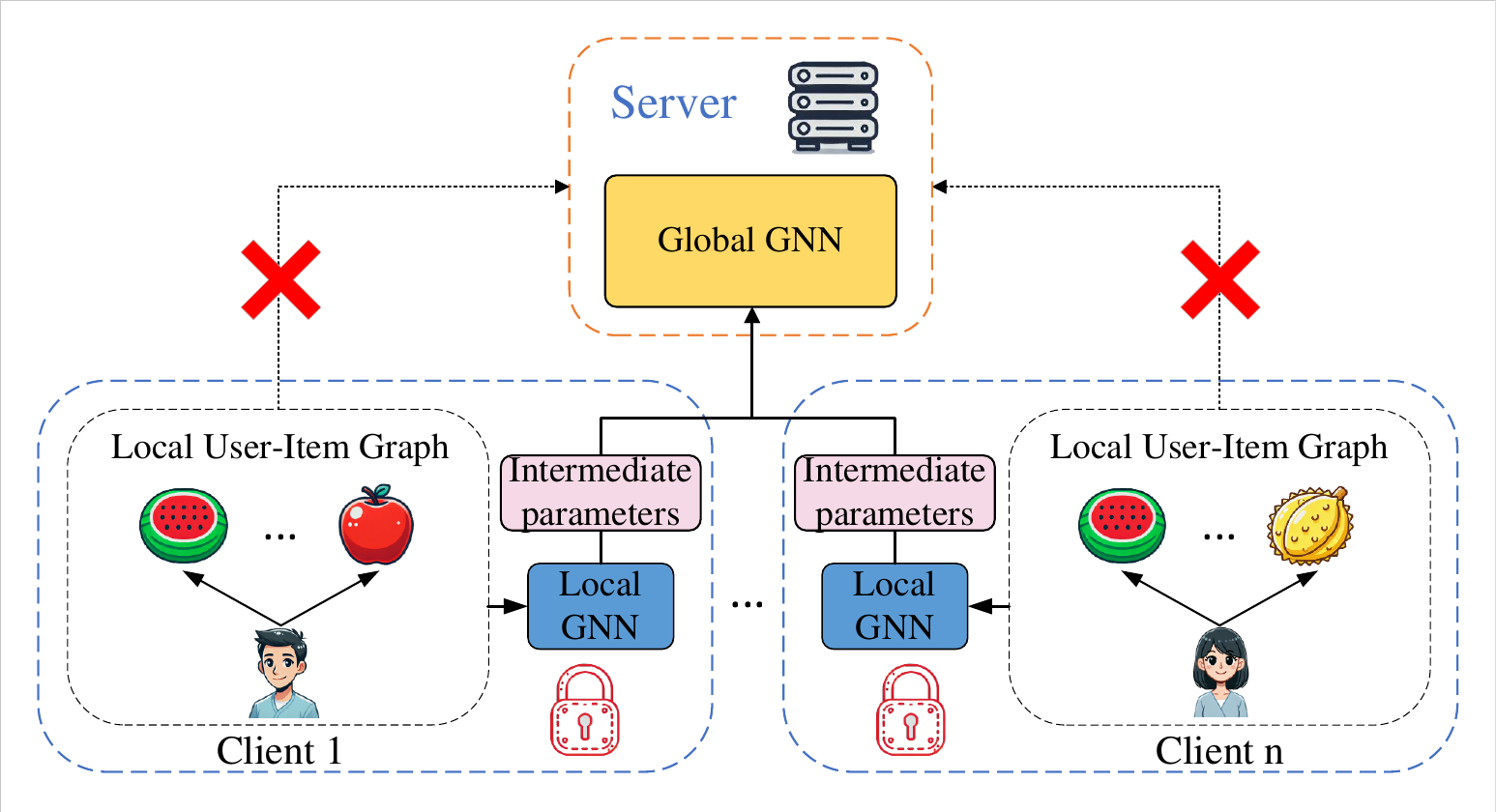}
	}
	\caption{Illustration of the centralized and federated versions of GNN-based recommendation.}
	\label{fig:Comparison of the centralized and federated versions of the GNN-based recommendation model}
	\Description{Comparison of the centralized and federated versions of the GNN-based recommendation model}
\end{figure*}
Graph neural network (GNN) is a deep learning model that operates on a graph-structured data, capturing relationships between nodes through a message-passing process. It can learn node representations by modeling complex relationships. Therefore, graph neural network (GNN) has been widely used in personalized recommendation algorithms because they are able to effectively capture higher-order interaction information, enhancing user and item representations. As a result, federating GNN-based recommendation algorithms have attracted considerable attention. In traditional centralized recommendation models, as shown in Figure~\ref{fig:GNN-based centralized recommendations}, users can directly send their original interaction data to the server, which builds a global user-item graph and trains a model. However, in a federated scenario, as shown in Figure~\ref{fig:GNN-based federated recommendation mode}, due to privacy concerns, users cannot directly send their interaction data to the server or other users. Instead, each user keeps a small local sub-graph containing his or her interaction behaviors, forming a bipartite graph between the user and the interacted items. We focus on this federated scenario, aiming to enable effective recommendation while preserving user privacy.

To address this issue, some GNN-based federated recommendation frameworks have been proposed~\cite{wu2022federated,luo2022personalized,hu2023privacy}. However, they face two main problems. Firstly, there is a risk of privacy leakage. These frameworks expand users' sub-graphs to obtain higher-order neighboring information by encrypting item-IDs or using group-wise concealing. Encrypting item-IDs relies on a third-party server, and group-wise concealing will be exposed to inference attacks. Secondly, they may suffer from poor performance. Due to privacy constraints, the graph convolution process is incomplete, resulting in a performance loss compared with the non-federated centralized counterparts.

In this paper, we propose a novel lossless federated graph recommendation framework called lossless and privacy-preserving graph convolution network (LP-GCN). Our LP-GCN fully executes the graph convolution process, including both forward propagation and backward propagation. Therefore, it can achieves performance equivalent to the corresponding non-federated centralized counterpart while effectively protecting user privacy. Specifically, we introduce a method combining hybrid-encrypted item-IDs and virtual item obfuscation, which enables the server to build a global user-item graph and assist participating clients with graph expansion. This method effectively protects user privacy, even in cases where the server colludes with some users. We propose an embedding-synchronization mechanism, where the embedding of a shared node is first computed on a specific client. This embedding is then synchronized via the server to other clients that also contain the same node, ensuring that no information is lost during the graph convolution process. Additionally, we provide theoretical proof and experimental validation showing that federating an existing algorithm with our LP-GCN achieves performance equivalent to the centralized counterpart. Furthermore, we conduct extensive experiments, including performance comparison, convergence studies, and communication cost analysis. The results demonstrate the rationality and effectiveness of our LP-GCN.
The main contributions of this paper are summarized as follows:
\begin{itemize}	
	\item We are the first to propose a lossless GNN-based federated recommendation framework called LP-GCN, which fully completes the graph convolution process using decentralized user-item sub-graphs while safeguarding privacy. It achieves performance equivalent to the non-federated centralized counterpart.
	
	\item 
	We propose a method combining hybrid-encrypted item-IDs and virtual item obfuscation to address the privacy and security issues in expanding user sub-graphs, which remains secure even in cases where the server colludes with some users.
	
	\item 
	We propose an embedding-synchronization mechanism that ensures a lossless graph convolution process, including both forward propagation and backward propagation. Additionally, we have validated this mechanism through both theoretical analysis and experimental verification.
	\item 
	We conduct extensive experiments on three real-world datasets, evaluating the recommendation performance and communication cost. The results show that our LP-GCN outperforms the existing methods while maintaining strong privacy protection.
\end{itemize}
The remainder of this paper is organized as follows: In Section~\ref{sec:related_work}, we review the related works. Section~\ref{sec:preliminaries} provides the preliminaries, including the problem definition, an introduction to the seminal GNN-based recommendation method LightGCN, its improved version LightGCN+, and a discussion of the existing encryption technologies. In Section~\ref{sec:LP-GCN}, we present a detailed description of our LP-GCN. Section~\ref{sec:analysis} provides an analysis of privacy protection and communication cost. In Section~\ref{sec:experiments}, we present empirical studies conducted on three datasets. Finally, we conclude the paper and discuss some promising future research directions in Section~\ref{sec:conclusions}.
\section{Related Work} \label{sec:related_work}
\subsection{Conventional Recommendation}
Conventional recommender systems typically do not take privacy protection into account in user behaviors modeling. Collaborative filtering (CF) is a widely used technique in recommender systems~\cite{covington2016deep,ying2018graph}.
A common paradigm in CF models is to represent users and items as embeddings, which are then used to reconstruct the  historical user-item interactions.
Earlier CF models such as matrix factorization (MF)~\cite{koren2009matrix,rendle2012bpr} map the ID of a user (or an item) to an embedding vector.
To improve user representations, FISM~\cite{kabbur2013fism} and SVD++~\cite{koren2008factorization} represent a target user's embedding by computing the weighted average of the ID embeddings from the user's previously interacted items.
Additionally, models like NCF~\cite{he2017neural} and LRML~\cite{tay2018latent} leverage neural networks to enhance behaviors modeling with a similar embedding component.
Non-graph methods can be seen as the one-hop version of graph-based methods. Specifically, representing the historical interactions as a user-item bipartite graph shows that performance improvement comes from encoding local neighborhoods (i.e., one-hop neighbors), which enhances the embedding learning process.
To further explore higher-order information, graph-based modeling methods have emerged. GCN~\cite{kipf2016semi} is one of the most effective convolution approaches and is widely studied in the community of recommender systems.
For example, models like NGCF~\cite{wang2019neural} and PinSage~\cite{ying2018graph} adapt GCN to exploit the user-item interaction graphs, effectively capturing collaborative filtering signals from higher-hop neighbors to enhance the recommendation performance.
Based on NGCF, LightGCN~\cite{he2020lightgcn} eliminates redundant components, i.e., the non-linear activation and feature transformation, resulting in better performance and establishing itself as a highly practical model.
KGAT~\cite{wang2019kgat} introduces attention mechanisms into graph convolution, leading to improved results.
GraphRec~\cite{fan2019graph} integrates a social graph with a user-item interaction graph, embedding social network information to enhance the recommendation performance.
Dual-LightGCN~\cite{huang2023dual} improves performance by shifting GNN aggregation from a single-scale space to a multi-scale space.
\subsection{General Federated Recommendation}
General federated recommendation methods often do not construct a global user-item bipartite graph because that will contradict the core principle, i.e., the user data should remain local. Non-GNN-based CF methods can roughly be divided into matrix factorization (MF) and neural networks (NN), both of which have garnered significant attention. Some recent works extend classical centralized MF methods to federated versions, such as FedMF~\cite{chai2020secure}, FedRec~\cite{lin2020fedrec}, and FedRec++~\cite{liang2021fedrec++} for recommendation with explicit feedback, and FedeRank~\cite{anelli2021federank} for that with implicit feedback. Similarly, to adapt deep learning-based methods to federated settings, FedNCF~\cite{jiang2022fedncf} and FedVAE~\cite{polato2021federated} have been developed, which are based on the NCF~\cite{he2017neural} and Mult-VAE~\cite{shenbin2020recvae}, respectively. FedNCF employs secure aggregation~\cite{bonawitz2017practical} to protect model updates from inference attacks by the server, thus safeguarding user privacy. As for FedVAE~\cite{polato2021federated}, it assumes that the noise introduced by dropout ensures differential privacy~\cite{kang2020input}, providing protection for model updates.
Recently, a generic federated recommendation framework named FR-FMSS~\cite{lin2022generic} is proposed, which leverages secret sharing~\cite{shamir1979share} and fake masks~\cite{lin2020fedrec}. This framework can federate various non-GNN-based methods, including those based on MF and deep learning. However, it does not address how to federate GNN-based methods or how to exploit high-order connectivity information in a privacy-preserving manner.
LightFR~\cite{zhang2023lightfr} focuses on improving the efficiency of the model by storing user and item vectors in a compact manner and optimizing similarity search during the inference phase. DFMR~\cite{yang2024discrete} prioritizes both efficiency and performance. Specifically, it addresses the item ranking task with heterogeneous implicit feedback by leveraging the differences between actions like views and purchases to enhance the recommendation performance.
Besides, there are some works focusing on personalized modeling to enhance the performance. PFedRec~\cite{zhang2023dual} addresses user-side personalization by introducing a bipartite personalization mechanism. FedRAP~\cite{li2023federated} further improves performance by applying a bipartite personalized algorithm to manage data heterogeneity from diverse user behaviors, while considering both unique and shared item information.
\subsection{GNN-based Federated Recommendation}
Recently, there are some initial attempts to combine GNN-based models with federated recommendation, aiming to build models that ensure privacy protection~\cite{wu2022federated,qu2023semi,hu2023privacy,yan2024federated,zhang2024gpfedrec}. GNN-based CF methods can exploit high-order information, which often lead to better performance compared with non-GNN-based methods. As a result, federating these methods has recently gained more attention. Although federating GNN-based methods is challenging, some researchers    propose various GNN-based federated recommendation methods.

Wu et al.~\cite{wu2022federated} propose the first GNN-based federated recommendation framework called FedPerGNN, which uses item-ID encryption to expand each user's local sub-graph for high-order neighbors. Users send encrypted item-IDs and embeddings to a third-party server, which returns neighboring users' embeddings for second-order modeling. However, it is of two limitations. Firstly, due to the involvement of a third-party server, privacy concerns arise, making it difficult to be deployed in real-world scenarios~\cite{qu2023semi,hu2023privacy}. Secondly, to protect privacy, fake items and local differential privacy (LDP) noise are introduced during gradient aggregation. As a result, the graph convolution process becomes inaccurate compared with the centralized counterpart, affecting both forward propagation and backward propagation, which leads to a decline in performance.

Based on FedPerGNN, PerFedRec~\cite{luo2022personalized} further incorporates user and item attributes by combining a user-side end-to-end local recommendation network with a server-side clustering-based aggregator, which generates embeddings and builds personalized models for each user, enhancing the overall recommendation performance. However, akin to that of FedPerGNN, it obtains high-order connectivity information through a third-party server, raising the similar privacy concerns.

FedGRec~\cite{li2022fedgrec} not only focuses on better exploiting indirect user-item interactions but also considers reducing time consumption by avoiding homomorphic encryption used in FedPerGNN. Specifically, it captures missing indirect interactions by storing the latent embeddings on both the local devices and server. Users update these embeddings locally, while the latent embeddings are synchronized only during communication between the server and clients.

Compared with FedPerGNN, SemiDFEGL~\cite{qu2023semi} eliminates the need for a third-party server, offering enhanced privacy protection. Additionally, it introduces a novel device-to-device collaborative learning mechanism to improve scalability and reduce the communication cost typical of conventional federated recommendation methods. Specifically, it stores the ego graph on the device while item data remains on the server. Each device generates embeddings for its ego graph and uploads them to the server. The server then clusters users and items, connecting users within each group through predicted interacted item nodes, forming a higher-order local sub-graph.

The above methods, though capable of capturing higher-order information through local sub-graph expansion on the client side, incur losses during the graph convolution process compared with the centralized counterparts, leading to performance degradation. P-GCN~\cite{hu2023privacy} can successfully achieved lossless forward propagation, meaning that its forward propagation is equivalent to that of the corresponding centralized counterpart. However, due to its privacy protection, the gradient back propagation cannot be fully completed. Moreover, though P-GCN uses group-wise concealing to protect privacy, it still suffers from attribute inference attacks. Our LP-GCN effectively addresses these two issues, i.e., it not only preserves privacy well but also fully completes the graph convolution process, including both forward and back propagation, achieving greater performance improvement and faster convergence.

Additionally, some similar works have introduced some extra information, such as social relationships in preference learning. For example, FesoG~\cite{liu2022federated} uses first-order connectivity (i.e., interacted items and social neighbors), where each client employs heterogeneous graph attention layers and relational graph aggregation layers to extract the user representations. FedGR~\cite{ma2023fedgr} utilizes social relationship graphs and item interaction graphs to learn the user and item representations, further enhancing performance. 
FedHGNN~\cite{yan2024federated} utilizes a heterogeneous information network (HIN), which captures rich semantics through meta-paths, to enhance performance. However, both methods introduce noise to protect privacy: FesoG and FedHGNN use pseudo items and LDP during item gradient aggregation, while FedGR applies noise injection. As a result, their graph convolution processes are not lossless compared with the centralized methods, leading to lower performance.

There are some relevant works in the field of GNN-based federated learning~\cite{zhang2021subgraph,wang2022federatedscope,tan2023federated,chen2024fedgl}. However, their tasks differ from those in GNN-based federated recommendation, making them unable to be applied to recommendation tasks directly. 
Additionally, in the cross-organization recommendation scenario, there are also relevant works~\cite{mai2023vertical,liu2023federated,han2024subgraph,tian2024privacy}. However, these works primarily focus on the cross-organization federated recommendation setting, where each organization holds a large sub-graph. As a result, in these works, each organization can naturally model high-order connectivity information locally.
However, in the cross-user federated recommendation scenario, which is the focus of this study, each user only holds their own user-item bipartite sub-graphs without high-order connectivity. Therefore, these methods cannot be applied to the cross-user federated recommendation problem as well.
\begin{table}[htbp]
	\caption{A brief summary of the existing GNN-based federated recommendation methods.}
	\label{tab:problems}
        \centering
        \fontsize{9.5}{10}\selectfont  %
\begin{tabular}{>{\centering\arraybackslash}p{2.3cm} >{\centering\arraybackslash}p{2.8cm} >{\centering\arraybackslash}p{2.9cm} >{\centering\arraybackslash}p{3.1cm} >{\centering\arraybackslash}p{2cm}}
		\toprule
		\multirow{2}{*}{Algorithm} & Remove the need of            & \multicolumn{2}{c}{Graph convolution} & \multirow{2}{*}{Lossless}                                      \\
          &    a third-party server                              & Forward propagation                     & Backward propagation      & 		                                                                           \\
		\midrule
		FedPerGNN \cite{wu2022federated}                     & $\times$                               & $\times$                  &$\times$                               &       $\times$\\
		PerFedRec \cite{luo2022personalized}                     & $\times$                               & $\times$                  & $\times$                               & $\times$                              \\
		FedGRec \cite{li2022fedgrec}                 & $\surd$                              & $\times$                  & $\times$                                & $\times$                              \\
		SemiDFEGL \cite{qu2023semi}                     & $\surd$                               & $\times$                  &$\times$                               & $\times$                            \\
		FeSoG \cite{liu2022federated}                     & $\surd$                               & $\times$                  & $\times$                               & $\times$                            \\
  		FedHGNN \cite{yan2024federated}                     & $\surd$                               & $\times$                  & $\times$                               & $\times$                            \\
		FedGR\cite{ma2023fedgr}                 & $\surd$                              & $\times$                  & $\times$                                & $\times$                             \\
		P-GCN \cite{hu2023privacy}                 & $\surd$                               & $\surd$                   & $\times$                                & $\times$                               \\
		Our LP-GCN                                & $\surd$                                & $\surd$                   & $\surd$                                & $\surd$ \\
		\bottomrule
	\end{tabular}
\end{table}

We briefly summarize the discussed works in Table~\ref{tab:problems}. In summary, existing GNN-based federated methods fail to protect user privacy well during the modeling process. Additionally, the need to introduce noise for privacy protection during the graph convolution process results in a non-lossless operation, leading to a decline in performance compared with the corresponding non-GNN-based centralized counterpart, which motivates us to propose a lossless GNN-based federated recommendation framework to address the above issues.
\section{Preliminaries}\label{sec:preliminaries}
In this section, we first define the problem. Then, we discuss the threat models. After that, we introduce the base model, i.e., LightGCN, and its enhanced version LightGCN+, which will be used as the backbone model for our LP-GCN. Finally, we introduce some encryption technology, which will be used later. 

Some notations and their explanations are shown in Table~\ref{tab:notations}.
\begin{table}
	\centering
	\caption{Notations and their explanations.}
	\label{tab:notations}
	
	\begin{tabular}{l|l} 
		\hline\hline
		
		$n$ & the number of users\\
		
		$m$ & the number of items\\
  
		  $u$ & the user ID \\
    
		$\mathcal{U}$ & the set of all users\\
  
		$i$ & the item ID \\

		$\mathcal{I}$ & the set of all items\\
  
		$\tilde{u}$ & the client $u$ who is denoted as a convolution-client\\

  		$\tilde{i}$ & the item $i$ which is denoted as a convolution-item\\
    
  		$\tilde{u}_{\tilde{i}}$ & the convolution-client that is responsible for calculating the embedding of the convolution-item $\tilde{i}$\\
    
		$\mathcal{I}_u$ & the set of items that are interacted by user $u$\\
			
            $\mathcal{I}_u'$ & the set of items that are virtually interacted by user $u$\\

		$\mathcal{U}_i$ & the set of users that interact with item $i$\\
  
  		$\mathcal{\tilde{U}}=\{ \tilde{u}\}$ & the set of all convolution-clients\\ 
    
            $\mathcal{\tilde{I}}=\{ \tilde{i}\}$ & the set of all convolution-items\\ 

            $\mathcal{\tilde{I}}_{\tilde{u}}$ & the set of all convolution-items of the convolution-client $\tilde{u}$\\ 
            
            $\boldsymbol{e}_{u'(u)}^{(l)}$ & the $l$-th layer's user embedding of user $u'$ stored in client $u$ \\

            $\boldsymbol{e}_{i(u)}^{(l)}$ & the $l$-th layer's embedding of item $i$ stored in client $u$ \\

            $\boldsymbol{g}_{u'(u)}^{(l)}$ & the $l$-th layer's gradient of the user embedding of user $u'$ stored in client $u$ \\

            $\boldsymbol{g}_{i(u)}^{(l)}$ & the $l$-th layer's gradient of the embedding of item $i$ stored in client $u$ \\
            
            $\mathcal{N}_{\tilde{u}}$ & convolution-item based neighboring users of client $\tilde{u}$, i.e., $\bigcup_{\tilde{i} \in \mathcal{\tilde{I}}_{\tilde{u}}} \left( \mathcal{U}_{\tilde{i}} \setminus \{\tilde{u}_{\tilde{i}}\} \right)$\\
            
            $S$ & a shared key generated by symmetric encryption\\
            
		$ID_{i(u)}$ & the ID of item $i$ which is stored in client $u$\\

            $S(ID_{i(u)})$ & the ciphertext of $ID_{i(u)}$ which is encrypted by the shared key $S$\\
		\hline
  		$T$                       & the iteration number                                                                    \\
		$d$                       & the number of latent dimensions                                                         \\
		$\gamma$                  & the learning rate                                                                       \\
		
		$L$ & the number of graph convolution layers\\
		
		$l\in \{0,1,...,L\}$ & the graph convolution layer ID\\

		\hline\hline
	\end{tabular}
\end{table}
\subsection{Problem Definition}
In this paper, we focus on GNN-based federated recommendation. In a centralized GNN-based recommender system, all users' data (i.e., historical interactions $\textbf{R} \in \{0,1\}^{n\times m}$) is collected by the server to build a global graph used for training the model, without considering privacy concerns. However, a GNN-based federated model aims to provide recommendation while protecting user privacy. Specifically, during the entire process, the server can not access any user's interaction data, nor can users see each other's interaction data.
From the perspective of graph structure, we compare the centralized and federated versions of a GNN-based recommendation model as illustrated in Figure~\ref{fig:Comparison of the centralized and federated versions of the GNN-based recommendation model}. In the centralized version, the server constructs a global graph by collecting all users' interaction data. However, in the federated version, considering the privacy of each user, the server could not collect the users' interaction data. Each client maintains a private sub-graph consisting of the interactions w.r.t. to the interacted items. Generally, the local GNN-based sub-models and the global model are constructed in the clients and server, respectively. Gradient aggregation, distribution, and other operations are completed through communication between the server and clients, allowing for the update of the local parameters on clients and the update of the global part on the server. Our objective is to rank the items $\mathcal{I} \backslash \mathcal{I}_u$ that each user $u$ has not interacted with before and generate a list of top-$K$ items that may be of interest to each user $u$. In this scenario, a GNN-based model can be built by transmitting the model parameters, i.e., the embeddings and gradients between the server and clients. 
\begin{figure*}[ht]
	\subfigure[Centralized version]{
		\label{fig:Centralized version}
		\centering
		\includegraphics[width=0.33\linewidth]{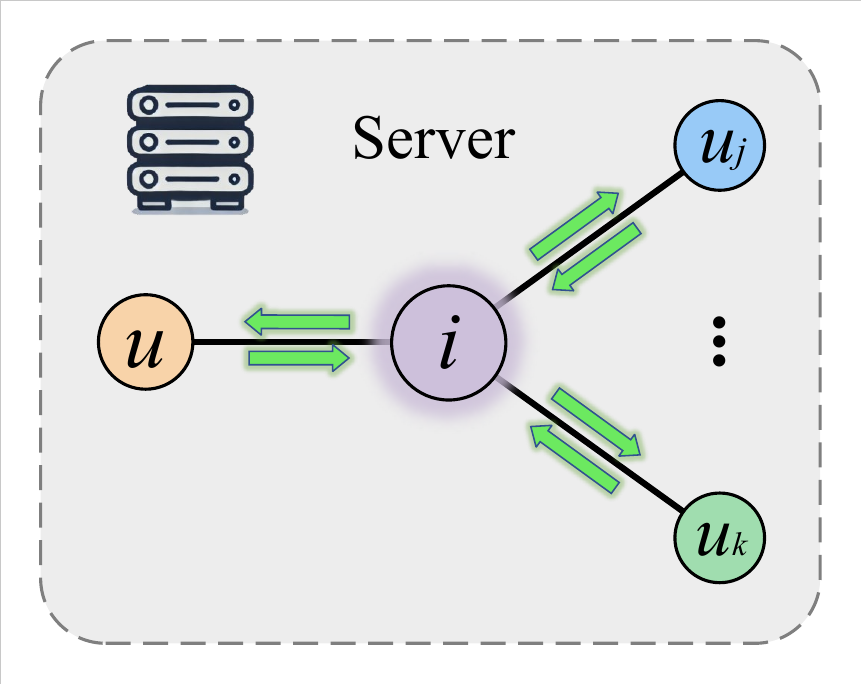}
	}%
	\subfigure[Federated version]{
		\label{fig:Federated version}
		\centering
		\includegraphics[width=0.377\linewidth]{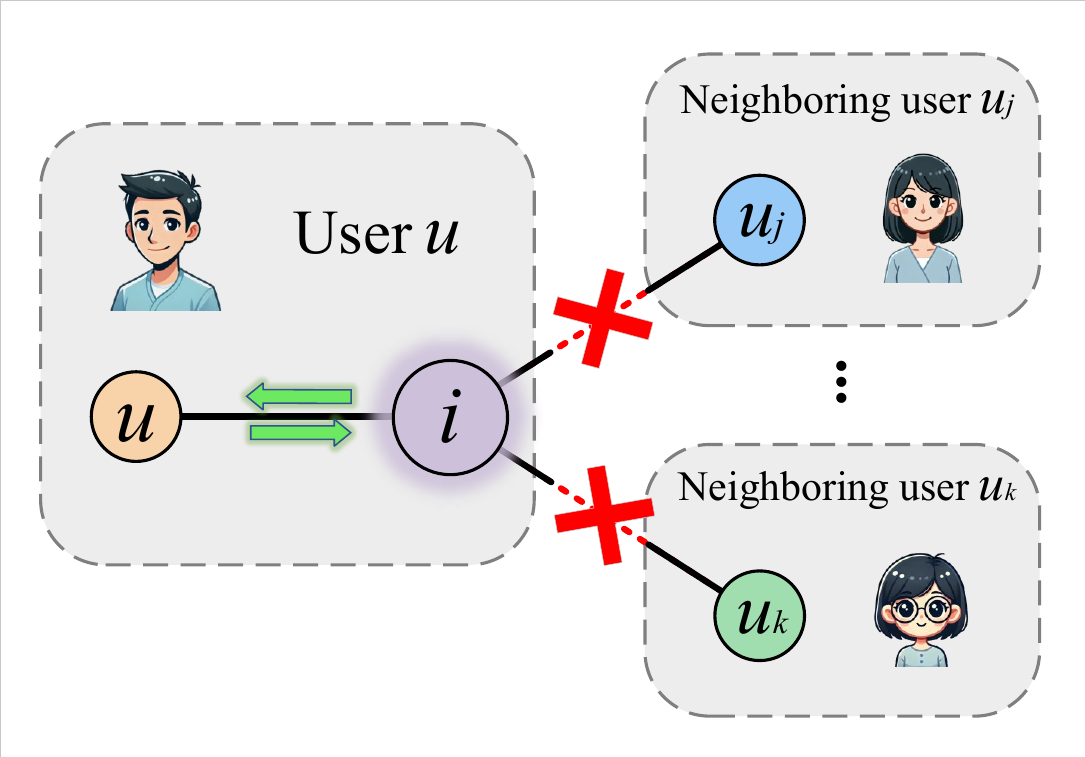}
	}
	\caption{Illustration of the centralized and federated versions of forward propagation and backward propagation w.r.t. item $i$.}
	\label{fig:Comparison of the centralized and federated versions of propagation of item $i$ node}
	\Description{Comparison of the centralized and federated versions of propagation of item $i$ node}
\end{figure*}
However, due to the fragmented nature of the local data and the privacy concerns, it becomes challenging to exploit the structural information as effectively as that in a fully centralized user-item graph when propagating the embeddings. Specifically, the most challenging issue is completing the propagation of item nodes within the local sub-graph. For an item node $i$, in the centralized version, both forward propagation and backward propagation of an item node $i$ can be handled by the server, as shown in Figure~\ref{fig:Centralized version}. However, in the federated version, item $i$ is stored in client $u_i$, the neighboring users' nodes are distributed across different clients. Hence, client $u_i$ cannot directly communicate with them due to privacy concerns, which results in the disruption of both forward propagation and backward propagation paths, as shown in Figure~\ref{fig:Federated version}. Therefore, it is very challenging to exploit the high-order user-item interactions without risking privacy leakage. 
\subsection{Threat models}
We assume a semi-honest threat model, like previous works~\cite{lin2022generic,PNSMF,hu2023privacy}, which means that the server and clients are honest but curious. They follow the protocol correctly but may try to learn some sensitive information from the data they receive.
Note that we do not consider the malicious threat model because both the server and clients benefit more from following the protocol honestly. Unlike the medical or financial industries, recommender systems often only need to address the semi-honest model to achieve adequate privacy protection, along with acceptable computational and communication cost~\cite{semi1,semi2}.

In our LP-GCN, it is crucial to protect the user-item interactions and the user embeddings. The user embeddings, in particular, need to be protected as they reflect user preferences and can lead to privacy leaks~\cite{hu2023privacy}. Each user's interactions with items should remain confidential and not be disclosed to the server or other users\footnote{We use the terms user and client interchangeably.}.

Under the semi-honest threat model, three typical attacks on machine learning models are reconstruction attacks, model inversion attacks, and membership inference attacks. We will discuss how our LP-GCN defends against these attacks in Section~\ref{sec:Privacy Analysis}.

\subsection{LightGCN}
LightGCN~\cite{he2020lightgcn} is a state-of-the-art GNN designed specifically for enhancing the item recommendation performance. It leverages the rich connectivity information inherent in a user-item interaction graphs to learn more precise representations of users and items, thus improving the recommendation performance. Next, we will introduce this model specifically from the following five aspects: initial embeddings, graph convolution operations, final embeddings, preference prediction, and training.

{\noindent \bf Initial Embeddings:} Each user $u$ and each item $i$ are represented by an initial embedding $U_{u}^{(0)} \in \mathbb{R}^{1\times d}$ and $V_{i}^{(0)} \in \mathbb{R}^{1\times d}$, respectively. These embeddings are essentially the raw features of the users and items.

{\noindent \bf Graph Convolution Operations:} LightGCN simplifies the traditional graph convolution process by removing the nonlinear activation functions and feature transformations. This is reflected in the iterative update formulas~\cite{he2020lightgcn},
\begin{equation} \label{eq:lightGCN compute user embedding}
	U_{u}^{(l)} = 
	\sum_{i \in \mathcal{I}_u} \frac{1}{\sqrt{|\mathcal{I}_u|} \sqrt{|\mathcal{U}_i|}} V_{i}^{(l-1)}
\end{equation}

\begin{equation} \label{eq:lightGCN compute item embedding}
	V_{i}^{(l)} = 
	\sum_{u \in \mathcal{U}_i} \frac{1}{\sqrt{|\mathcal{U}_i|} \sqrt{|\mathcal{I}_u|}} U_{u}^{(l-1)}
\end{equation}
Here, $\mathcal{I}_u$ denotes the set of items interacted with by user $u$, and $\mathcal{U}_i$ denotes the set of users who have interacted with the item $i$. This aggregation process aims to refine the embeddings by incorporating the neighborhood information directly, thus capturing the collaborative filtering effects more naturally and efficiently.

{\noindent \bf Final Embeddings:} After $L$ layers of convolution, the final embeddings of the users and items are computed as a weighted sum of all layer embeddings,
\begin{equation} \label{eq:combine}
	U_{u} = \sum_{l=0}^{L} \alpha_l U_{u}^{(l)},\quad
	V_{i} = \sum_{l=0}^{L} \alpha_l V_{i}^{(l)}
\end{equation}
where $\alpha_l$ is the weight assigned to the embeddings for the $l$-th layer, typically set to $1/(L+1)$ to ensure a uniform influence across all layers.

{\noindent \bf Preference Prediction:} The preference of a user for an item is predicted by the inner product of their final embeddings,
\begin{equation} \label{eq:pred}
	\hat{y}_{ui} = U_{u} \cdot V_{i}
\end{equation}
{\noindent \bf Training:} The model parameters, primarily the initial embeddings, are trained using the Bayesian personalized ranking (BPR) loss,
\begin{equation} \label{eq:loss}
	\mathcal{L}_{BPR}
	= - \sum_{u \in \mathcal{U}} \sum_{i \in \mathcal{I}_u} \sum_{j \notin \mathcal{I}_u}
	\Big(
	\text{ln }\sigma(\hat{y}_{ui} - \hat{y}_{uj}) + \lambda \textbf{Reg}
	\Big)
\end{equation}
where $\textbf{Reg}=\|U_{u}^{(0)}\|^2 + \|V_{i}^{(0)}\|^2 + \|V_{j}^{(0)}\|^2$ is the regularization term, and $\lambda$ is the regularization coefficient.

\subsection{LightGCN+}\label{LightGCN+}
FISM (factored item similarity models)~\cite{kabbur2013fism} is a matrix factorization (MF) method, which constructs a user's embedding from his or her interacted items' embeddings to enhance the recommendation performance.
Inspired from FISM, P-GCN~\cite{hu2023privacy} also constructs a user's embedding from his or her interacted items' embeddings via Eq.(\ref{eq:construct_user_embedding}) to enhance the performance.

Inspired from FISM and P-GCN, we only make the following improvement to the Initial Embeddings in LightGCN.

We initialize the two sets of item embeddings, denoted as $W^{(0)} \in \mathbb{R}^{m\times d}$ and $V^{(0)} \in \mathbb{R}^{m\times d}$, where $m$ is the number of items. Note that $W^{(0)}$ is used to construct the user embeddings via Eq.(\ref{eq:construct_user_embedding}), while the $V^{(0)}$ is used to represent the item embeddings. The embedding of user $u$ and item $i$ is denoted as $U_{u}^{(0)} \in \mathbb{R}^{1\times d}$ and $V_{i}^{(0)} \in \mathbb{R}^{1\times d}$, respectively. 
\begin{equation} \label{eq:construct_user_embedding}
	U_{u}^{(0)} = \frac{1}{\sqrt{|\mathcal{I}_u|}} \sum_{i \in \mathcal{I}_u} W_{i}
\end{equation}
\subsection{Encryption Technology}

\subsubsection{Symmetric encryption~\cite{yassein2017comprehensive}}Symmetric encryption, also known as symmetric key encryption, uses the same key for both encryption and decryption of data. This means that both the sender and the receiver must have access to the same secret key to ensure secure communication. Symmetric encryption algorithms are generally faster and more efficient for processing large amounts of data compared with asymmetric encryption. Common symmetric encryption algorithms include advanced encryption standard (AES), data encryption standard (DES), and triple DES (3DES).
One major drawback of symmetric encryption is the key distribution problem. Since both parties must share the same key, securely distributing and managing these keys can be challenging, especially over insecure channels.

\subsubsection{Asymmetric encryption~\cite{yassein2017comprehensive}}Asymmetric encryption, also known as public-key encryption, uses a pair of keys—a public key and a private key. The public key is used to encrypt the data, while the private key is used to decrypt it. This approach eliminates the key distribution problem inherent in symmetric encryption, as the public key can be openly shared while the private key remains confidential.
Asymmetric encryption is widely used for securing communications over the internet, including email encryption and digital signatures. Common algorithms include RSA (rivest-shamir-adleman), ECC (elliptic curve cryptography), and DSA (digital signature algorithm). Although it provides stronger security mechanisms, asymmetric encryption is computationally more intensive and slower compared with symmetric encryption.
\subsubsection{Hybrid encryption~\cite{zhang2021overview}} Considering the characteristics of both symmetric and asymmetric encryption, while symmetric encryption is advantageous due to its high computational speed, making it ideal for encrypting large amounts of data, it is not as secure as asymmetric encryption. The use of the same key for both encryption and decryption in symmetric algorithms introduces potential security risks. However, protecting the key can enhance security. A practical approach is to use an asymmetric algorithm to secure the key and then use a symmetric algorithm to encrypt the message. This hybrid encryption technique is becoming increasingly widespread, and our framework will also incorporate a hybrid encryption technology.
\section{Our Solution: LP-GCN} \label{sec:LP-GCN}
\begin{figure*}[ht]
	\centering
	\includegraphics[width=0.9\linewidth]{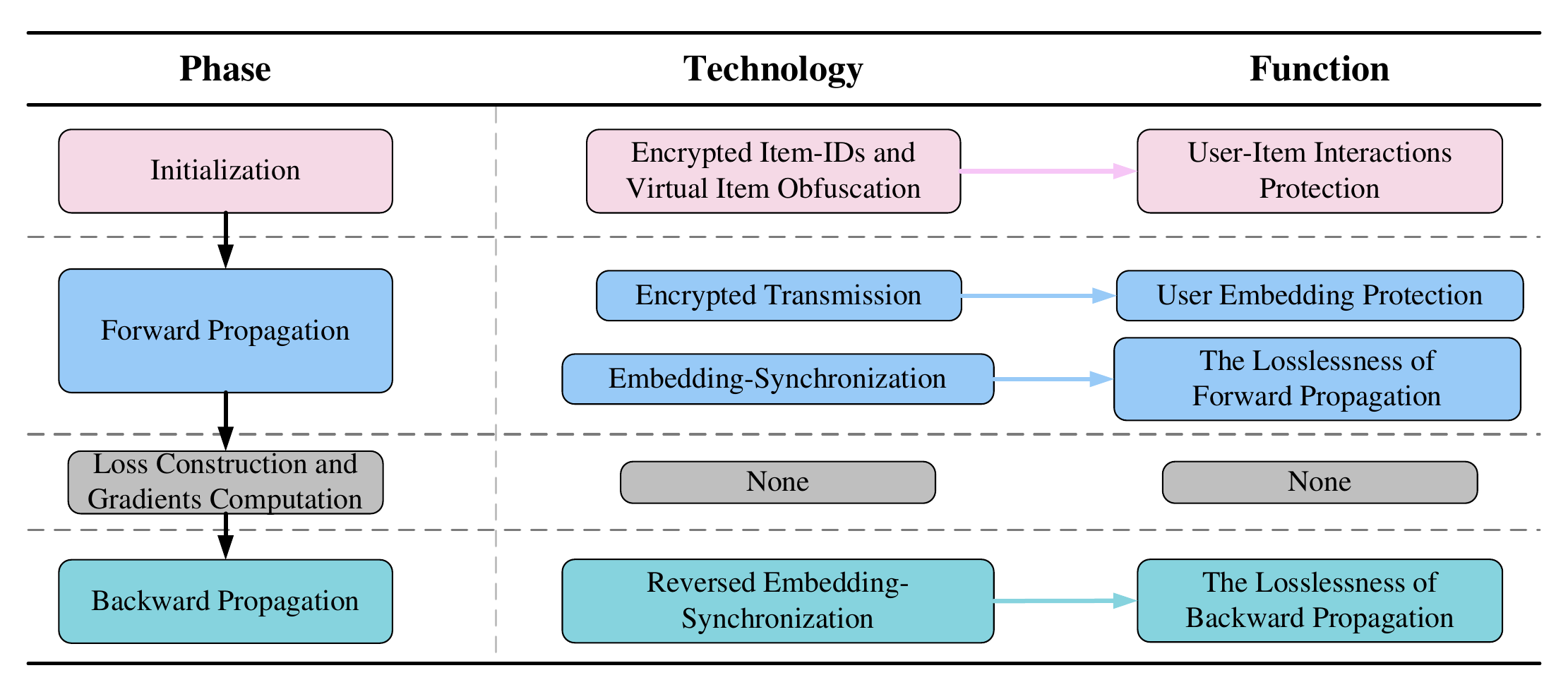}
	\caption{The overview of our LP-GCN.}
	\label{fig:Overview of our LP-GCN}
	\Description{Overview of our LP-GCN}
\end{figure*}

In our LP-GCN, a server coordinates the model training process, while each client stores his or her sensitive user-item interaction data locally. We build a GNN-based recommendation model by transmitting the model parameters between the server and clients. We show the overview of our LP-GCN in Figure~\ref{fig:Overview of our LP-GCN}. Specifically, the training process is divided into four phases: Initialization, Forward Propagation, Loss Construction and Gradients Computation, and Backward Propagation. In each phase, we employ some specific techniques to ensure the lossless nature of the graph convolution process while protecting user privacy.

Next, we will give the overall framework of our LP-GCN, and then provide a detailed introduction to each module in the order of their construction, named (i) Initialization (Section~\ref{sec:Initialization}), (ii) Forward Propagation (Section~\ref{sec:Forward Propagation}), (iii) Loss Construction and Gradients Computation (Section~\ref{sec:Loss Construction and Gradients Computation}), (iv) Backward Propagation (Section~\ref{sec:Backward Propagation}), and (v) Parameter Update and Prediction (Section~\ref{sec:Parameter Update and Prediction}).

\subsection{The Overall Framework of Our LP-GCN}\label{sec:The Overall Framework of Our LP-GCN}
\begin{algorithm}
\caption{Lossless and Privacy-Preserving Graph Convolution Network (LP-GCN)}
\label{LP-GCN}
\begin{algorithmic}[1] 
\STATE Initialization(), i.e., Algorithm~\ref{Initialization}
\FOR{ $ t=1, 2, \ldots, T$ }
    \FOR{each convolution-client $\tilde{u} \in \tilde{\mathcal{U}}$ in parallel}
    \STATE  Sends the $0$-th layer's convolution-item embeddings $\{\boldsymbol{e}^{(0)}_{\tilde{i}(\tilde{u})} : \tilde{i} \in \tilde{\mathcal{I}}_{\tilde{u}}\}$ to the server.
    \ENDFOR
    \STATE The server receives all items' embeddings of the $0$-th layer $\{\boldsymbol{e}_{\tilde{i}}^{(0)} : \tilde{i} \in \mathcal{I}\}$ from all convolution-clients and distributes $\boldsymbol{e}_{\tilde{i}}^{(0)}$ to clients $\{u' : u' \in \mathcal{U}_{\tilde{i}} \backslash \{\tilde{u}_{\tilde{i}}\}\}$ (i.e., the users who has interacted with item $\tilde{i}$ exclude $\tilde{u}$ where $\tilde{i}$ locates).
    \STATE Forward Propagation, i.e., Algorithm~\ref{Forward Propagation}.
    \STATE Loss Construction and Gradients Computation, i.e., Algorithm~\ref{Loss Construction and Gradients Computation}.
    \STATE Backward Propagation, i.e., Algorithm~\ref{Backward Propagation}.
    \FOR{each client $u \in \mathcal{U}$ in parallel}
    \STATE Updates the user embedding $\boldsymbol{e}^{(0)}_{u(u)} \leftarrow \boldsymbol{e}^{(0)}_{u(u)} - \gamma \boldsymbol{g}^{(0)}_{u(u)}$.
    \IF{$u \in \tilde{\mathcal{U}}$}
    \STATE Updates the convolution-item embeddings $\boldsymbol{e}^{(0)}_{i(u)} \leftarrow \boldsymbol{e}^{(0)}_{i(u)} - \gamma \boldsymbol{g}^{(0)}_{i(u)}, i \in \tilde{\mathcal{I}}_u$.
    \ENDIF
    \ENDFOR
\ENDFOR
\end{algorithmic}
\end{algorithm}

\begin{figure*}
	\centering
	\includegraphics[width=0.99\linewidth]{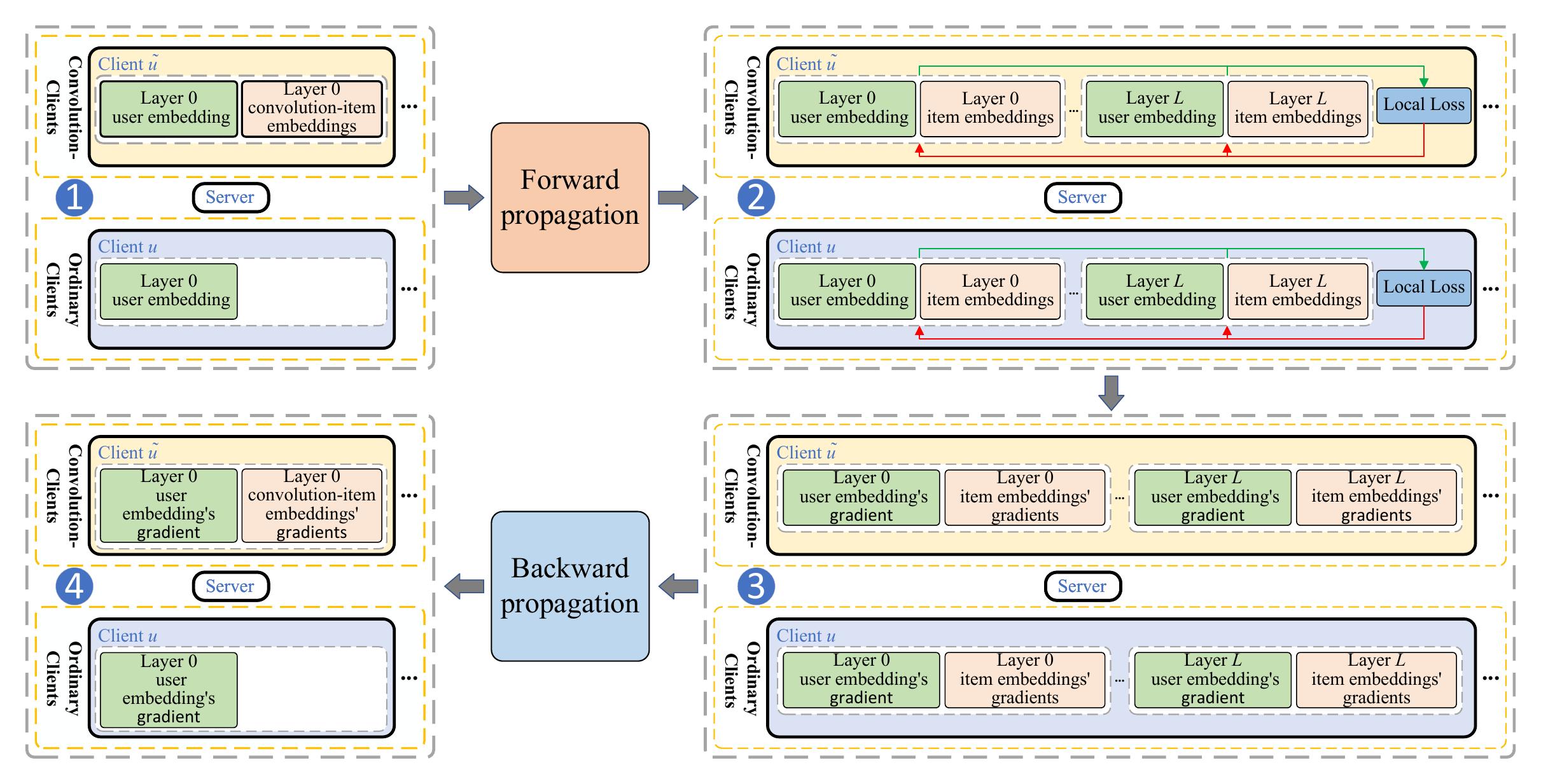}
	\caption{\textbf{The overall framework of our LP-GCN}, which consists of four states. In state 1, each client has the $0$-th user embeddings, and the convolution-items in convolution-clients have the $0$-th item embeddings, while ordinary items do not have any embeddings. After forward propagation, the model moves to state 2, where all clients have user and item embeddings of all layers. Subsequently, clients use these embeddings to construct local losses (i.e., the green lines) and compute gradients for all layers' embeddings (i.e., the red lines) and then transitions to state 3. Finally, through backward propagation, all the intermediate gradients would be propagated back to the learnable parameters, as shown in state 4.}
	\label{fig:The overall framework of our LP-GCN}
	\Description{The overall framework of our LP-GCN}
\end{figure*}
We show the framework of our LP-GCN by using Algorithm~\ref{LP-GCN} and Figure~\ref{fig:The overall framework of our LP-GCN}. Before training begins, all users encrypt the IDs of the items they have interacted with using a common symmetric key $S$ and upload them to the server, which allows the server to construct a global graph of encrypted item-IDs, ensuring that the user privacy is preserved. Based on the global graph, the server will classify the clients into convolution-clients and ordinary clients. Within the convolution-clients, some items are designated as the convolution-items, which are used to compute the item embeddings of the next layer. And the remaining items are treated as the ordinary items. Note that the items in ordinary clients are all ordinary items. We show how clients and items are classified in Algorithm~\ref{Preprocessing for handling item-IDs in the server}. Subsequently, the user embeddings of all clients and the item embeddings of the convolution-items are initialized as learnable parameters, and then the training process starts.

We show the model training process of our LP-GCN at each iteration in Figure~\ref{fig:The overall framework of our LP-GCN}. There are four states for the clients. In state 1, right after initialization (i.e., lines 1-6 in Algorithm~\ref{LP-GCN}), each client has the $0$-th user embeddings, and the convolution-items have the $0$-th item embeddings, while ordinary items do not have any embeddings. After forward propagation (i.e., line 7 in Algorithm~\ref{LP-GCN}), the model moves to state 2, where all clients have user and item embeddings of all layers. Subsequently, clients use these embeddings to calculate local losses and compute gradients for all layers (i.e., line 8 in Algorithm~\ref{LP-GCN}) and moves to state 3. Note that the gradients from layers $1$ to $L$ are intermediate gradients, which need to be back propagated to layer $0$ using the chain rule. Finally, through backward propagation (i.e., line 9 in Algorithm~\ref{LP-GCN}), all the intermediate gradients would be propagated back to the learnable parameters which corresponds to the state 4. We can then update the model parameters in each client (i.e., lines 10-15 in Algorithm~\ref{LP-GCN}). This training process repeats until the model converges.
\subsection{Initialization}\label{sec:Initialization}
\begin{algorithm}
\caption{Initialization}
\label{Initialization}
\begin{algorithmic}[1] 
\FOR{each client $u$ in parallel}
    \STATE Generates a pair of keys by asymmetric encryption, including a private key $PR_u$ and a public key $PB_u$.
    \STATE Sends the public key $PB_u$ to the server.
\ENDFOR

\STATE The server collects the public keys of all clients $\{PB_u : u \in \mathcal{U}\}$ and sends them to client $u_c$ (i.e., one user randomly chosen by the server).

\STATE The client $u_c$ generates a shared key $S$ by symmetric encryption and encrypts $S$ with each of the public keys, acquiring the ciphertexts $\{{PB_u}(S) : u \in \mathcal{U}\}$ which are then sent to the server.

\STATE The server receives the ciphertexts $\{{PB_u}(S) : u \in \mathcal{U}\}$ and distributes each ${PB_u}(S)$ to client $u$.

\FOR{each client $u \in \mathcal{U}$ in parallel}
    \STATE Receives ${PB_u}(S)$ and decrypts it with the private key $PR_{u}$,  so as to acquire the shared key $S$.
    \STATE Generates $\alpha$ virtual items $\{i' : i' \notin \mathcal{I}_u\}$ denoted as $\mathcal{I}_u'$.
\ENDFOR
\FOR{each client $u \in \mathcal{U}$ in parallel}
    \STATE Encrypts item-IDs $\{ ID_{i(u)} : i \in \mathcal{I}_{u} \cup \mathcal{I}_u'\}$ with the shared key $S$, in order to acquire the ciphertexts of the item-IDs $\{S(ID_{i(u)}) : i \in \mathcal{I}_{u} \cup \mathcal{I}_u'\}$ and then sends them to the server.
\ENDFOR
\STATE The server performs preprocessing for handling the item-IDs in the server (i.e., Algorithm~\ref{Preprocessing for handling item-IDs in the server}), so as to split the clients and items.

\FOR{each client $\tilde{u} \in \tilde{\mathcal{U}}$ in parallel}
    \STATE Verifies whether each $u' \in \mathcal{U}_{\tilde{i}} \backslash \{\tilde{u}_{\tilde{i}}\}$ has really interacted with each convolution-item $\tilde{i} \in \tilde{\mathcal{I}}_{\tilde{u}}$.
\ENDFOR
\FOR{each client $u \in \mathcal{U}$ in parallel}
    \STATE Initializes $\boldsymbol{e}^{(0)}_{u(u)}$ (i.e., the $0$-th layer's user embedding in client $u$).
    \IF{$u \in \tilde{\mathcal{U}}$}
        \STATE Initializes $\{\boldsymbol{e}^{(0)}_{\tilde{i}(u)} : \tilde{i} \in \tilde{\mathcal{I}}_u\}$ (i.e., the set of the $0$-th layer's convolution-item embeddings in client $u$) and then sends them to the server.
    \ENDIF
\ENDFOR

\end{algorithmic}
\end{algorithm}

\begin{algorithm}
\caption{Preprocessing for handling the item-IDs in the server}
\label{Preprocessing for handling item-IDs in the server}
\begin{algorithmic}[1]
\STATE Receives the ciphertexts of the item-IDs from all clients $\{S(ID_{i(u)}) : u \in \mathcal{U}, i \in \mathcal{I}_{u} \cup \mathcal{I}_u'\}$, which are used to construct a global graph with the encrypted item-IDs (including real items and virtual items).
\STATE Selects a client group where the combined interactions of its users cover all items. The clients in this group are called candidate convolution-clients denoted as $\overline{\mathcal{U}}$.

\FOR{each $i \in \mathcal{I}$}
\STATE Randomly chooses one client $\tilde{u}_i$ $(\tilde{u}_i \in \overline{\mathcal{U}} \cap \{\mathcal{U}_i \cup {\mathcal{U}}_i'\})$ and considers item $i$ in client $\tilde{u}_i$ as a convolution-item, denoted as $\tilde{i}$.
\ENDFOR

\STATE A client that contains convolution-item is considered as a convolution client, and the set they form is denoted as $\tilde{\mathcal{U}}$. The other clients are called ordinary clients denoted as $ \mathcal{U} \backslash \tilde{\mathcal{U}}$.

\FOR{each $\tilde{u} \in \tilde{\mathcal{U}}$}
\STATE Sends $\tilde{\mathcal{I}}_{\tilde{u}}$ (i.e., the set of all convolution-items of the convolution-client $\tilde{u}$ which are chosen by the server) to the convolution-client $\tilde{u}$.
\ENDFOR
\end{algorithmic}
\end{algorithm}

\begin{figure*}[ht]
	\subfigure[User-item matrix perspective]{
		\label{fig:User-item matrix perspective}
		\centering
		\includegraphics[width=0.4\linewidth]{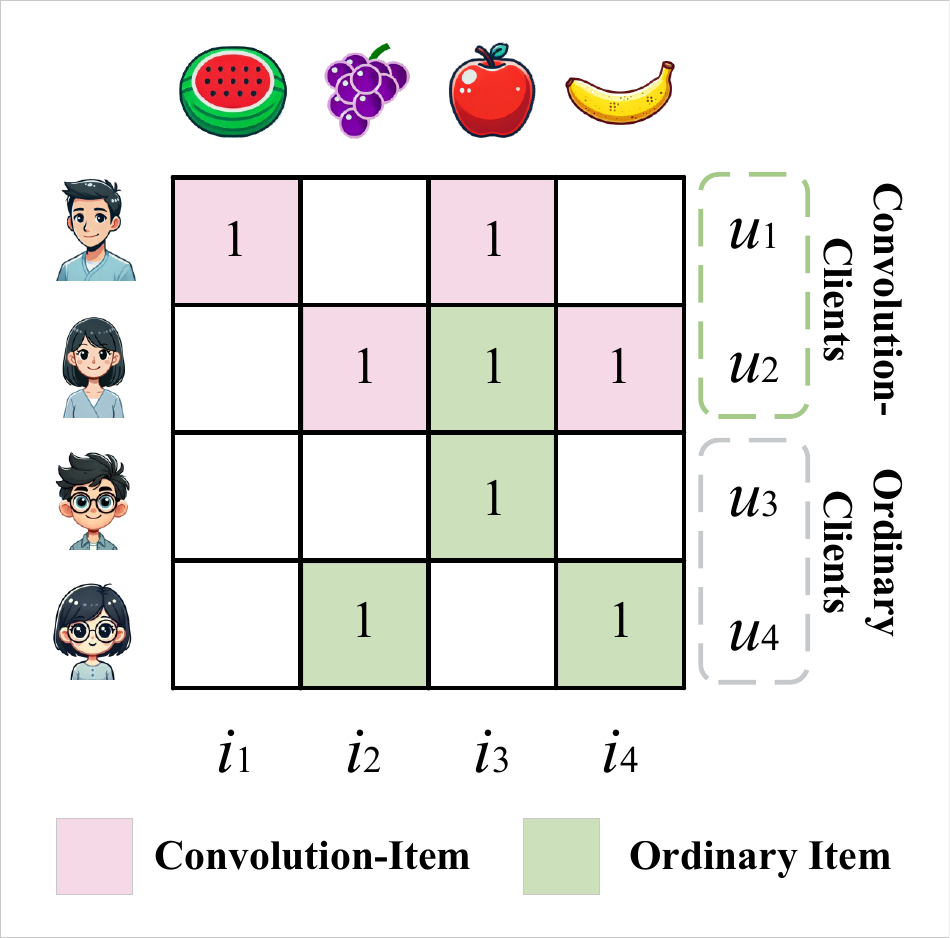}
	}%
	\subfigure[Local sub-graph perspective]{
		\label{fig:local sub-graph perspective}
		\centering
		\includegraphics[width=0.463\linewidth]{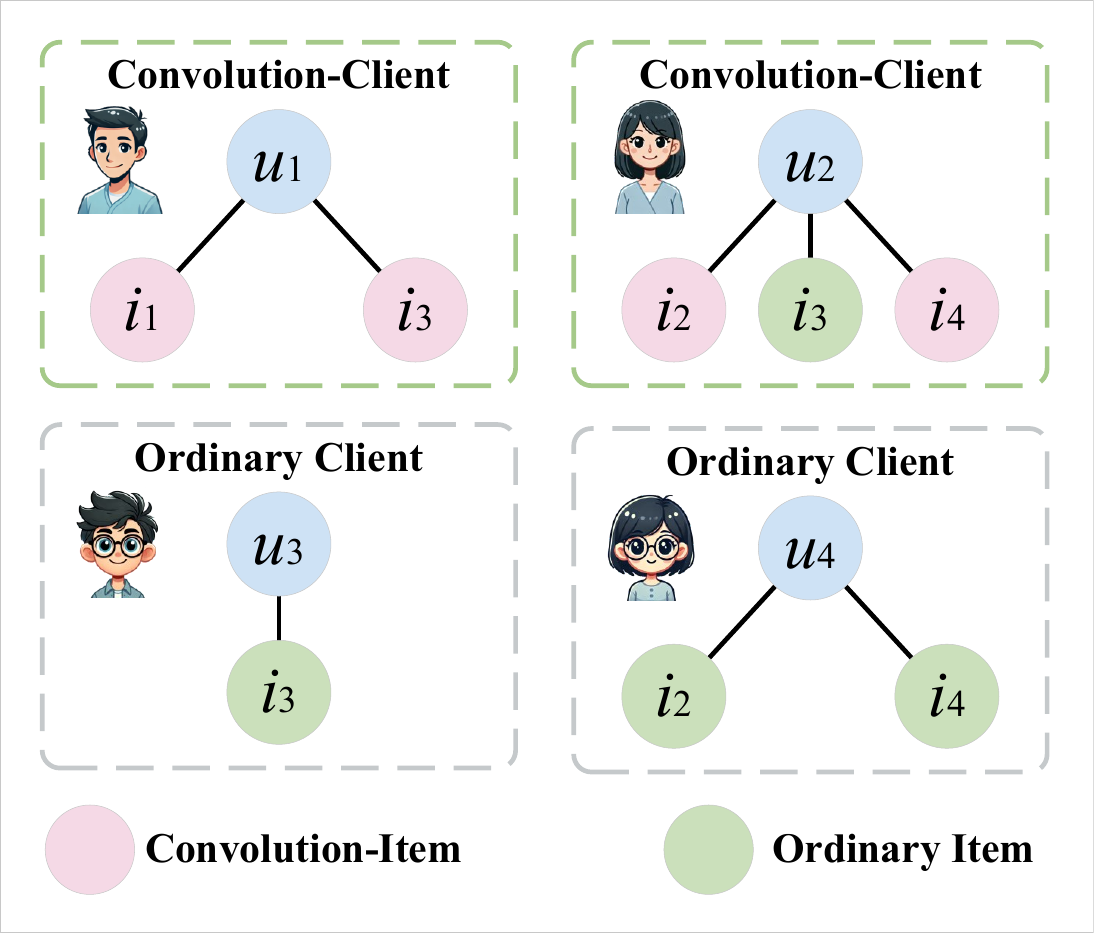}
	}
	\caption{An example of splitting clients and items.}
	\label{fig:one example of spliting clients and items}
	\Description{one example of splitting clients and items}
\end{figure*}

For doing pre-training preparations, we show the process in Algorithm~\ref{Initialization}. Firstly (lines 1-9 in Algorithm~\ref{Initialization}), we expect all clients to possess a common shared key $S$ generated through symmetric encryption, for which the server does not know. Specifically, each client $u$ generates a pair of keys including a private key $PR_u$ and a public key $PB_u$ using asymmetric encryption and then sends the public key $PB_u$ to the server. Subsequently, the server collects the public keys from all clients $\{PB_u : u \in \mathcal{U}\}$ and sends them to a randomly chosen client $u_c$. The client $u_c$ generates a shared key $S$ by symmetric encryption and encrypts $S$ with all the public keys, producing the ciphertexts $\{{PB_u}(S) : u \in \mathcal{U}\}$ which are then sent to the server. The server receives these ciphertexts and distributes each ${PB_u}(S)$ to the corresponding client $u$. Each client $u$ then decrypts ${PB_u}(S)$ with its private key $PR_{u}$,  acquiring the shared key $S$. 
Secondly (line 10 in Algorithm~\ref{Initialization}), each client $u$ generates some virtual items $\{i' : i' \notin \mathcal{I}_u\}$ denoted as $\mathcal{I}_u'$, where the total number of virtual items is $\alpha$, i.e., $\alpha = | \mathcal{I}_u' |$. After this step, all clients possess both virtual and real items, while the server remains unaware of which items are virtual, thus preserving user privacy. The level of privacy protection are determined by the parameters $\alpha$.
Thirdly (line 13 in Algorithm~\ref{Initialization}), each client $u$ encrypts the IDs of all items including virtual items (i.e., $\{ ID_{i(u)} : i \in \mathcal{I}_{u} \cup \mathcal{I}_u'\}$, where $ID_{i(u)}$ represents the ID of item $i$ stored in client $u$) using the shared key $S$, acquiring the ciphertexts of item-IDs $\{S(ID_{i(u)}) : i \in \mathcal{I}_{u} \cup \mathcal{I}_u' \}$, and then sends them to the server. 
Fourthly (Algorithm~\ref{Preprocessing for handling item-IDs in the server}), the server collects the encrypted item-IDs from all clients, which are used to construct a global graph of encrypted item-IDs, and splits the clients into convolution-clients and ordinary clients, and the items into convolution-items and ordinary items. Specifically (line 1 in Algorithm~\ref{Preprocessing for handling item-IDs in the server}), with symmetric encryption, using a shared key $S$ to encrypt the same text results in identical ciphertexts. Thus, if client $u_1$ and client $u_2$ both encrypt the ID of a common item $i_1$ using the common shared key $S$ and send the ciphertext to the server, the server can recognize the common item by its ciphertext. However, the server does not know the actual ID of item $i_1$, preserving user privacy. This allows the server to construct a global graph, where item-IDs are ciphertexts containing both real and virtual items.
Subsequently,we will introduce how to split the clients and items (lines 2-6 in Algorithm~\ref{Preprocessing for handling item-IDs in the server}). The server selects a group of clients that satisfies the following condition: the union of items they have interacted with equals the entire item set. 
While there are many such user groups, we aim to find one where the size of the group $k$ is as small as possible, as $k$ significantly impacts the communication cost, which will be discussed in detail in Section~\ref{sec:experiments}.
This user group is called candidate convolution-clients denoted as $\overline{\mathcal{U}}$.
According to the definition, for any item $i$, there is at least one client among the candidate convolution-clients who has interacted with it. Therefore, for each item $i$, we can randomly select one client $\tilde{u}_i$ $(\tilde{u}_i \in \overline{\mathcal{U}} \cap \{\mathcal{U}_i \cup {\mathcal{U}}_i'\})$ and mark it as a convolution-item in the client $\tilde{u}_i$, denoted as $\tilde{i}$. Note that item $i$ will be considered as an ordinary item in all other clients and the virtual item can be chosen as the convolution-item as well. A client that contains the convolution-item is considered as a convolution client, and the set they form is convolution-clients denoted as $\tilde{\mathcal{U}}$. The other clients are called ordinary clients denoted as $ \mathcal{U} \backslash \tilde{\mathcal{U}}$.
To better illustrate this, we provide an example in Figure~\ref{fig:one example of spliting clients and items}. The union of the items interacted with by user $u_1$ and $u_2$ equals the set of all items, so they are chosen as the group of candidate convolution-clients (or candidate convolution-users). Users $u_1$, $u_3$, and $u_4$ could also form another candidate convolution-clients group, but we prefer the one with the fewest users. Each item is then marked as a convolution-item in a specific convolution-client. For example, the apple with the label $i_3$ is marked as the convolution-item in the convolution-client $u_1$, but considered as an ordinary item in the convolution-client $u_2$. After that, the server will inform each convolution-client of the convolution-item set (line 8 in Algorithm~\ref{Preprocessing for handling item-IDs in the server}).
Fifthly (lines 16-18 in Algorithm~\ref{Initialization}), each convolution-client $\tilde{u}$ verifies whether each $u' \in \mathcal{U}_{\tilde{i}} \backslash \{\tilde{u}_{\tilde{i}}\}$ has really interacted with each convolution-item $\tilde{i} \in \tilde{\mathcal{I}}_{\tilde{u}}$. For example, a convolution-client $\tilde{u}$ contains a convolution-item $\tilde{i}$. The server knows that the client $u'$ is one of the clients that have interacted with the item $i$, but it doesn't know whether the item $i$ is a virtual item, and neither does the client $\tilde{u}$. The client $\tilde{u}$ then asks for the client $u'$ whether the item $i$ is a virtual item through the server. In response, the client $u'$ encrypts the answer with the public key of the client $u'$ and then sends the ciphertext to the client $\tilde{u}$ through the server. Since the communication is secured by asymmetric encryption, the server cannot access the contents of the message, ensuring that the user privacy is protected.
Finally (lines 19-24 in Algorithm~\ref{Initialization}), each client $u$ initializes the $0$-th user embedding, and each convolution-client $\tilde{u}$ initialize the $0$-th convolution-item embeddings, corresponding to the state 1 in Figure~\ref{fig:The overall framework of our LP-GCN}.
\subsection{Forward Propagation}\label{sec:Forward Propagation}
\begin{algorithm}
\caption{Forward Propagation}
\label{Forward Propagation}
\begin{algorithmic}[1]
\FOR{$l = 0, 1, \ldots, L-1$}
    \FOR{each client $u \in \mathcal{U}$ in parallel}
        \STATE Computes $\boldsymbol{e}_{u(u)}^{(l+1)}$ (i.e., the $(l+1)$-th layer's user embedding in client $u$) via Eq.(\ref{eq:user_embedding_convolution}).
        \STATE Encrypts $\boldsymbol{e}_{u(u)}^{(l)}$ (i.e., the $l$-th layer's user embedding in client $u$) with the shared key $S$, acquiring $S(\boldsymbol{e}_{u(u)}^{(l)})$ (i.e., the ciphertext of $\boldsymbol{e}_{u(u)}^{(l)}$) and sends it to the server.
    \ENDFOR
    \STATE The server receives $\{S(\boldsymbol{e}_{u(u)}^{(l)}) : u \in \mathcal{U}\}$ (i.e., the ciphertexts of all user embeddings), and sends $\{S(\boldsymbol{e}_{u'(u')}^{(l)}) : u' \in \mathcal{N}_{\tilde{u}}\}$ (i.e., the convolution-item based neighboring users' embeddings of client $\tilde{u}$) to each convolution-client $\tilde{u} \in \tilde{\mathcal{U}}$.
    \FOR{each client $\tilde{u} \in \tilde{\mathcal{U}}$ in parallel}
        \STATE Receives $\{S(\boldsymbol{e}_{u'(u')}^{(l)}) : u' \in \mathcal{N}_{\tilde{u}}\}$ and decrypts them with the shared key $S$, acquiring $\{\boldsymbol{e}_{u'(u')}^{(l)} : u' \in \mathcal{N}_{\tilde{u}}\}$.
        \STATE Computes $\{\boldsymbol{e}_{\tilde{i}(\tilde{u}_{\tilde{i}})}^{(l+1)} : \tilde{i} \in \tilde{\mathcal{I}}_{\tilde{u}}\}$ (i.e., the $(l+1)$-th layer's convolution-item embeddings in client $\tilde{u}$) via Eq.(\ref{eq:compute_convolution_item_embedding}) and sends them to the server.
    \ENDFOR
    \STATE The server receives all the item embeddings of $(l+1)$-th layer $\{\boldsymbol{e}_i^{(l+1)} : i \in \mathcal{I}\}$ from all convolution-clients and distributes each $\boldsymbol{e}_i^{(l+1)}$ to clients $\{u' : u' \in \mathcal{N}_{\tilde{u}_i}\}$ (i.e., the neighboring users of client $\tilde{u}_i$ when focusing on item $i$).
    
    \FOR{each client $u \in \mathcal{U}$ in parallel}
        \STATE Receives $\{\boldsymbol{e}_{i'}^{(l+1)} : i' \in \{\mathcal{I}_u \cup \mathcal{I}_u'\}\backslash \tilde{\mathcal{I}}_u\}$ (i.e., the $(l+1)$-th layer's ordinary item embeddings).
    \ENDFOR
\ENDFOR
\end{algorithmic}
\end{algorithm}

\begin{figure*}[ht]
	\centering
	\includegraphics[width=0.99\linewidth]{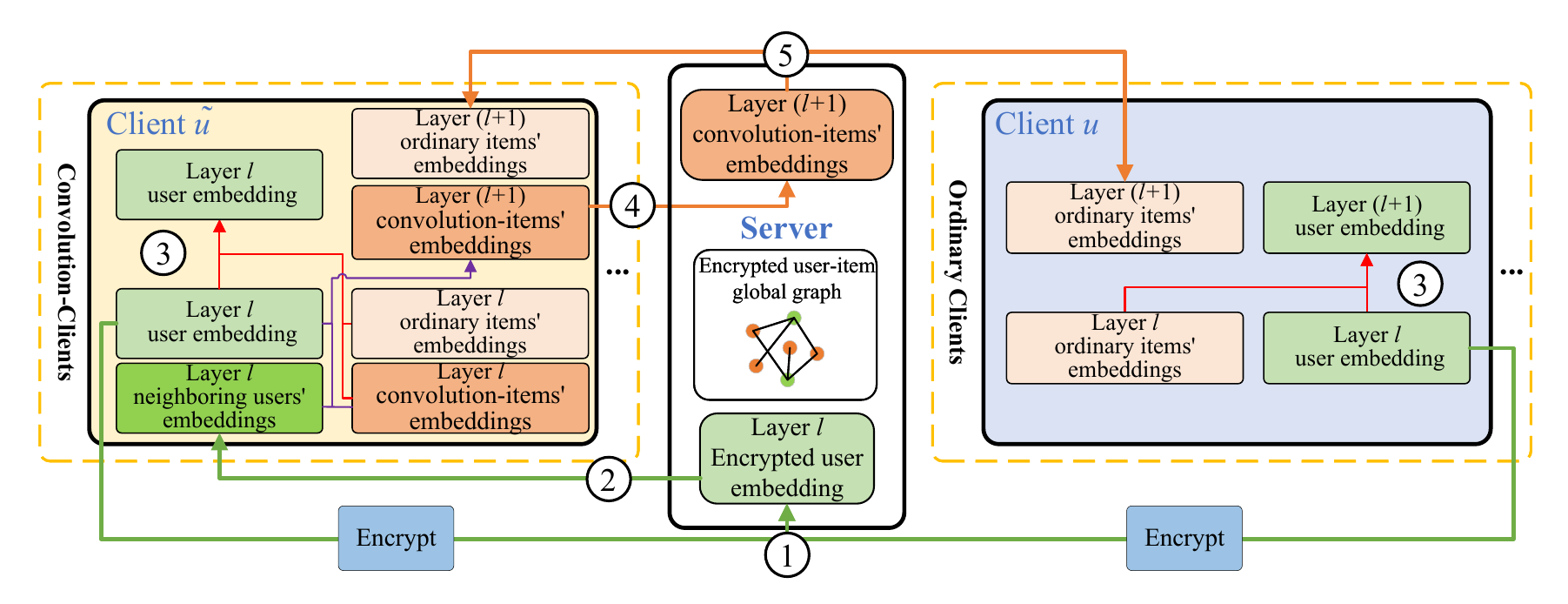}
	\caption{\textbf{Illustration of forward propagation.} Each client encrypts the user embedding with the shared key $S$, and then sends it to the server. Subsequently, the server sends the embeddings of neighboring users to each convolution-client. Then all clients compute the user embedding of layer $(l+1)$. In parallel, convolution-clients compute the embeddings for the convolution-items at layer $(l+1)$, and then for each convolution-item's embedding, it will be shared with the other clients that interact with that item.}
	\label{fig:Illustration of Forward Propagation}
	\Description{Illustration of Forward Propagation}
\end{figure*}
After initialization, we know that each client lacks the $0$-th ordinary item embeddings, thus we need to deal with it before introducing forward propagation. (lines 3-6 in Algorithm~\ref{LP-GCN}) Each convolution-client sends the $0$-th convolution-item embeddings to the server, then the server will acquire the $0$-th embeddings of all items. According to the definition, we know that the convolution-item set equals to the whole item set $\mathcal{I}$. Then the server distributes each item embedding $\boldsymbol{e}_i^{(0)}$ to the other clients according to the global graph of the encrypted item-IDs. For example, item $i_1$ is interacted with by the clients $u_1$, $u_2$, $u_3$, and if item $i_1$ is denoted as a convolution-item in the client $u_1$, and an ordinary item in the clients $u_2$ and $u_3$. When the server receives the $0$-th layer's item embedding of $i_1$ from the client $u_1$, then it will distribute it to the client $u_2$ and the client $u_3$ according to the global graph of the encrypted item-IDs. 

Next, we will introduce forward propagation by using Algorithm~\ref{Forward Propagation} and Figure~\ref{fig:Illustration of Forward Propagation}. Forward propagation aims to make each client acquire user embedding and item embeddings of all layers. Now, we discuss when each client has the $l$-th user embedding and item embeddings (i.e., convolution-clients have the $l$-th convolution-item and ordinary item embeddings, while ordinary clients only have the $l$-th ordinary item embeddings), how to acquire the $(l+1)$-th user embedding and item embeddings. Firstly (lines 2-5 in Algorithm~\ref{Forward Propagation}), each client $u$ computes the $(l+1)$-th user embedding locally,
\begin{equation} \label{eq:user_embedding_convolution}
\boldsymbol{e}_{u(u)}^{(l+1)} = AGG_u\left(\boldsymbol{e}_{u(u)}^{(l)}, \{\boldsymbol{e}_{i'(u)}^{(l)} : i'\in \mathcal{I}_u\}\right)
\end{equation}
where $\boldsymbol{e}_{u(u)}^{(l+1)}$ is the user embedding for client $u$ at layer $(l+1)$, $AGG_u$ is the update function for the user node, determining how the embedding is updated, and $\{\boldsymbol{e}_{i'(u)}^{(l)} : i'\in \mathcal{I}_u\}$ represents the set of item embeddings at layer $l$ for all items $i'$ that user $u$ has really interacted with.
By comparing Eq.(\ref{eq:user_embedding_convolution}) with the centralized version, i.e., Eq.(\ref{eq_prove_fed:user_embedding_convolution}), we can conclude that the forward propagation process for the user nodes is lossless.
Subsequently, each client encrypts the $l$-th layer's user embedding with the shared key $S$, and then sends it to the server. Secondly (line 6 in Algorithm~\ref{Forward Propagation}), the server receives the ciphertexts of the user embedding from all clients. After that, for each convolution-client, the server sends the convolution-item based neighboring users' embeddings to it according to the global graph of the encrypted item-IDs. Specifically, if user $u_1$ and user $u_2$ have interacted with a same item 
$i$, then they are considered as neighboring users with respect to the item $i$. Here, for each convolution-client $\tilde{u}$, the server needs to send the embeddings of the convolution-item based neighboring users, which means if user $u$ has only interacted with some ordinary items of the convolution-client $\tilde{u}$, then there is no need to send the user embedding of user $u$ to the convolution-client $\tilde{u}$. Moreover, since the server does not possess the shared key $S$, it cannot access the user embeddings, thereby protecting user privacy.
Thirdly (lines 7-10 in Algorithm~\ref{Forward Propagation}), each convolution-client $\tilde{u}$ receives the set of the ciphertext of the convolution-item based neighboring users' embeddings, decrypts them with the shared key $S$. And then for each convolution-item, it computes the $(l+1)$-th item embeddings,
\begin{equation} \label{eq:compute_convolution_item_embedding}
\boldsymbol{e}_{\tilde{i}(\tilde{u}_{\tilde{i}})}^{(l+1)} = AGG_i (\boldsymbol{e}_{\tilde{i}(\tilde{u}_{\tilde{i}})}^{(l)}, \boldsymbol{e}_{\tilde{u}_{\tilde{i}}(\tilde{u}_{\tilde{i}})}^{(l)}, \{\boldsymbol{e}_{u'(\tilde{u}_{\tilde{i}})}^{(l)} : u' \in \mathcal{U}_{\tilde{i}} \backslash \{\tilde{u}_{\tilde{i}}\} \})
\end{equation}
where $\boldsymbol{e}_{\tilde{i}(\tilde{u}_{\tilde{i}})}^{(l+1)}$ is the $(l+1)$-th embedding of convolution-item $\tilde{i}$ in client $\tilde{u}_{\tilde{i}}$, $AGG_i$ is the update function for the item node, determining how the embedding is updated, $\boldsymbol{e}_{u'(\tilde{u}_{\tilde{i}})}^{(l)}$ is the $l$-th user embedding from the neighboring user $u'$, $\{\boldsymbol{e}_{u'(\tilde{u}_{\tilde{i}})}^{(l)} : u' \in \mathcal{U}_{\tilde{i}} \backslash \{\tilde{u}_{\tilde{i}}\} \}$ represents the set of embeddings of neighboring users when focusing on item $\tilde{i}$, and $\mathcal{U}_{\tilde{i}} \backslash \{\tilde{u}_{\tilde{i}}\}$ represents the users who has truly interacted with item $\tilde{i}$ exclude $\tilde{u}$ where $\tilde{i}$ locates.
By comparing Eq.(\ref{eq:compute_convolution_item_embedding}) with the centralized version, i.e., Eq.(\ref{eq_prove_cen:compute_convolution_item_embedding}), we can conclude that the forward propagation process for the item nodes is lossless.
After computing the embeddings of the convolution-items, each convolution-client $\tilde{u}$ sends them to the server. Item embeddings are not sensitive parameters, which do not need to be encrypted. 
Fourthly (lines 11-14 in Algorithm~\ref{Forward Propagation}), the server receives all the item embeddings of the $(l+1)$-th layer, then distributes them to all clients to synchronize the ordinary item embeddings of the $(l+1)$-th layer. 
Note that all clients receive the embeddings of virtual items as well though they will not be used to compute the user embedding of the next layer.

To sum up, in the process of forward propagation, we start with the user embeddings at the $l$-th layer for all clients, along with the item embeddings. Note that a same item has the same item embedding across different clients. Next, each client locally computes the user embedding of the $(l+1)$-th layer. Concurrently, the convolution-clients compute all the item embeddings of the $(l+1)$-th layer. The $(l+1)$-th embedding of each item $i$ is computed once by its convolution-client and then shared with other clients that interact with that item. Finally, we obtain the user and item embeddings at layer $(l+1)$ for all clients, with the item embeddings being consistent across different clients for a same item. For the above process, we only need to set $l = 0, 1, \ldots, L-1$. In this way, all clients can obtain the user and item embeddings from layers $0$ to $L$. 
\begin{figure*}[ht]
	\centering
	\includegraphics[width=0.98\linewidth]{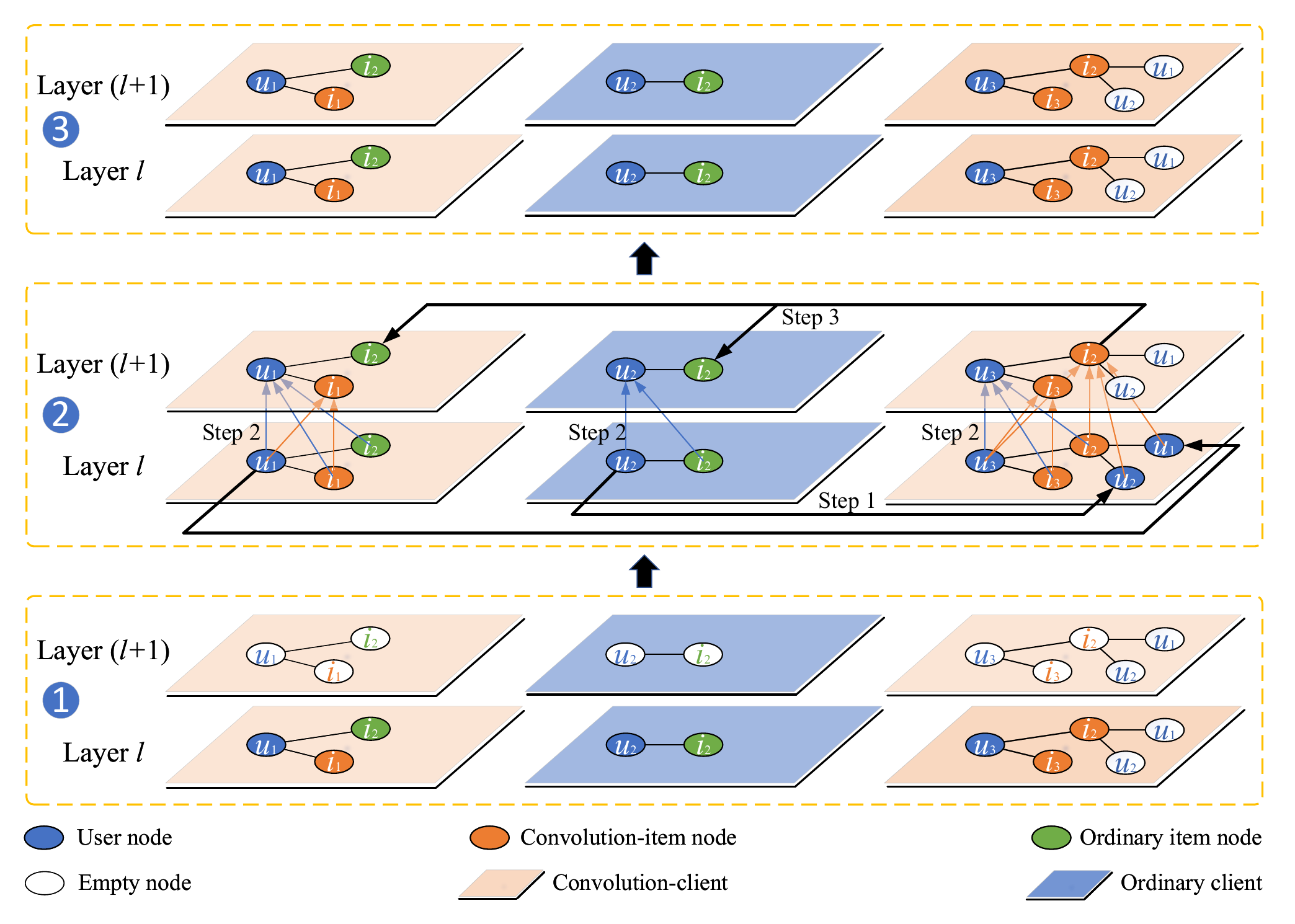}
	\caption{\textbf{An example of forward propagation.} There are three users ($u_1$, $u_2$, $u_3$) and three items ($i_1$, $i_2$, $i_3$). Clients $u_1$ and $u_3$ are the convolution-clients, while $u_2$ is an ordinary client. In $u_1$, $i_1$ is convolution-item, and in $u_3$, $i_2$ and $i_3$ are considered as the convolution-items. Initially, all clients have embeddings of layer $l$. Firstly, $u_1$ and $u_2$ send their user embeddings to $u_3$ for computing the embedding of $i_2$. Then, all clients perform graph convolution to obtain the embeddings for the user and convolution-item nodes at layer $(l+1)$. Finally, $u_3$ synchronizes the $(l+1)$-th embedding of $i_2$ with the clients that interacted with $i_2$, and all clients obtain the embeddings of the layer $(l+1)$.}
	\label{fig:An Example of Forward Propagation}
	\Description{An Example of Forward Propagation}
\end{figure*}
To better understand the process of forward propagation, let's consider a simple example, which is illustrated in Figure~\ref{fig:An Example of Forward Propagation}. There are three users and three items, labeled as $u_1$, $u_2$, $u_3$, and $i_1$, $i_2$, $i_3$. Clients $u_1$ and $u_3$ are the convolution-clients, while client $u_2$ is an ordinary client. In client $u_1$, $i_1$ is considered as the convolution-item; and in $u_3$, $i_2$ and $i_3$ are considered as the convolution-items. Initially, all clients only have the embedding of layer $l$. At the beginning, clients $u_1$ and $u_2$ send their user embeddings to $u_3$ for computing the embedding of item $i_2$, i.e., step 1 of the stage 2 in Figure~\ref{fig:An Example of Forward Propagation}. Then, all clients perform graph convolution locally to obtain the embeddings of the user nodes and the convolution-item nodes at layer $(l+1)$. After that, the convolution-client $u_3$ synchronizes the layer $(l+1)$-th embedding of $i_2$ with other clients that have interacted with $i_2$. Finally, all clients obtain the embeddings of all nodes at layer $(l+1)$.

\subsection{Loss Construction and Gradients Computation}\label{sec:Loss Construction and Gradients Computation}

\begin{algorithm}
\caption{Loss Construction and Gradients Computation}
\label{Loss Construction and Gradients Computation}
\begin{algorithmic}[1]
\FOR{each client $u \in \mathcal{U}$ in parallel}
    \STATE Computes the final user embedding $\boldsymbol{e}_{u(u)}$ via Eq.(\ref{eq:construct_final_user_embedding}).
    \STATE Computes the final item embeddings $\boldsymbol{e}_{i(u)}$ via Eq.(\ref{eq:construct_final_item_embedding}).
    \STATE Computes the predicted value $\{\hat{y}_{ui(u)}: i \in \mathcal{I}_u \}$ via Eq.(\ref{eq:predicted_score}).
    \STATE Computes the local training loss $Loss(u)_{(u)}$ via Eq.(\ref{eq:compute one of the local train loss}) and Eq.(\ref{eq:compute the local train loss}).
    \STATE Computes the gradients of all layers via Eq.(\ref{eq:computes user gradients of all layers}) and Eq.(\ref{eq:computes item gradients of all layers}).
    \STATE Sends the last layers' ordinary item-embedding gradients $\{\boldsymbol{g}_{i(u)}^{(L)} : i \in \mathcal{I}_u \backslash \tilde{\mathcal{I}}_u\}$ to the server.
\ENDFOR

\STATE The server receives all the item embedding gradients of the last layer $L$ $\{\boldsymbol{g}_{i(u')}^{(L)}: i \in \mathcal{I}, u' \in \mathcal{U}_i \backslash \tilde{u}_i\}$, aggregates them via Eq.(\ref{eq:aggregate item embedding gradient}) and distributes each $\boldsymbol{g}_i^{(L)}(i \in \mathcal{I})$ to client $\tilde{u}_i$.
\FOR{each client $\tilde{u} \in \tilde{\mathcal{U}}$ in parallel}
    \STATE Receives the gradients $\{\boldsymbol{g}_i^{(L)}: i \in \tilde{\mathcal{I}}_{\tilde{u}}\}$ and adds each $\boldsymbol{g}_i^{(L)}(i \in \tilde{\mathcal{I}}_{\tilde{u}})$ to the corresponding local gradient $\boldsymbol{g}_{i(\tilde{u})}^{(L)}$ via Eq.(\ref{eq:add item gradient}).
\ENDFOR
\end{algorithmic}
\end{algorithm}
For how to construct a loss and compute the gradients, we show the process in Algorithm~\ref{Loss Construction and Gradients Computation}. Firstly (lines 2-3 in Algorithm~\ref{Loss Construction and Gradients Computation}), each client locally computes the final user embedding and item embeddings,
\begin{equation} \label{eq:construct_final_user_embedding}
\begin{aligned}
\boldsymbol{e}_{u(u)} = f_u(\boldsymbol{e}_{u(u)}^{(0)}, \ldots, \boldsymbol{e}_{u(u)}^{(L)})
\end{aligned}
\end{equation}
\begin{equation} \label{eq:construct_final_item_embedding}
\begin{aligned}
\boldsymbol{e}_{i(u)} = f_i(\boldsymbol{e}_{i(u)}^{(0)}, \ldots, \boldsymbol{e}_{i(u)}^{(L)})
\end{aligned}
\end{equation}
where $\boldsymbol{e}_{u(u)}$ represents the final user embedding stored in client $u$, $\boldsymbol{e}_{u(u)}^{(l)}$ represents the user embedding of the $l$-th
 layer which is stored in client $u$; $\boldsymbol{e}_{i(u)}$ represents the final embedding of item $i$ stored in client $u$, $\boldsymbol{e}_{i(u)}^{(l)}$ represents the embedding of item $i$ at layer $l$ which is stored in client $u$; and $f_u$ and $f_i$ are the functions used for computing the user embedding and the item embedding, respectively. 
 
 Secondly (line 4 in Algorithm~\ref{Loss Construction and Gradients Computation}), each client $u$ computes the predicted score,
 \begin{equation} \label{eq:predicted_score}
\hat{y}_{ui(u)} = f_{ui}(\boldsymbol{e}_{u(u)}, \boldsymbol{e}_{i(u)}) 
\end{equation}
where $\hat{y}_{ui(u)}$ represents the predicted score of user $u$ to item $i$, and $f_{ui}$ denotes the prediction rule.

Thirdly (line 5 in Algorithm~\ref{Loss Construction and Gradients Computation}), each client $u$ computes the local training loss $Loss(u)_{(u)}$,
 \begin{equation} \label{eq:compute one of the local train loss}
\text{loss}(u, i)_{(u)} = h(y_{ui(u)}, \hat{y}_{ui(u)})
\end{equation}

 \begin{equation} \label{eq:compute the local train loss}
\text{Loss}(u)_{(u)} = \sum_{(u,i)\in P_u} \text{loss}(u, i)_{(u)}
\end{equation}
where $\text{loss}(u, i)_{(u)}$ denotes the loss of user $u$ for item $i$ which is stored in client $u$, $y_{ui(u)}$ represents the ground truth of user $u$ for item $i$, and $h$ could be any loss function, such as square loss and BPR loss. Note that if it is BPR loss, there will be an additional input, which is the predicted score of a negative item and will be explained in Section~\ref{sec:LP-GCN (LightGCN+)}. $P_u$ represents the set of training pairs w.r.t. user $u$.

Fourthly (line 6 in Algorithm~\ref{Loss Construction and Gradients Computation}), each client $u$ computes the gradients of all layers,
 \begin{equation} \label{eq:computes user gradients of all layers}
\boldsymbol{g}_{u(u)}^{(l)} = \frac{\partial \text{Loss}(u)_{(u)}}{\partial \boldsymbol{e}_{u(u)}^{(l)}} \quad (l = 0, 1, \ldots, L)
\end{equation}
 \begin{equation} \label{eq:computes item gradients of all layers}
\boldsymbol{g}_{i(u)}^{(l)} = \frac{\partial \text{Loss}(u)_{(u)}}{\partial \boldsymbol{e}_{i(u)}^{(l)}} \quad (l = 0, 1, \ldots, L ; i \in \mathcal{I}_u)
\end{equation}
where $\boldsymbol{g}_{u(u)}^{(l)}$ represents the $l$-th gradient of the user embedding which is stored in client $u$, and $\boldsymbol{g}_{i(u)}^{(l)}$ represents the $l$-th gradient of item $i$ also stored in client $u$.

After computing the gradients, we could know the state of all clients which are shown in state 3 of Figure~\ref{fig:The overall framework of our LP-GCN}. Note that when $l$ is not equal to 0, $\boldsymbol{g}_{u(u)}^{(l)}$ and $\boldsymbol{g}_{i(u)}^{(l)}$ are intermediate gradients, which need to be backward propagated until $l$ reaches 0.

Finally (lines 7-11 in Algorithm~\ref{Loss Construction and Gradients Computation}), we need to prepare for backward propagation. Each client $u$ sends the gradients of the ordinary item embeddings at the $L$-th layer $\{\boldsymbol{g}_{i(u)}^{(L)} : i \in \mathcal{I}_u \backslash \tilde{\mathcal{I}}_u\}$ to the server. Hence, the server obtains the gradients of all item embeddings at the last layer $L$. After that, for each item $i$, the server aggregates them,
 \begin{equation} \label{eq:aggregate item embedding gradient}
\boldsymbol{g}_{i}^{(L)} = \sum_{u' \in \mathcal{U}_i \backslash \tilde{u}_i} \boldsymbol{g}_{i(u')}^{(L)}
\end{equation}
where $\mathcal{U}_i \backslash \tilde{u}_i$ represents the set of users who have interacted with the item $i$ except the convolution-client where the item $i$ is marked as the convolution-item. $\boldsymbol{g}_{i(u')}^{(L)}$ is the gradient of the item $i$ at the $L$-th layer which is stored in client $u'$.

After aggregating the gradients of all item embeddings at the $L$-th layer, for each item $i$, the server will distribute its gradient $\boldsymbol{g}_i^{(L)}$ to the client $\tilde{u}_{\tilde{i}}$ where item $i$ is considered as a convolution-item denoted as $\tilde{i}$. Then the client $\tilde{u}_{\tilde{i}}$ would add $\boldsymbol{g}_i^{(L)}$ to the corresponding local gradient $\boldsymbol{g}_{\tilde{i}(\tilde{u}_{\tilde{i}})}^{(L)}$,
 \begin{equation} \label{eq:add item gradient}
\boldsymbol{g}_{\tilde{i}(\tilde{u}_{\tilde{i}})}^{(L)}=\boldsymbol{g}_{\tilde{i}(\tilde{u}_{\tilde{i}})}^{(L)}+\boldsymbol{g}_i^{(L)}
\end{equation}
where $\boldsymbol{g}_{\tilde{i}(\tilde{u}_{\tilde{i}})}^{(L)}$ represents the local gradient of item $\tilde{i}$ stored in client $\tilde{u}_{\tilde{i}}$, and $\boldsymbol{g}_i^{(L)}$ is the gradient of item $i$ from the server.

Besides, when considering virtual items, Eq.(\ref{eq:aggregate item embedding gradient}) needs to be calculated in the convolution-client $\tilde{u}_{\tilde{i}}$, where only the gradients of real items are aggregated, while virtual items are excluded.
\subsection{Backward Propagation}\label{sec:Backward Propagation}
\begin{algorithm}
\caption{Backward Propagation}
\label{Backward Propagation}
\begin{algorithmic}[1]
\FOR{$l = L-1, L-2, \ldots, 0$}
    \FOR{each client $\tilde{u} \in \tilde{\mathcal{U}}$ in parallel}
        \STATE Computes the gradient of the user embedding at the $l$-th layer $\boldsymbol{g}_{\tilde{u}(\tilde{u})}^{(l)}$ via Eq.(\ref{eq:con_client compute user embedding gradient}).
        \STATE Computes the gradients of the convolution-item embeddings at the $l$-th layer $\{\boldsymbol{g}_{\tilde{i}(\tilde{u})}^{(l)}:\tilde{i} \in \mathcal{\tilde{I}}_{\tilde{u}}\}$ via Eq.(\ref{eq:con_client compute convolution item embedding gradient}).
        \STATE Computes the gradients of the ordinary item embeddings at the $l$-th layer $\{\boldsymbol{g}_{i({\tilde{u}})}^{(l)}:i \in \mathcal{I}_{\tilde{u}} \backslash \tilde{\mathcal{I}}_{\tilde{u}}\}$ via Eq.(\ref{eq:con_client compute ordinary item embedding gradient}).
        \STATE Computes the gradients of the neighboring users' embeddings at the $l$-th layer $\{\boldsymbol{g}^{(l)}_{u'(\tilde{u})} : u' \in \mathcal{N}_{\tilde{u}}\}$ via Eq.(\ref{eq:con_client compute neighbor user embedding gradient}).
        \STATE Sends the gradients of the ordinary item embeddings at the $l$-th layer $\{\boldsymbol{g}^{(l)}_{i'(\tilde{u})} : i' \in \mathcal{I}_{\tilde{u}} \backslash \tilde{\mathcal{I}}_{\tilde{u}}\}$ and the gradients of the neighboring users' embeddings at the $l$-th layer $\{\boldsymbol{g}^{(l)}_{u'(\tilde{u})} : u' \in \mathcal{N}_{\tilde{u}} \}$ to the server.
    \ENDFOR    
    \FOR{each client $u \in \mathcal{U} \backslash \tilde{\mathcal{U}}$ in parallel}
        \STATE Computes the gradient of the user embedding at the $l$-th layer $\boldsymbol{g}_{u(u)}^{(l)}$ via Eq.(\ref{eq:ord_client compute user embedding gradient}).
        \STATE Computes the gradients of the ordinary item embeddings at the $l$-th layer $\{\boldsymbol{g}_{i(u)}^{(l)}:i \in \mathcal{I}_u\}$ via Eq.(\ref{eq:ord_client compute ord_item embedding gradient}).
        \STATE Sends the gradients of the ordinary item embeddings at the $l$-th layer $\{\boldsymbol{g}^{(l)}_{i(u)}: i \in \mathcal{I}_u \}$ to the server.
    \ENDFOR
    \STATE The server receives the gradients of the ordinary item embeddings at the $l$-th layer from all clients, i.e., $\{\boldsymbol{g}^{(l)}_{i(u)} : u \in \mathcal{U}, i \in \mathcal{I}_u\}$, aggregates them via Eq.(\ref{eq:server aggregate item embedding gradient}), acquiring $\{\boldsymbol{g}^{(l)}_{i} : i \in \mathcal{I}\}$ and distributes each $\boldsymbol{g}^{(l)}_{i}(i \in \mathcal{I})$ to client $\tilde{u}_{\tilde{i}}$.
    \STATE The server receives the gradients of the neighboring users' embeddings at the $l$-th layer from all clients, i.e., $\{\boldsymbol{g}^{(l)}_{u'(u)} : u \in \tilde{\mathcal{U}}, u' \in \mathcal{N}_u\}$, aggregates them via Eq.(\ref{eq:server aggregate neighbor user embedding gradients}), acquiring $\{\boldsymbol{g}^{(l)}_{u'} : u' \in \mathcal{U}\}$ and distributes each $\boldsymbol{g}^{(l)}_{u'}(u' \in \mathcal{U})$ to client $u'$.
    \FOR{each client $u \in \mathcal{U}$ in parallel}
        \STATE Receives the gradient of user embedding at the $l$-th layer $\boldsymbol{g}^{(l)}_{u}$ and  adds it to the gradient of local user embedding $\boldsymbol{g}^{(l)}_{u(u)}$ via Eq.(\ref{eq:client add user embedding gradients}). 
        \IF{$u \in \tilde{\mathcal{U}}$}
            \STATE Receives the gradients of item embeddings at the $l$-th layer $\{\boldsymbol{g}^{(l)}_{i} : i \in \tilde{\mathcal{I}}_u\}$ and adds each $\boldsymbol{g}^{(l)}_{i}(i \in \tilde{\mathcal{I}}_u)$ to the gradient of local item embedding $\boldsymbol{g}^{(l)}_{\tilde{i}(u)}$ via Eq.(\ref{eq:con_client add item embedding gradients}).
        \ENDIF
    \ENDFOR
\ENDFOR
\end{algorithmic}
\end{algorithm}
\begin{figure*}[htbp]
	\centering
	\includegraphics[width=0.99\linewidth]{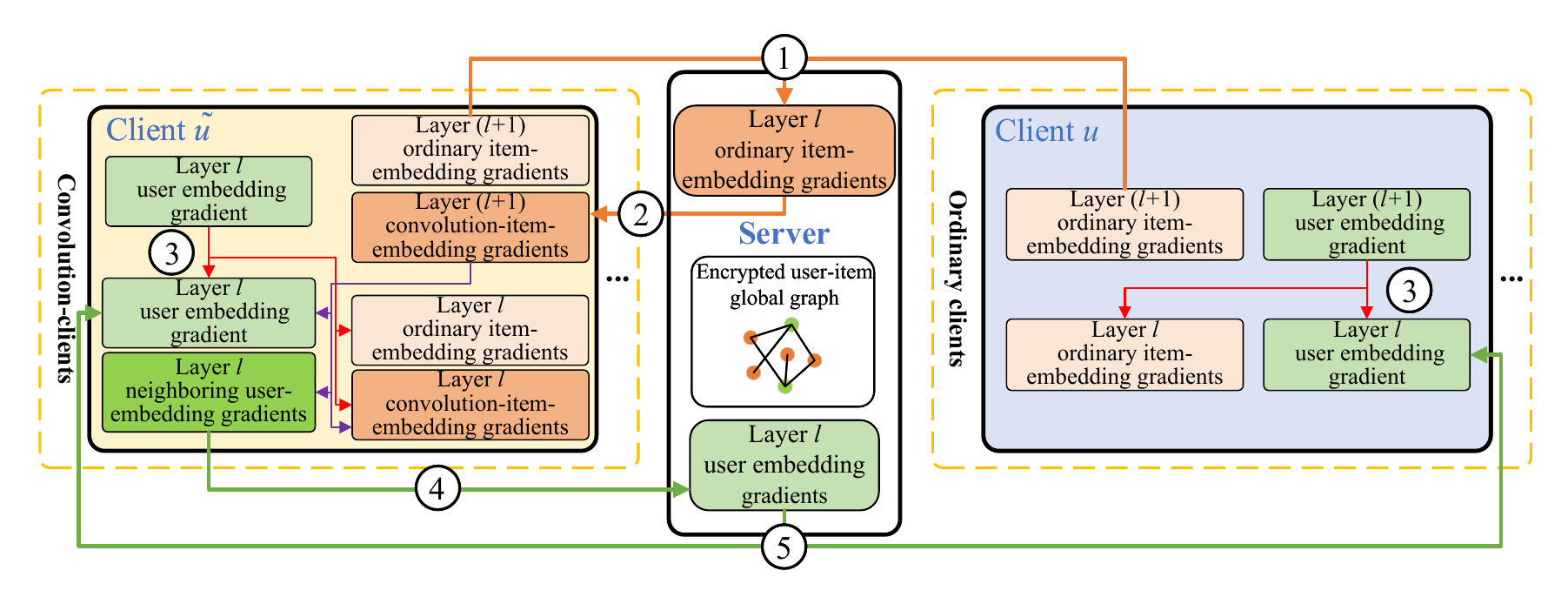}
	\caption{\textbf{Illustration of backward propagation,} which is essentially the reverse of forward propagation. Each client sends the gradients of the ordinary item embedding at the $(l+1)$-th to the server for aggregation. Then the server sends the gradient of each ordinary item to its corresponding client. Each client locally computes the gradients for the nodes at layer $l$ based on the gradients of the nodes at layer $(l+1)$. Next, the convolution-clients send the the gradients of the neighboring users' embeddings at the $l$-th layer to the server for aggregation, which are then sent back to the corresponding clients.}
	\label{fig:Illustration of Backward Propagation}
	\Description{Illustration of Backward Propagation}
\end{figure*}
 The back propagation process is essentially the reverse of forward propagation. Before starting back propagation, we know that the nodes (including user and item nodes) in all clients have the gradients from layers $0$ to $(L-1)$, with layers 1 to $(L-1)$ containing intermediate gradients that need to be propagated back to layer 0. However, at layer $L$, the gradients of ordinary items in all clients have already been aggregated by the server and sent back to the corresponding convolution-items, meaning only user nodes and convolution-item nodes have gradients at this layer.
 
 Now we will show how to propagate gradients from layers $(l+1)$ to $l$ in detail. Specifically, we will explain how the gradients are calculated in both convolution-clients and ordinary clients, and how they are transmitted through the server by using Algorithm~\ref{Backward Propagation} and Figure~\ref{fig:Illustration of Backward Propagation}.
 
 In the convolution-clients, the program proceeds as follows (lines 2-8 in Algorithm~\ref{Backward Propagation}):
 
 Firstly, each convolution-client $\tilde{u}$ computes the gradient of the user embedding at the $l$-th layer $\boldsymbol{g}^{(l)}_{\tilde{u}(\tilde{u})}$,
\begin{equation} \label{eq:con_client compute user embedding gradient}
\boldsymbol{g}^{(l)}_{\tilde{u}(\tilde{u})} = \frac{\partial \boldsymbol{g}^{(l+1)}_{\tilde{u}(\tilde{u})}}{\partial \boldsymbol{e}^{(l)}_{\tilde{u}(\tilde{u})}} + \sum_{i' \in \tilde{\mathcal{I}}_{\tilde{u}}} \frac{\partial \boldsymbol{g}^{(l+1)}_{i'(\tilde{u})}}{\partial \boldsymbol{e}^{(l)}_{\tilde{u}(\tilde{u})}} + \frac{\partial \text{Loss}(\tilde{u})_{(\tilde{u})}}{\partial \boldsymbol{e}^{(l)}_{\tilde{u}(\tilde{u})}}
\end{equation}
where $\boldsymbol{g}^{(l+1)}_{\tilde{u}(\tilde{u})}$ represents the gradient of the $(l+1)$-th layer user embedding for the convolution-client $\tilde{u}$, and $\tilde{\mathcal{I}}_{\tilde{u}}$ represents the set of all convolution-items of the convolution-client $\tilde{u}$. According to forward propagation, $\boldsymbol{e}^{(l)}_{\tilde{u}(\tilde{u})}$ is utilized to compute $\boldsymbol{e}^{(l+1)}_{\tilde{u}(\tilde{u})}$(i.e., user embedding of next layer) and $\{\boldsymbol{e}_{\tilde{i}(\tilde{u})}^{(l+1)}:\tilde{i} \in \tilde{\mathcal{I}}_{\tilde{u}}\}$ (i.e., embeddings of the convolution-item set of next layer). Therefore, for backward propagation, $\frac{\partial \boldsymbol{g}^{(l+1)}_{\tilde{u}(\tilde{u})}}{\partial \boldsymbol{e}^{(l)}_{\tilde{u}(\tilde{u})}}$ and 
$\sum_{i' \in \tilde{\mathcal{I}}_{\tilde{u}}} \frac{\partial \boldsymbol{g}^{(l+1)}_{i'(\tilde{u})}}{\partial \boldsymbol{e}^{(l)}_{\tilde{u}(\tilde{u})}}$ respectively correspond to these computations. $\frac{\partial \text{Loss}(\tilde{u})_{(\tilde{u})}}{\partial \boldsymbol{e}^{(l)}_{\tilde{u}(\tilde{u})}}$ has been calculated before, shown in Eq.(\ref{eq:computes user gradients of all layers}).

Secondly, each convolution-client $\tilde{u}$ computes the gradients of the convolution-item embeddings at the $l$-th layer $\{\boldsymbol{g}_{\tilde{i}(\tilde{u})}^{(l)}:\tilde{i} \in \mathcal{\tilde{I}}_{\tilde{u}}\}$,
\begin{equation} \label{eq:con_client compute convolution item embedding gradient}
\boldsymbol{g}^{(l)}_{\tilde{i}(\tilde{u})} = \frac{\partial \boldsymbol{g}^{(l+1)}_{\tilde{u}(\tilde{u})}}{\partial \boldsymbol{e}^{(l)}_{\tilde{i}(\tilde{u})}} + \frac{\partial \boldsymbol{g}^{(l+1)}_{\tilde{i}(\tilde{u})}}{\partial \boldsymbol{e}^{(l)}_{\tilde{i}(\tilde{u})}} + \frac{\partial \text{Loss}(\tilde{u})_{(\tilde{u})}}{\partial \boldsymbol{e}^{(l)}_{\tilde{i}(\tilde{u})}} \quad (\tilde{i} \in \mathcal{\tilde{I}}_{\tilde{u}})
\end{equation}
where $\boldsymbol{g}^{(l)}_{\tilde{i}(\tilde{u})}$ represents the gradient of the $l$-th layer embedding of convolution-item $\tilde{i}$ stored in client $\tilde{u}$. According to forward propagation, $\boldsymbol{e}^{(l)}_{\tilde{i}(\tilde{u})}$ is utilized to compute $\boldsymbol{e}^{(l+1)}_{\tilde{u}(\tilde{u})}$ (i.e., user embedding of next layer) and $\boldsymbol{e}_{\tilde{i}(\tilde{u})}^{(l+1)}$ (i.e., item embedding of next layer). Therefore, for backward propagation, $\frac{\partial \boldsymbol{g}^{(l+1)}_{\tilde{u}(\tilde{u})}}{\partial \boldsymbol{e}^{(l)}_{\tilde{i}(\tilde{u})}}$ and $\frac{\partial \boldsymbol{g}^{(l+1)}_{\tilde{i}(\tilde{u})}}{\partial \boldsymbol{e}^{(l)}_{\tilde{i}(\tilde{u})}}$ respectively correspond to these computations. $\frac{\partial \text{Loss}(\tilde{u})_{(\tilde{u})}}{\partial \boldsymbol{e}^{(l)}_{\tilde{i}(\tilde{u})}} $ has been calculated before, shown in Eq.(\ref{eq:computes item gradients of all layers}).

Thirdly, each convolution-client $\tilde{u}$ computes the gradients of the ordinary item embeddings at the $l$-th layer $\{\boldsymbol{g}_{i(u)}^{(l)}:i \in \mathcal{I}_{\tilde{u}} \backslash \tilde{\mathcal{I}}_{\tilde{u}}\}$,
\begin{equation} \label{eq:con_client compute ordinary item embedding gradient}
\boldsymbol{g}^{(l)}_{i(\tilde{u})} = \frac{\partial \boldsymbol{g}^{(l+1)}_{\tilde{u}(\tilde{u})}}{\partial \boldsymbol{e}^{(l)}_{i(\tilde{u})}} + \frac{\partial \text{Loss}(\tilde{u})_{(\tilde{u})}}{\partial \boldsymbol{e}^{(l)}_{i(\tilde{u})}} \quad (i \in \mathcal{I}_{\tilde{u}} \backslash \tilde{\mathcal{I}}_{\tilde{u}})
\end{equation}
According to forward propagation, $\boldsymbol{e}^{(l)}_{i(\tilde{u})}$ is only utilized to compute $\boldsymbol{e}^{(l+1)}_{i(\tilde{u})}$ (i.e., user embedding of next layer). Therefore, for backward propagation, $\frac{\partial \boldsymbol{g}^{(l+1)}_{\tilde{u}(\tilde{u})}}{\partial \boldsymbol{e}^{(l)}_{i(\tilde{u})}}$ corresponds to this computation. $\frac{\partial \text{Loss}(\tilde{u})_{(\tilde{u})}}{\partial \boldsymbol{e}^{(l)}_{\tilde{i}(\tilde{u})}} $ has been calculated before, shown in Eq.(\ref{eq:computes item gradients of all layers}).

Fourthly, each convolution-client $\tilde{u}$ computes the gradients of the neighboring users' embeddings at the $l$-th layer $\{\boldsymbol{g}^{(l)}_{u'(\tilde{u})} : u' \in \mathcal{N}_{\tilde{u}}\}$,
\begin{equation} \label{eq:con_client compute neighbor user embedding gradient}
\boldsymbol{g}^{(l)}_{u'(\tilde{u})} = \sum_{i' \in \{\mathcal{\tilde{I}}_{\tilde{u}} \cap \mathcal{I}_{u'}\}} \frac{\partial \boldsymbol{g}^{(l+1)}_{i'(\tilde{u})}}{\partial \boldsymbol{e}^{(l)}_{u'(\tilde{u})}}
\end{equation}
where $\mathcal{N}_{\tilde{u}}$ represents the set of convolution-item based neighboring users of the convolution-client $\tilde{u}$. $\boldsymbol{g}^{(l)}_{u'(\tilde{u})}$ is the $l$-th embedding gradient of neighboring user $u'$ stored in client $\tilde{u}$. According to forward propagation, $\boldsymbol{e}^{(l)}_{u'(\tilde{u})}$ is utilized to compute the set of convolution-item embeddings of the next layer $\{\boldsymbol{e}_{\tilde{i}(\tilde{u})}^{(l+1)}: \tilde{i} \in \mathcal{\tilde{I}}_{\tilde{u}} \cap \mathcal{I}_{u'}\}$. Therefore, for backward propagation, $\frac{\partial \boldsymbol{g}^{(l+1)}_{i'(\tilde{u})}}{\partial \boldsymbol{e}^{(l)}_{u'(\tilde{u})}}$ corresponds to this computation.

Finally, each convolution-client $\tilde{u}$ sends the gradients of the ordinary item embeddings at the $l$-th layer and the gradients of the convolution-item based neighboring users' embeddings at the $l$-th layer to the server.

Meanwhile, like the convolution-client, in ordinary clients, the program proceeds as follows (lines 9-12 in Algorithm~\ref{Backward Propagation}): 

Firstly, each ordinary client $u$ computes the gradient of the user embedding at the $l$-th layer $\boldsymbol{g}_{u(u)}^{(l)}$,
\begin{equation} \label{eq:ord_client compute user embedding gradient}
\boldsymbol{g}^{(l)}_{u(u)} = \frac{\partial \boldsymbol{g}^{(l+1)}_{u(u)}}{\partial \boldsymbol{e}^{(l)}_{u(u)}} + \frac{\partial \text{Loss}(u)_{(u)}}{\partial \boldsymbol{e}^{(l)}_{u(u)}}
\end{equation}

Secondly, each ordinary client $u$ computes the gradients of the ordinary item embeddings at the $l$-th layer $\{\boldsymbol{g}_{i(u)}^{(l)}:i \in \mathcal{I}_u\}$,
\begin{equation} \label{eq:ord_client compute ord_item embedding gradient}
\boldsymbol{g}^{(l)}_{i(u)} = \frac{\partial \boldsymbol{g}^{(l+1)}_{u(u)}}{\partial \boldsymbol{e}^{(l)}_{i(u)}} + \frac{\partial \text{Loss}(u)_{(u)}}{\partial \boldsymbol{e}^{(l)}_{i(u)}} \quad (i \in \mathcal{I}_u)
\end{equation}

Finally, each ordinary client $u$ sends the gradients of the ordinary item embeddings at the $l$-th layer to the server.

Then the server proceeds the program as follows (lines 14-15 in Algorithm~\ref{Backward Propagation}):

Firstly, the server receives the gradients of the ordinary item embeddings at the $l$-th layer from all clients, i.e., $\{\boldsymbol{g}^{(l)}_{i(u)} : u \in \mathcal{U}, i \in \mathcal{I}_u \backslash \mathcal{\tilde{I}}_{u}\}$, and aggregates them,
\begin{equation} \label{eq:server aggregate item embedding gradient}
\boldsymbol{g}^{(l)}_i = \sum_{u \in \mathcal{U}} { \boldsymbol{g}^{(l)}_{i(u)}}
\end{equation}
where $\boldsymbol{g}^{(l)}_{i(u)}$ represents the gradient of item $i$ at the $l$-th layer which is stored in client $u$. Then the server distributes each $\boldsymbol{g}^{(l)}_i(i \in \mathcal{I})$ to client $\tilde{u}_{\tilde{i}}$.

Secondly, the server receives the gradients of the neighboring users' embeddings at the $l$-th layer from convolution-clients, i.e., $\{\boldsymbol{g}^{(l)}_{u'(u)} : u \in \tilde{\mathcal{U}}, u' \in \mathcal{N}_u\}$, and aggregates them,
\begin{equation} \label{eq:server aggregate neighbor user embedding gradients}
\boldsymbol{g}^{(l)}_{u'} = \sum_{u \in \tilde{\mathcal{U}}} { \boldsymbol{g}^{(l)}_{u'(u)}}
\end{equation}
acquiring $\{\boldsymbol{g}^{(l)}_{u'} : u' \in \mathcal{U}\}$ and distributes each $\boldsymbol{g}^{(l)}_{u'}(u' \in \mathcal{U})$ to client $u'$.

At the last step of backward propagation (lines 16-21 in Algorithm~\ref{Backward Propagation}), each client $u$ (including convolution-client and ordinary client) receives the gradient of user embedding at the $l$-th layer $\boldsymbol{g}^{(l)}_{u}$ and  adds it to the gradient of local user embedding $\boldsymbol{g}^{(l)}_{u(u)}$,
\begin{equation} \label{eq:client add user embedding gradients}
\boldsymbol{g}_{u(u)}^{(l)}=\boldsymbol{g}_{u(u)}^{(l)}+\boldsymbol{g}_u^{(l)}
\end{equation}

At the same time, each convolution-client $\tilde{u}$ receives the gradients of item embeddings at the $l$-th layer $\{\boldsymbol{g}^{(l)}_{i} : i \in \tilde{\mathcal{I}}_{\tilde{u}}\}$ and adds each $\boldsymbol{g}^{(l)}_{i}(i \in \tilde{\mathcal{I}}_{\tilde{u}})$ to the gradient of local item embedding $\boldsymbol{g}^{(l)}_{\tilde{i}(u)}$,
\begin{equation} \label{eq:con_client add item embedding gradients}
\boldsymbol{g}_{\tilde{i}(u)}^{(l)}=\boldsymbol{g}_{\tilde{i}(u)}^{(l)}+\boldsymbol{g}_i^{(l)}
\end{equation}

Besides, when considering virtual items, Eq.(\ref{eq:server aggregate item embedding gradient}) and Eq.(\ref{eq:server aggregate neighbor user embedding gradients}) need to be performed in client side. The aggregation process is similar, with the only difference being that the gradients related to virtual items are excluded, while the gradients of real items are retained. Because the client side can determine which gradients are related to virtual items.

To sum up, in the process of backward propagation, at layer $(l+1)$, all clients, including both convolution-clients and ordinary clients, have the gradients of user embeddings. Ordinary items do not have embedding gradients, whereas the convolution-items in convolution-clients do have embedding gradients. Next, the convolution-clients locally compute the $l$-th gradients for user embedding, embeddings of all items (including ordinary items and the convolution-items), and the convolution-item based neighboring users' embeddings. Ordinary clients also locally calculate the gradients for user embeddings and ordinary items, both at layer $l$. Then, all clients send the gradients of ordinary items at layer $l$ to the server for aggregation, which are then forwarded to the respective convolution-clients for gradient addition. Convolution-clients send the gradients of neighboring users at layer $l$ to the server for aggregation, which are then sent to the corresponding clients for gradient addition. Finally, at layer $l$, all clients, including both convolution-clients and ordinary clients, have gradients of user embedding. Ordinary items do not have embedding gradients, whereas the convolution-items in the convolution-clients do have embedding gradients. When $l$ = $L$-1, it corresponds to the initial state of gradient back propagation as shown in Figure~\ref{fig:The overall framework of our LP-GCN}. For the above process, we only need to set $l = 0, 1, \ldots, L-1$, allowing gradients to be eventually propagated back to layer 0.
\begin{figure*}[b]
	\centering
	\includegraphics[width=0.98\linewidth]{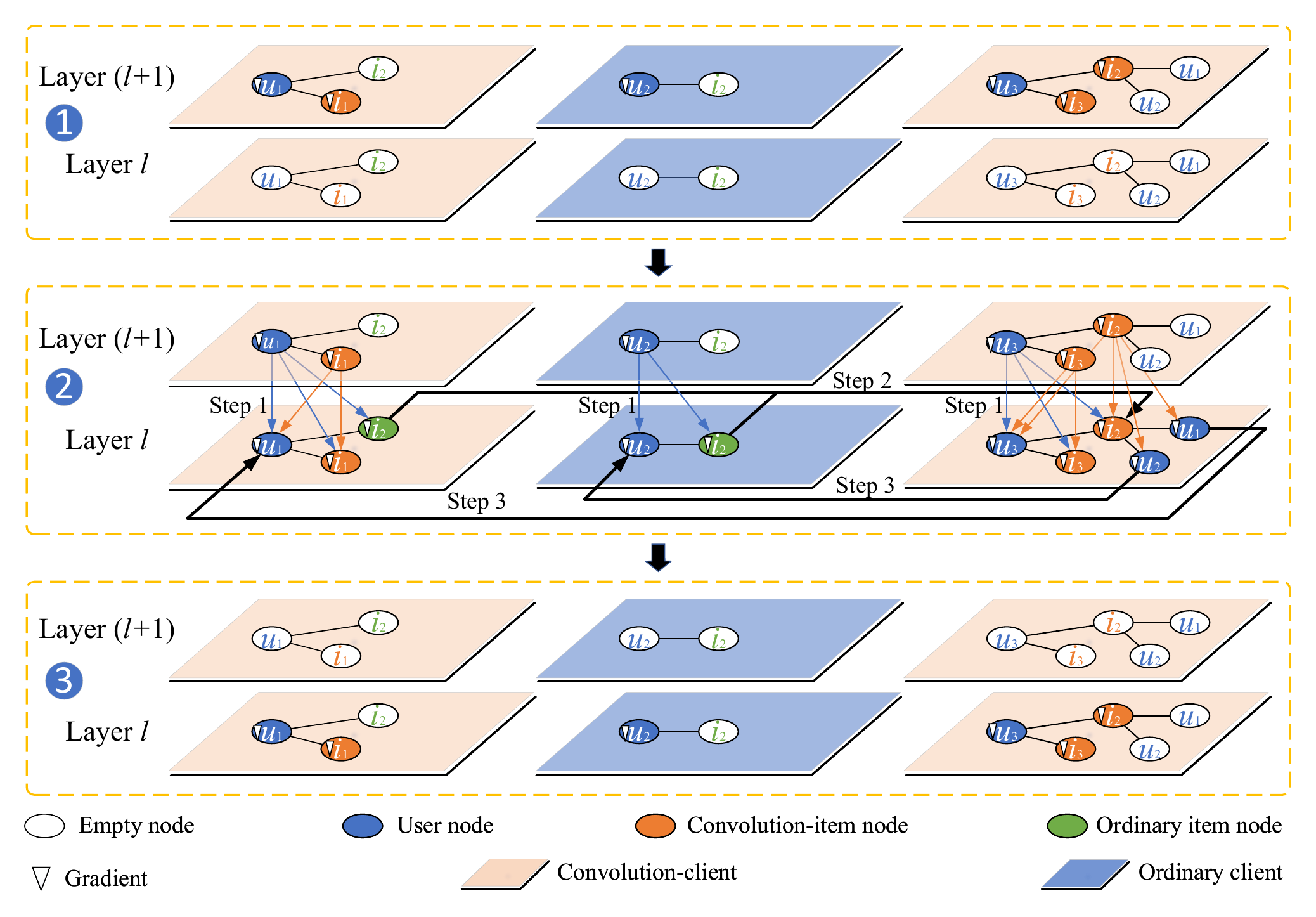}
	\caption{\textbf{An example of backward propagation.} There are three users ($u_1$, $u_2$, $u_3$) and three items ($i_1$, $i_2$, $i_3$). Clients $u_1$ and $u_3$ are convolution-clients, while $u_2$ is an ordinary client. In $u_1$, $i_1$ is a convolution-item, and in $u_3$, $i_2$ and $i_3$ are the convolution-items. First, Each client locally computes the gradients for the nodes at layer $l$ based on the gradients of the nodes at layer $(l+1)$. Then $u_1$ and $u_2$ send the gradient of the ordinary item $i_2$ to $u_3$. Meanwhile, $u_3$ sends the the gradients of the neighboring users' embeddings (i.e., user nodes $u_1$ and $u_2$ of layer $l$) to their respective clients.}
	\label{fig:An Example of Backward Propagation}
	\Description{An Example of Backward Propagation}
\end{figure*}

Backward propagation is essentially the reverse process of forward propagation. To clearly explain the backward propagation process, let's continue with the example of forward propagation we discussed earlier shown in Figure~\ref{fig:An Example of Forward Propagation}. Assume there are three users, $u_1$, $u_2$, and $u_3$, and three items, $i_1$, $i_2$, and $i_3$. In this setup, $i_1$ for $u_1$ and $i_2$ and $i_3$ for $u_3$ are marked as the convolution-items, while the remaining item, $i_2$ for $u_1$ and $u_2$, is considered as an ordinary item. Initially depicted in state 1 of Figure~\ref{fig:An Example of Backward Propagation}. The $(l+1)$-th layer contains gradients for user embeddings and the convolution-items. Subsequently, all clients locally compute the gradients for layer $l$ based on the gradients from the $(l+1)$-th layer, as shown in step 1 of the state 2 of Figure~\ref{fig:An Example of Backward Propagation}. Then, the gradients for ordinary items are sent back to the corresponding convolution-items for aggregation. Specifically, the gradients for $i_2$ from $u_1$ and $u_2$ are sent to client $u_3$ and accumulated there, as shown in step 2 of the state 2 of Figure~\ref{fig:An Example of Backward Propagation}. Finally, the convolution-clients send the neighboring users’ embedding gradients back to the respective clients. For instance, the client $u_3$ sends the gradients of $u_1$ and $u_2$ back to $u_1$ and $u_2$, respectively, and these are added on, as depicted in step 3 of the state 2 of Figure~\ref{fig:An Example of Backward Propagation}. As illustrated in state 3 of Figure~\ref{fig:An Example of Backward Propagation}, the gradients from layer $(l+1)$ are successfully propagated back to layer $l$. This process effectively demonstrates how gradients are back propagated through the network in a comprehensible manner.

\subsection{Parameter Update and Prediction}\label{sec:Parameter Update and Prediction}
(lines 10-15 in Algorithm~\ref{LP-GCN}) After forward propagation and backward propagation, we can obtain the final gradients of the training parameters, and then the clients can perform parameter update locally.

As for prediction phase, we know that the server has access to all item embeddings, which are non-sensitive information. These embeddings are then distributed to all clients, who can locally compute the dot product between their own user embeddings and the item embeddings to make predictions.

\subsection{Comprehensive Performance}\label{sec:Comprehensive Performance}
The comprehensive performance of our LP-GCN depends on the accuracy, privacy preservation, and efficiency. Formally, we can represent it as follows,
\begin{equation} \label{eq:comprehensive performance}
CPS (\lambda_1, \lambda_2) = Acc + Pri (\lambda_1) - Eff (\lambda_2)
\end{equation}
where $CPS$ stands for the comprehensive performance score. $Acc$ represents the score of accuracy, which is a constant. $Pri (\lambda_1)$ denotes the score of the level of privacy preservation, and $Eff (\lambda_2)$ represents the score of efficiency.

The parameter $\lambda_1$ in Eq.(\ref{eq:comprehensive performance}) is related to the number of virtual items (i.e., the parameter $\alpha$). As the value of $\alpha$ increases, $Pri (\lambda_1)$ rises, leading to stronger privacy protection. Similarly, $\lambda_2$ is influenced by both the number of virtual items $\alpha$ and the number of convolution clients $k$, which will be further discussed in Section~\ref{sec:experiments}. As the values of $\alpha$ and $k$ increase, $Eff (\lambda_2)$ also increases, resulting in lower efficiency.

\subsection{Proof}
We provide the detailed theoretical proof of losslessness in appendix~\ref{sec:Proof}, where we compare each phase of our LP-GCN with its corresponding centralized version to demonstrate the equivalence of the graph convolution process, thereby proving that our proposed federated framework is lossless.

\subsection{LP-GCN (LightGCN+)}\label{sec:LP-GCN (LightGCN+)}
\begin{algorithm}
\caption{LP-GCN (LightGCN+)}
\label{LP-GCN (LightGCN+)}
\begin{algorithmic}[1] 
\STATE Initialization(), i.e., Algorithm~\ref{Initialization}
\FOR{ $ t=1, 2, \ldots, T$ }
    \FOR{each convolution-client $\tilde{u} \in \tilde{\mathcal{U}}$ in parallel}
    \STATE  Sends the $0$-th layer's convolution-item embeddings $\{\boldsymbol{e}^{(0)}_{\tilde{i}(\tilde{u})} : \tilde{i} \in \tilde{\mathcal{I}}_{\tilde{u}}\}$ to the server.
    \ENDFOR
    \STATE The server receives all items' embeddings of the $0$-th layer $\{\boldsymbol{e}_{\tilde{i}}^{(0)} : \tilde{i} \in \mathcal{I}\}$ from all convolution-clients and distributes $\boldsymbol{e}_{\tilde{i}}^{(0)}$ to clients $\{u' : u' \in \mathcal{U}_{\tilde{i}} \backslash \{\tilde{u}_{\tilde{i}}\}\}$ (i.e., the users who has interacted with item $\tilde{i}$ excluding $\tilde{u}$ where $\tilde{i}$ locates).
    \STATE Forward Propagation, i.e., Algorithm~\ref{Forward Propagation}.
    \STATE Loss Construction and Gradients Computation, i.e., Algorithm~\ref{Loss Construction and Gradients Computation}.
    \STATE Backward Propagation, i.e., Algorithm~\ref{Backward Propagation}.
    \FOR{each client $u \in \mathcal{U}$ in parallel}
    \STATE \hl{Computes the gradients of embeddings, acquiring $\{\nabla \boldsymbol{w}_i: i \in \mathcal{I}_{u} \cup \mathcal{I}_u'\}$ and then sends them to the server.}
    \IF{$u \in \tilde{\mathcal{U}}$}
    \STATE Updates the convolution-item embeddings $\boldsymbol{e}^{(0)}_{i(u)} \leftarrow \boldsymbol{e}^{(0)}_{i(u)} - \gamma \boldsymbol{g}^{(0)}_{i(u)}(i \in \tilde{\mathcal{I}}_u)$.
    \ENDIF
    \ENDFOR
    \STATE \hl{The server receives the gradients of $\nabla\boldsymbol{W}$, updates them, and then sends $\{\boldsymbol{W}_i: i\in \mathcal{I}_{u} \cup \mathcal{I}_u'\}$ (i.e., the embeddings of all items that user $u$ has interacted with, including the virtual items) to each client $u$.}
\ENDFOR
\end{algorithmic}
\end{algorithm}

\begin{algorithm}
\caption{Initialization (LightGCN+)}
\label{Initialization(LightGCN+)}
\begin{algorithmic}[1] 
\FOR{each client $u$ in parallel}
    \STATE Generates a pair of keys by asymmetric encryption, including a private key $PR_u$ and a public key $PB_u$.
    \STATE Sends the public key $PB_u$ to the server.
\ENDFOR

\STATE The server collects the public keys of all clients $\{PB_u : u \in \mathcal{U}\}$ and sends them to client $u_c$ (i.e., one user randomly chosen by the server).

\STATE The client $u_c$ generates a shared key $S$ by symmetric encryption and encrypts $S$ with each of the public keys, acquiring the ciphertexts $\{{PB_u}(S) : u \in \mathcal{U}\}$ which are then sent to the server.

\STATE The server receives the ciphertexts $\{{PB_u}(S) : u \in \mathcal{U}\}$ and distributes each ${PB_u}(S)$ to client $u$.

\FOR{each client $u \in \mathcal{U}$ in parallel}
    \STATE Receives ${PB_u}(S)$ and decrypts it with the private key $PR_{u}$,  so as to acquire the shared key $S$.
    \STATE Generates $\alpha$ virtual items $\{i' : i' \notin \mathcal{I}_u\}$ denoted as $\mathcal{I}_u'$.
\ENDFOR
\FOR{each client $u \in \mathcal{U}$ in parallel}
    \STATE Encrypts item-IDs $\{ ID_{i(u)} : i \in \mathcal{I}_{u} \cup \mathcal{I}_u'\}$ with the shared key $S$, in order to acquire the ciphertexts of the item-IDs $\{S(ID_{i(u)}) : i \in \mathcal{I}_{u} \cup \mathcal{I}_u'\}$ and then sends them to the server.
\ENDFOR
\STATE The server performs preprocessing for handling the item-IDs in the server (i.e., Algorithm~\ref{Preprocessing for handling item-IDs in the server}), so as to split the clients and items.
\FOR{each client $\tilde{u} \in \tilde{\mathcal{U}}$ in parallel}
    \STATE Verifies whether each $u' \in \mathcal{U}_{\tilde{i}} \backslash \{\tilde{u}_{\tilde{i}}\}$ has really interacted with each convolution-item $\tilde{i} \in \tilde{\mathcal{I}}_{\tilde{u}}$.
\ENDFOR
\STATE \hl{The server initializes another item embedding matrix $\boldsymbol{W} \in \mathbb{R}^{m\times d}$ according to the encrypted item-IDs, and then sends $\{\boldsymbol{W}_i: i\in \mathcal{I}_{u} \cup \mathcal{I}_u'\}$ (i.e., the embeddings of all items) to each client $u$.}
\STATE \hl{The server obtains the degree of each item according to the global graph of the encrypted item-IDs and then informs each client of the degree of each item it interacts with.}
\FOR{each client $u \in \mathcal{U}$ in parallel}
    \STATE \hl{Constructs $\boldsymbol{e}^{(0)}_{u(u)}$ (i.e., the $0$-th layer's user embedding in client $u$) using $\{\boldsymbol{W}_i: i\in \mathcal{I}_{u} \cup \mathcal{I}_u'\}$.}
    \IF{$u \in \tilde{\mathcal{U}}$}
        \STATE Initializes $\{\boldsymbol{e}^{(0)}_{\tilde{i}(u)} : \tilde{i} \in \tilde{\mathcal{I}}_u\}$ (i.e., the set of the $0$-th layer's convolution-item embeddings in client $u$) and then sends them to the server.
    \ENDIF
\ENDFOR
\end{algorithmic}
\end{algorithm}

In this section, we will instantiate our GNN-based federated recommendation framework LP-GCN, using LightGCN+ as the backbone, which has been mentioned in Section~\ref{LightGCN+}.

We will discuss the changes that occur during the instantiation process in the order introduced by the framework. The forward propagation and backward propagation processes remain largely unchanged, but there are some significant changes in Initialization and Loss Construction and Gradients Computation. Below, we will provide a detailed introduction and discuss the reasons for these changes.

There are three changes in Initialization.
Firstly (line 19 in Algorithm~\ref{Initialization(LightGCN+)}), the server needs to initialize another item embedding matrix denoted $\boldsymbol{W} \in \mathbb{R}^{m\times d}$, and then sends $\{\boldsymbol{W}_i: i\in \mathcal{I}_{u} \cup \mathcal{I}_u'\}$ (i.e., the embeddings of all items that user $u$ has interacted with, including the virtual items used to protect users' privacy) to each client $u$ for constructing the user embedding later.
Secondly (line 20 in Algorithm~\ref{Initialization(LightGCN+)}), the server also needs to distribute the degrees of all items to all clients for the graph convolution operation later.
Thirdly (line 22 in Algorithm~\ref{Initialization(LightGCN+)}), all clients construct the user embedding by item embedding via Eq.(\ref{eq:construct_user_embedding}). Note that we don't use the embedding of virtual items.

\begin{algorithm}[t]
\caption{Loss Construction and Gradients Computation (LightGCN+)}
\label{Loss Construction and Gradients Computation (LightGCN+)}
\begin{algorithmic}[1]
\STATE \hl{The server samples some training pairs for each client $u$ denoted $P_u$ and sends all layers' item embeddings of the negative items $\{\boldsymbol{e}_{i'}^{(l)} : i' \in \mathcal{I}_u^{neg};l = 0, 1, \ldots, L\}$ to each client $u$.}
\FOR{each client $u \in \mathcal{U}$ in parallel}
    \STATE Computes the final user embedding $\boldsymbol{e}_{u(u)}$ via Eq.(\ref{eq:construct_final_user_embedding}).
    \STATE Computes the final item embeddings $\boldsymbol{e}_{i(u)}$ via Eq.(\ref{eq:construct_final_item_embedding}).
    \STATE Computes the predicted value $\{\hat{y}_{ui(u)}: i \in \mathcal{I}_u \}$ via Eq.(\ref{eq:predicted_score}).
    \STATE Computes the local training loss $Loss(u)_{(u)}$ via Eq.(\ref{eq:compute one of the local train loss}) and Eq.(\ref{eq:compute the local train loss}).
    \STATE Computes the gradients of all layers via Eq.(\ref{eq:computes user gradients of all layers}) and Eq.(\ref{eq:computes item gradients of all layers}).
    \STATE Sends the last layers' ordinary item-embedding gradients $\{\boldsymbol{g}_{i(u)}^{(L)} : i \in \mathcal{I}_u \backslash \tilde{\mathcal{I}}_u\}$ to the server.
    \STATE \hl{Sends the embedding gradients of negative items $\{\boldsymbol{g}_{i(u)}^{(l)}:l = 0, 1, \ldots, L ; i \in \mathcal{I}_u^{neg}\}$ to the server.}
\ENDFOR
\STATE The server receives all the item embedding gradients of the last layer $L$ $\{\boldsymbol{g}_{i(u')}^{(L)}: i \in \mathcal{I}, u' \in \mathcal{U}_i \backslash \tilde{u}_i\}$, aggregates them via Eq.(\ref{eq:aggregate item embedding gradient}) and distributes each $\boldsymbol{g}_i^{(L)}(i \in \mathcal{I})$ to client $\tilde{u}_i$.
\STATE \hl{The server receives the embedding gradients of all negative items, aggregates them, and then distributes $\{\boldsymbol{g}_i^{(l)}:i \in \mathcal{I};l = 0, 1, \ldots, L \}$ to client $\tilde{u}_i$.}
\FOR{each client $\tilde{u} \in \tilde{\mathcal{U}}$ in parallel}
    \STATE Receives the gradients $\{\boldsymbol{g}_i^{(L)}: i \in \tilde{\mathcal{I}}_{\tilde{u}}\}$ and adds each $\boldsymbol{g}_i^{(L)}(i \in \tilde{\mathcal{I}}_{\tilde{u}})$ to the corresponding local gradient $\boldsymbol{g}_{i(\tilde{u})}^{(L)}$ via Eq.(\ref{eq:add item gradient}).
    \STATE \hl{Receives the embedding gradients of the negative items $\{\boldsymbol{g}_i^{(L)}: i \in \tilde{\mathcal{I}}_{\tilde{u}};l = 0, 1, \ldots, L \}$ and adds each $\boldsymbol{g}_i^{(l)}(i \in \tilde{\mathcal{I}}_{\tilde{u}};l = 0, 1, \ldots, L )$ to the corresponding local gradient $\boldsymbol{g}_{i(\tilde{u})}^{(l)}$.}
\ENDFOR
\end{algorithmic}
\end{algorithm}

In Forward Propagation (Algorithm~\ref{Forward Propagation}) , we just need to concrete the formulas, i.e., replacing Eq.(\ref{eq:user_embedding_convolution}) and Eq.(\ref{eq:compute_convolution_item_embedding}) with Eq.(\ref{eq:lightGCN compute user embedding}) and Eq.(\ref{eq:lightGCN compute item embedding}), respectively.
In Loss Construction and Gradients Computation (Algorithm~\ref{Loss Construction and Gradients Computation (LightGCN+)}), there are four major changes.
Firstly (line 1 in Algorithm~\ref{Loss Construction and Gradients Computation (LightGCN+)}), the server samples training pairs in the BPR manner for each client $u$ denoted as $P_u$ and sends the item embeddings of all negative items at each layer $\{\boldsymbol{e}_{i'}^{(l)} : i' \in \mathcal{I}_u^{neg};l = 0, 1, \ldots, L\}$ to each client $u$.
Secondly (line 9 in Algorithm~\ref{Loss Construction and Gradients Computation (LightGCN+)}), each client $u$ needs to compute the embedding gradients of the negative items $\{\boldsymbol{g}_{i(u)}^{(l)}:l = 0, 1, \ldots, L ; i \in \mathcal{I}_u^{neg}\}$ and sends them to the server.
Thirdly (line 12 in Algorithm~\ref{Loss Construction and Gradients Computation (LightGCN+)}), the server receives the embedding gradients of all negative items, aggregates them in the same manner as it does for other items, and then distributes $\{\boldsymbol{g}_i^{(l)}:i \in \mathcal{I};l = 0, 1, \ldots, L \}$ to client $\tilde{u}_i$.
Fourthly (line 15 in Algorithm~\ref{Loss Construction and Gradients Computation (LightGCN+)}), each convolution-client $\tilde{u}$ receives the embedding gradients of the negative items $\{\boldsymbol{g}_i^{(L)}: i \in \tilde{\mathcal{I}}_{\tilde{u}}; l = 0, 1, \ldots, L \}$ and adds each $\boldsymbol{g}_i^{(l)}(i \in \tilde{\mathcal{I}}_{\tilde{u}}; l = 0, 1, \ldots, L )$ to the corresponding local gradient $\boldsymbol{g}_{i(\tilde{u})}^{(l)}$.

After backward propagation, the gradients back are propagated to the user embeddings of all clients, which need to be backward propagated further. Firstly (line 11 in Algorithm~\ref{LP-GCN (LightGCN+)}), each client $u$ computes the gradients,
\begin{equation} \label{eq:clients compute item embedding gradient}
\nabla \boldsymbol{w}_{i(u)} = \frac{\partial \boldsymbol{g}^{(0)}_{u(u)}}{\partial \boldsymbol{w}_{i(u)}} \quad (i \in \mathcal{I}_u)
\end{equation}
where $\nabla \boldsymbol{w}_{i(u)}$ represents the gradient of item $\boldsymbol{w}_{i(u)}$. Note that $\boldsymbol{w}_{i(u)}$ is the embedding of item $i$, which comes from the item embedding matrix $\boldsymbol{W}$. Then each client $u$ acquires $\{\nabla \boldsymbol{w}_i: i \in \mathcal{I}_u\}$. Next, each client $u$ randomly assigns a value for each virtual item $i \in \mathcal{I}_u'$, acquiring $\{\nabla \boldsymbol{w}_i: i \in \mathcal{I}_{u} \cup \mathcal{I}_u'\}$
and then sends them to the server.
Secondly (line 16 in Algorithm~\ref{LP-GCN (LightGCN+)}), the server receives the gradients $\nabla\boldsymbol{W}$, updates them, and then sends $\{\boldsymbol{W}_i: i\in \mathcal{I}_{u} \cup \mathcal{I}_u'\}$ (i.e., the embeddings of all items that user $u$ has interacted with, including the virtual items) to each client $u$.

\section{Theoretical Analysis}\label{sec:analysis}
\subsection{Privacy Analysis}\label{sec:Privacy Analysis}
In a GNN-based federated recommendation model, each user’s local sub-graph represents sensitive user information that must be safeguarded. Additionally, the user embeddings in our LP-GCN, which hold sensitive details about users and can be used to predict their item preferences, should also be protected.
Additionally, some studies have shown that in matrix factorization based recommendation models, user embeddings can be inferred from the gradients, leading to privacy leaks~\cite{chai2020secure,lin2022generic}. Although this issue has not yet been discovered in GNN-based recommendation models, we have taken steps to protect the gradients to enhance the security of our LP-GCN.

We explore three typical attacks on machine learning models: reconstruction attacks, model inversion attacks, and membership inference attacks.

Firstly, we focus on reconstruction attacks during the model training phase. In these attacks, an entity such as a server or another client tries to reconstruct the original data from other clients. Specifically, they attempt to recover the items users have interacted with based on the data they receive.
We first consider the case where the server is the attacker. Since the item-IDs that clients send to the server are encrypted by a common shared key $S$ that is unknown to the server, the server can not reconstruct the data, i.e., the items that users have interacted with.
As for the case where the attacker is a client $u$. Since all of a user's original data is stored locally and the user does not send his or her data to other users directly, there will be no privacy leaks.

Secondly, we consider model inversion attacks to our LP-GCN, where an attacker
tries to extract the training data from the trained model.  We first consider the case where the server is the attacker.
Since a user's embedding is transmitted through encryption, the server cannot decrypt it as it does not have the shared key $S$. Additionally, for the gradient of each node, backward propagation is performed by aggregating the gradients from other clients in the server before sending it back to a specific client, the server only knows a part of the gradient. Therefore, the gradient is also protected.
As for the case where the attacker is a client $u$. Each convolution-client $\tilde{u}$ will receive the convolution-item based neighboring users' embeddings. However, it doesn't know the origin of each embedding.

Finally, we focus on the membership-inference attacks, where an attacker tries to determine whether a specific piece of data is included in the model's training set. 
Regardless of whether the attacker is a server or a client, the precondition for carrying out the attack is to obtain a user's embedding. However, as analyzed in the model inversion attacks, a user's embedding is well protected, making it resistant to member-inference attacks.

Additionally, in the processes of Loss Construction and Gradients Computation and parameter update, all operations are performed locally without any data transmission, so the aforementioned attack methods are ineffective in these processes.

Moreover, in our LP-GCN, each client contains a subset of virtual items, so even if the server colludes with some individual clients, it can only infer a portion of the virtual items. Unless the server colludes with the majority of the convolution-clients, it is difficult to infer the items a user has interacted with, which also means that our LP-GCN is more secure compared with the existing GNN-based federated algorithms.

In summary, our LP-GCN framework can effectively resist common attacks on machine learning models while safeguarding user privacy. Even in cases where the server colludes with certain clients, it ensures a robust level of privacy protection for GNN-based recommendation algorithms.
\subsection{Communication Cost}
As described in the model introduction in Section~\ref{sec:LP-GCN}, each node computes only once and then synchronizes by transmitting data to different clients. Therefore, our LP-GCN does not generate additional computational load compared with the centralized counterpart, which means we just need to transmit necessary embeddings for graph convolution. Next, we will analyze the communication cost of each client and the server in detail.

In the initialization phase, the communication cost mainly involves sharing the key $S$ and constructing the global graph of the encrypted item-IDs. However, compared with the model training phase, it could be ignored.

At each iteration of  training, our LP-GCN successively conducts forward propagation and backward propagation.
Firstly, we focus on forward propagation. 
Each convolution-client needs to receive the embeddings of the neighboring users and ordinary items, and send the embeddings of the convolution-items as well as its own user embedding, which takes $O((|\mathcal{N}_{\tilde{u}}|+|\mathcal{I}_{\tilde{u}}|+|\mathcal{I}'_{\tilde{u}}|+1)dL)$. 
Each ordinary client needs to send its user embedding and receives the item embeddings, taking $O((|\mathcal{I}_u|+|\mathcal{I}'_u|+1)dL)$. 
Owing to both ordinary clients and convolution-clients sending and receiving data through the server, the server takes $O((\sum_{\tilde{u} \in \tilde{\mathcal{U}}} (|\mathcal{N}_{\tilde{u}}|+|\mathcal{I}_{\tilde{u}}|+|\mathcal{I}'_{\tilde{u}}|+1)+\sum_{u \in \mathcal{U} \backslash \tilde{\mathcal{U}}}(|\mathcal{I}_u|+|\mathcal{I}'_u|+1))dL)$. 
As for backward propagation, since it is the reverse process of forward propagation, the communication cost is the same.
Therefore, the overall communication cost of an ordinary client $u$, a convolution-client $\tilde{u}$, and the server is $O((|\mathcal{I}_u|+|\mathcal{I}'_u|+1)dL)$, $O((|\mathcal{N}_{\tilde{u}}|+|\mathcal{I}_{\tilde{u}}|+|\mathcal{I}'_{\tilde{u}}|+1)dL)$, and $O((\sum_{\tilde{u} \in \tilde{\mathcal{U}}} (|\mathcal{N}_{\tilde{u}}|+|\mathcal{I}_{\tilde{u}}|+|\mathcal{I}'_{\tilde{u}}|+1)+\sum_{u \in \mathcal{U} \backslash \tilde{\mathcal{U}}}(|\mathcal{I}_u|+|\mathcal{I}'_u|+1))dL)$, respectively.

During the model inference phase, the server sends the embeddings of the items that each user $u$ has not interacted with. The user then calculates the prediction values locally to get recommended items. The communication cost in this process is negligible compared with the training phase.

To sum up, the overall communication cost of an ordinary client $u$, a convolution-client $\tilde{u}$, and the server is $O((|\mathcal{I}_u|+|\mathcal{I}'_u|+1)dL)$, $O((|\mathcal{N}_{\tilde{u}}|+|\mathcal{I}_{\tilde{u}}|+|\mathcal{I}'_{\tilde{u}}|+1)dL)$, and $O((\sum_{\tilde{u} \in \tilde{\mathcal{U}}} (|\mathcal{N}_{\tilde{u}}|+|\mathcal{I}_{\tilde{u}}|+|\mathcal{I}'_{\tilde{u}}|+1)+\sum_{u \in \mathcal{U} \backslash \tilde{\mathcal{U}}}(|\mathcal{I}_u|+|\mathcal{I}'_u|+1))dL)$, respectively.

\section{Experiments} \label{sec:experiments}
We conduct experiments on three public datasets to study the effectiveness and efficiency of our LP-GCN. We focus on the following five questions:
\begin{itemize}
    \item RQ1: Is our LP-GCN lossless, meaning whether it can fully complete both forward propagation and backward propagation?
    \item RQ2: How is the performance of our LP-GCN compared with the existing baselines?
    \item RQ3: How are the the convergence rates of P-GCN, LightGCN and our LP-GCN?
    \item RQ4: How are the convergence rates of P-GCN, LightGCN, and our LP-GCN (LightGCN+)?
    \item RQ5: How does the number of convolution-clients, i.e., $k$, affect the communication cost?
\end{itemize}
\subsection{Datasets and Evaluation Metrics}

\definecolor{lightgray}{rgb}{0.9, 0.9, 0.9}

\begin{table}[htbp]
    \caption{Statistics of the evaluation datasets, including the number of users ($n$), items ($m$), interactions ($|\mathcal{R}|$), and data density ($|\mathcal{R}|/n/m$).}
    \label{tab:dataset}
    \centering
    \setlength{\tabcolsep}{4pt} %
    \renewcommand{\arraystretch}{1.2} 
    \begin{tabular}{L{2.5cm} C{1.5cm} C{1.5cm} C{1.5cm} C{1.5cm}}
        \toprule
        Dataset & $n$ & $m$ & $|\mathcal{R}|$ & Density \\
        \midrule
        Gowalla       & 29,858 & 40,981   & 1,027,370  & 0.084\% \\
        Yelp2018        & 31,668 & 38,048   & 1,561,406  & 0.130\% \\
        Amazon-Book   & 52,643 & 91,599   & 2,984,108  & 0.062\% \\
        \bottomrule
    \end{tabular}
\end{table}

We use the same evaluation datasets (including 80-10-10 training-validation-test split) as those in P-GCN~\cite{hu2023privacy}, namely Gowalla, Yelp2018 and Amazon-Book.
The statistics of the evaluation datasets are shown in Table~\ref{tab:dataset}.
In evaluation, each model will generate a ranked list of items for each user by sorting all items that the user has not interacted with.
Following LightGCN~\cite{he2020lightgcn} and P-GCN~\cite{hu2023privacy}, we employ two commonly used evaluation metrics, namely recall and normalized discounted cumulative gain (NDCG), with the default number of recommended items set to $K=20$.
In other words, we utilize recall@20 and NDCG@20 to evaluate the top-20 items in each ranked list.
\subsection{Baselines and Parameter Settings}\label{sec:baseline}
In our experiments, the primary comparison is with P-GCN, a state-of-the-art GNN-based federated recommendation model, and LightGCN~\cite{he2020lightgcn}, which serves as the backbone of P-GCN~\cite{hu2023privacy}.
In addition, we also include the existing methods for item recommendation, along with their federated versions.
The details of the baselines are provided below.
\begin{itemize}
    \item BPR~\cite{rendle2012bpr}: This is an early classic method that uses matrix factorization to solve the item recommendation problem, with its core optimized w.r.t. the pairwise ranking loss in
    Eq.(\ref{eq:loss}).
    \item NCF~\cite{he2017neural}: This is a centralized version of FedNCF~\cite{jiang2022fedncf}. 
    It is a classic recommendation model, which combines neural networks with collaborative filtering, using nonlinear capabilities to enhance the recommendation performance through user and item embeddings.
    For a fair comparison, we follow~\cite{wang2019neural} and use the BPR loss when implementing a multilayer perceptron (MLP) with two hidden layers, i.e., 2$d$ → $d$, where $d$ represents the embedding size.
    \item Mult-VAE~\cite{shenbin2020recvae}: This is a centralized version of FedVAE~\cite{polato2021federated}.
    It leverages variational autoencoders to model complex user-item interactions by using an encoder to map interactions to a latent space and a decoder to reconstruct them.
    We adopt the same model architecture as described in~\cite{shenbin2020recvae}, specifically $600 \rightarrow 200 \rightarrow 600$, and search the best value of $\beta$ within the set $\{0.2, 0.4, 0.6, 0.8\}$.
    \item LightGCN~\cite{he2020lightgcn}: It is a simplified and efficient graph convolutional network designed for item recommendation, addressing the complexity and computational overhead of traditional GCNs while maintaining high recommendation performance.
    It is also the backbone model of P-GCN~\cite{hu2023privacy} and our LP-GCN.
    \item FedMF~\cite{chai2020secure}: It is a classic federated model that uses traditional matrix factorization as its backbone. By adopting the BPR loss in training, it becomes a federated version of BPR.
    
    \item FedNCF~\cite{jiang2022fedncf}: It is an advanced federated learning model that extends the NCF~\cite{he2017neural} framework to a decentralized environment, aiming to leverage the power of neural networks for collaborative filtering while ensuring data privacy by keeping user data local to their devices.
    \item FedPerGNN~\cite{wu2022federated}: It is a classic federated learning model that extends GNN to a federated and personalized setting.
    
    \item P-GCN~\cite{hu2023privacy}: It is a state-of-the-art GNN-based federated method for item recommendation, which is also the primary comparison method of our LP-GCN.
\end{itemize}
We do not include FedVAE~\cite{polato2021federated} as a baseline because that paper has demonstrated that FedVAE performs similarly to the non-federated version, i.e., Mult-VAE. Therefore, the performance of FedVAE can be inferred from the performance of Mult-VAE.
We do not include FedGRec~\cite{li2022fedgrec} and PerFedRec~\cite{luo2022personalized} as baselines because their experimental results show that the performance of these methods is inferior to their centralized counterpart (i.e., LightGCN). Our LP-GCN is equivalent to the centralized counterpart, i.e., it is lossles, as proven both theoretically and experimentally in Section~\ref{sec:Losslessness result}, which indicates that our LP-GCN outperforms both FedGRec and PerFedRec.

As same with P-GCN, we fix the embedding size as 64 for all models and optimize the model parameters using Adam with a learning rate of 0.001. We set the batch size to 500 users for Mult-VAE, and to 2048 samples for BPR, NCF, LightGCN, and LightGCN+. Following P-GCN, the batch size is fixed as 100 users for all federated models.
For the regularization coefficient $\lambda$, we search for the best value from $\{1e^{-2}, 1e^{-3}, ..., 1e^{-6}\}$, and the number of layers from $\{1, 2, 3, 4\}$ for GNN-based models. For our LP-GCN, all hyperparameters are set to match the backbone model. Other parameters of the baselines are configured according to P-GCN.
Additionally, we set the number of epochs to 1000, which is sufficient for all models to converge. As same with LightGCN, we employ an early stopping strategy, i.e., terminating training if recall@20 on the validation set does not improve for 50 consecutive epochs.

The source codes of our LP-GCN, scripts and datasets
are available at \url{https://github.com/SZU-GW/LP-GCN} (available after paper acceptance).
\subsection{Results}\label{sec:result}
\subsubsection{\textbf{RQ1: Losslessness}}\label{sec:Losslessness result}
\begin{figure}
	\centering
 	\subfigure{ \includegraphics[width=0.3\linewidth]{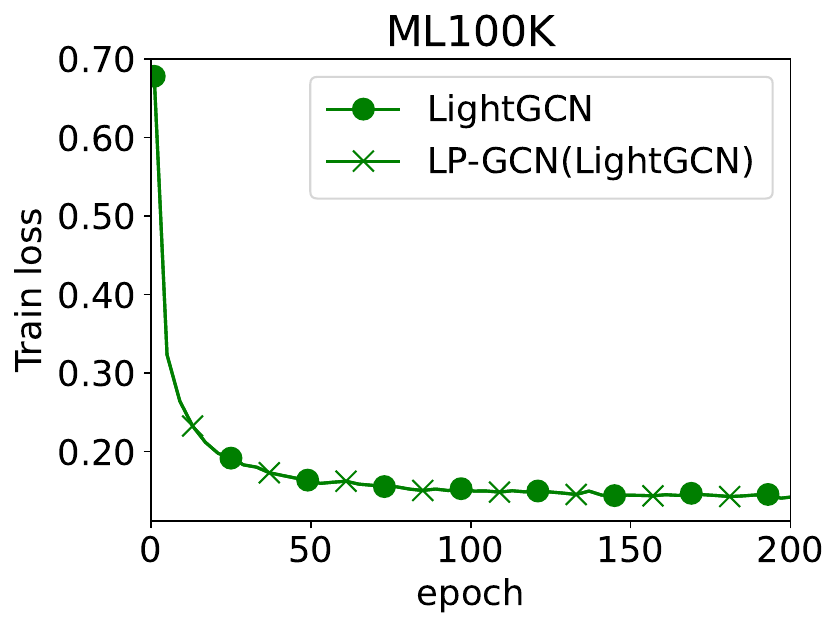} }
	\subfigure{ \includegraphics[width=0.3\linewidth]{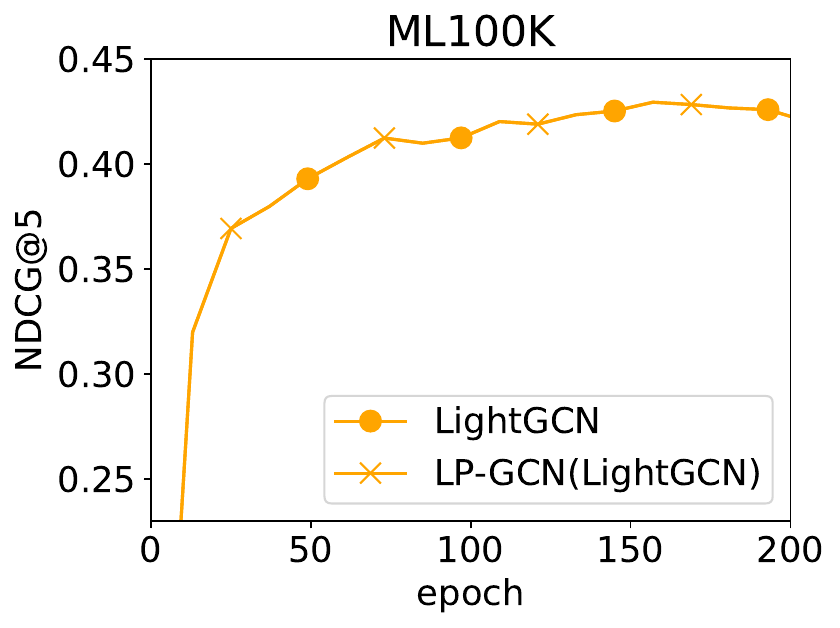} }
	\subfigure{ \includegraphics[width=0.3\linewidth]{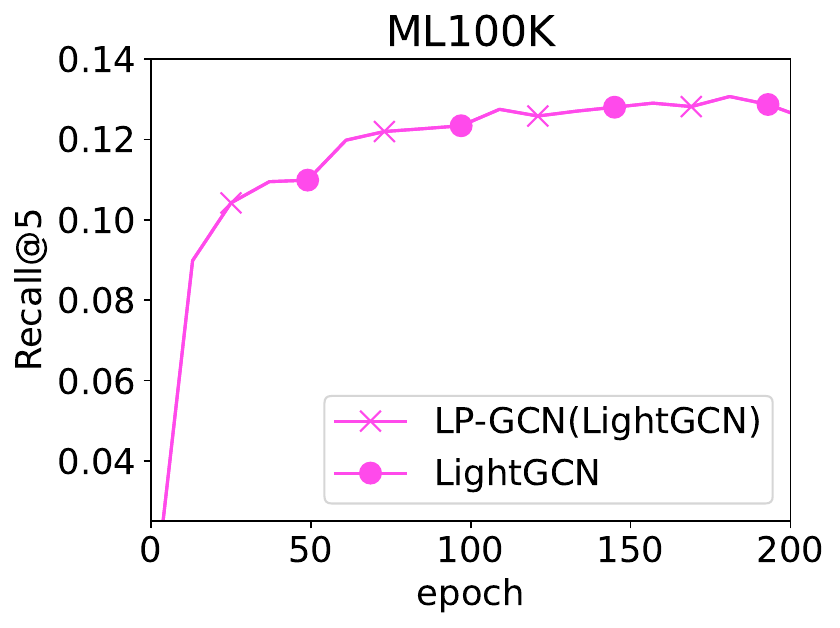} }

	\caption{Equivalence between the centralized version LightGCN and our federated version LP-GCN (LightGCN).}
	\label{fig:Lossless proved}
\end{figure}

%
Note that the focus of this experiment is not on performance, but to study whether our LP-GCN can fully complete both forward propagation and backward propagation, achieving graph convolution without loss. Since whether our LP-GCN is lossless is independent of the datasets and evaluation metrics, we choose a small dataset, ML100K. 

We report the training process in Fig~\ref{fig:Lossless proved}, from which we can draw the following observations: the training loss, NDCG@5 and Recall@5 of our LP-GCN (LightGCN) and LightGCN almost overlap, indicating that their training processes are identical, which confirms that our LP-GCN performs graph convolution without loss. Additionally, the final user embeddings and item embeddings are identical, as can be verified through the public code of our LP-GCN.
\subsubsection{\textbf{RQ2: Performance compared with the baselines}}

\begin{table}[htbp]
	\caption{Recommendation performance of the compared baselines and our LP-GCN (LightGCN+) on Gowalla, Yelp2018 and Amazon-Book. The best and second-best results of the federated recommendation methods are highlighted in bold and underlined, respectively.}
 	\label{tab:performance}
\centering
\resizebox{\textwidth}{!}{%
\begin{tabular}{c|c|cc|cc|cc}
\toprule
 & Dataset & \multicolumn{2}{c|}{Gowalla} & \multicolumn{2}{c}{Yelp2018} & \multicolumn{2}{|c}{Amazon-Book} \\
 \midrule
                        & Method                      & Recall@20 & NDCG@20 & Recall@20 & NDCG@20 & Recall@20 & NDCG@20 \\
\midrule
\multirow{5}{*}{Centralized} 
& BPR & 0.1670{\tiny$\pm$0.0018} & 0.1264{\tiny$\pm$0.0015} & 0.0527{\tiny$\pm$0.0006} & 0.0424{\tiny$\pm$0.0005} &0.0339{\tiny$\pm$0.0003} & 0.0258{\tiny$\pm$0.0003} \\
& NCF & 0.1685{\tiny$\pm$0.0015} & 0.1264{\tiny$\pm$0.0017} & 0.0468{\tiny$\pm$0.0005} & 0.0380{\tiny$\pm$0.0005} &0.0357{\tiny$\pm$0.0002} & 0.0272{\tiny$\pm$0.0001} \\
& Mult-VAE & 0.1761{\tiny$\pm$0.0014} & 0.1345{\tiny$\pm$0.0007} & 0.0607{\tiny$\pm$0.0005} & 0.0485{\tiny$\pm$0.0005} &0.0416{\tiny$\pm$0.0008} & 0.0321{\tiny$\pm$0.0005} \\
& LightGCN & 0.1894{\tiny$\pm$0.0014} & 0.1464{\tiny$\pm$0.0010} & 0.0627{\tiny$\pm$0.0006} & 0.0512{\tiny$\pm$0.0005} &0.0405{\tiny$\pm$0.0001} & 0.0314{\tiny$\pm$0.0001} \\
& LightGCN+ & 0.1937{\tiny$\pm$0.0002} & 0.1487{\tiny$\pm$0.0002} & 0.0653{\tiny$\pm$0.0002} & 0.0528{\tiny$\pm$0.0002} &0.0435{\tiny$\pm$0.0002} & 0.0332{\tiny$\pm$0.0002} \\
\midrule
\multirow{5}{*}{Federated} 
& FedMF & 0.1632{\tiny$\pm$0.0024} & 0.1243{\tiny$\pm$0.0016} & 0.0530{\tiny$\pm$0.0005} & 0.0431{\tiny$\pm$0.0003} &0.0322{\tiny$\pm$0.0003} & 0.0247{\tiny$\pm$0.0002} \\
& FedNCF & 0.1615{\tiny$\pm$0.0014} & 0.1209{\tiny$\pm$0.0012} & 0.0457{\tiny$\pm$0.0008} & 0.0366{\tiny$\pm$0.0007} &0.0327{\tiny$\pm$0.0003} & 0.0251{\tiny$\pm$0.0003} \\
& FedPerGNN & 0.1663{\tiny$\pm$0.0008} & 0.1244{\tiny$\pm$0.0003} & 0.0538{\tiny$\pm$0.0004} & 0.0429{\tiny$\pm$0.0004} &0.0340{\tiny$\pm$0.0003} & 0.0256{\tiny$\pm$0.0002} \\
& P-GCN & \underline{0.1882{\tiny$\pm$0.0005}} & \underline{0.1444{\tiny$\pm$0.0002}} & \underline{0.0623{\tiny$\pm$0.0004}} & \underline{0.0504{\tiny$\pm$0.0002}} & \underline{0.0409{\tiny$\pm$0.0003}} & \underline{0.0313{\tiny$\pm$0.0002}} \\
& \textbf{LP-GCN (LightGCN+)} & \textbf{0.1937{\tiny$\pm$0.0002}} & \textbf{0.1487{\tiny$\pm$0.0002}} & \textbf{0.0653{\tiny$\pm$0.0002}} & \textbf{0.0528{\tiny$\pm$0.0002}} & \textbf{0.0435{\tiny$\pm$0.0002}} & \textbf{0.0332{\tiny$\pm$0.0002}} \\

\bottomrule
\end{tabular}%
}
\end{table}

Considering the impact of randomness, we select five different random seeds to repeat the experiments, and then report the average performance with standard deviation. We show the recommendation performance of our LP-GCN and the compared baselines in Table~\ref{tab:performance}, from which we can observe:
\begin{itemize}
    \item Among the centralized non-federated methods, the best non-GNN-based method is Mult-VAE, while the best GNN-based method is LightGCN+. The latter consistently outperforms the former across all three datasets because GNN-based methods can utilize the high-order connectivity information in the user-item interaction graph to enhance the representation learning of users and items, which improves the recommendation performance. 
    \item Comparing the federated methods with their corresponding centralized counterparts, we can find that the federated versions typically perform worse. The reasons are as follows: For non-GNN-based methods, the performance loss is often due to the noise added to the gradients during aggregation on the server for privacy protection. In contrast, for GNN-based methods, the loss not only comes from the gradient aggregation process but also depends on whether the graph convolution process, including forward propagation and backward propagation, is as complete as in the centralized counterpart.
    FedPerGNN is a classic GNN-based federated method, but it is not lossless in both graph convolution and gradient aggregation. 
    P-GCN uses secure aggregation during gradient aggregation, making this part lossless. However, the graph convolution process is still not lossless. Therefore, the current best GNN-based federated model, i.e., P-GCN, performs better than FedPerGNN but is still worse than its centralized counterpart, LightGCN+.
    \item Among all federated methods, our LP-GCN (LightGCN+) consistently outperforms the existing methods across all three datasets. We have already discussed why P-GCN performs better than FedPerGNN. The reason our LP-GCN (LightGCN+) surpasses P-GCN is that, although they both use the same centralized model LightGCN+, P-GCN can only complete forward propagation and fails to fully achieve backward propagation. In contrast, our LP-GCN (LightGCN+) not only completes forward propagation but also overcomes the challenges of backward propagation, achieving lossless graph convolution. As a result, it is lossless as compared with the centralized counterpart. This will be validated again in subsequent experiments on hyperparameters and convergence.
\end{itemize}
\subsubsection{\textbf{RQ3: Performance compared with GNN-based methods at different layers}}

\begin{figure*}[ht]
	\centering
	\includegraphics[width=0.99\linewidth]{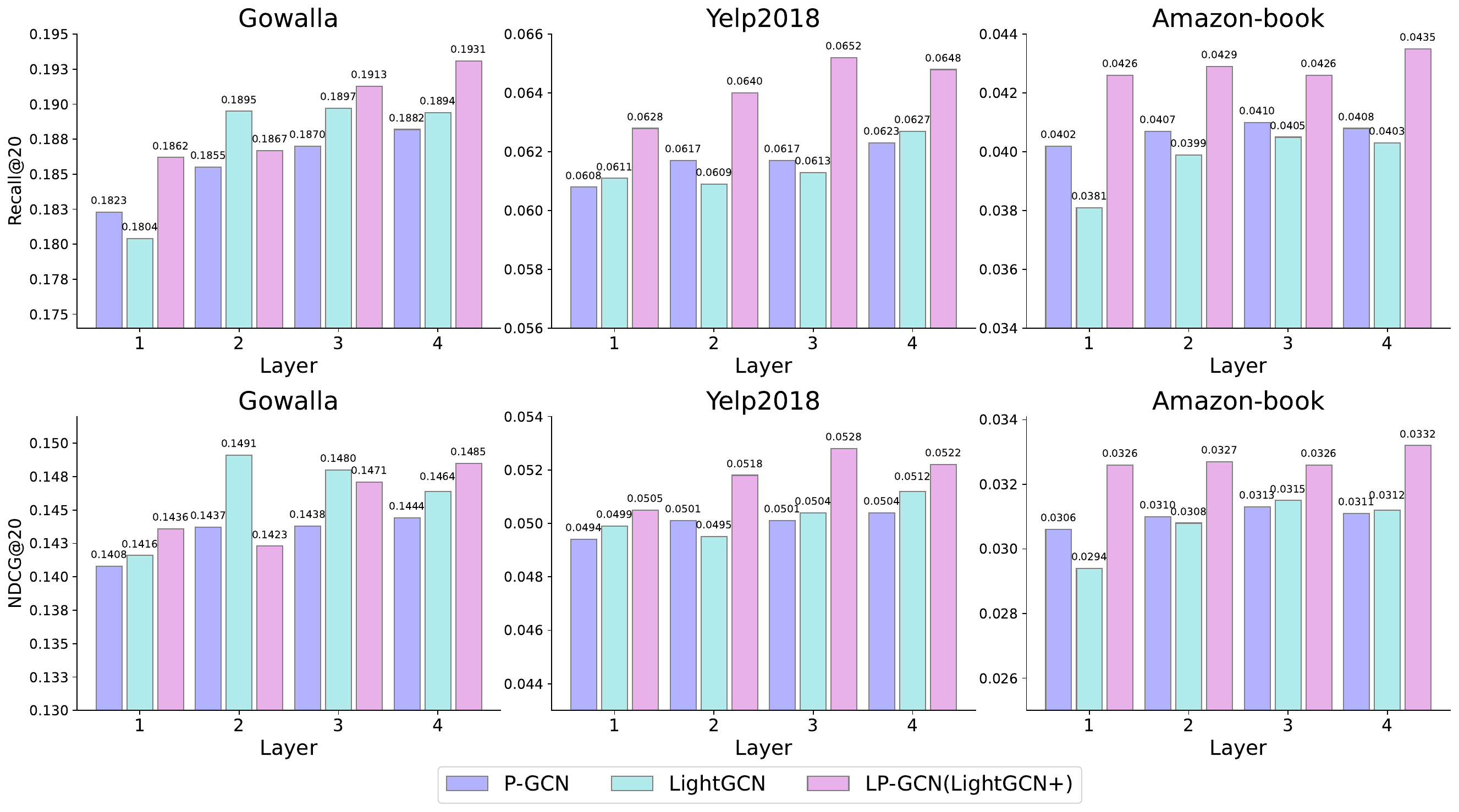}
	\caption{Recommendation performance of P-GCN, LightGCN and our LP-GCN (LightGCN+) at different layers.}
	\label{fig:different_layer}
	\Description{the interaction between the server and each client.}
\end{figure*}

Taking randomness into account, we use five different random seeds to repeat the experiments and report the average performance.
Since our LP-GCN (LightGCN+) and LightGCN+ are equivalent, we only present LP-GCN (LightGCN+). Figure~\ref{fig:different_layer} shows the recommendation performance of P-GCN, LightGCN and our LP-GCN (LightGCN+) from which we can have the following observations:
\begin{itemize}
    \item At any layer, the performance of LightGCN+ (i.e., our LP-GCN (LightGCN+)) is better than that of LightGCN across all three datasets. This improvement is due to the item-based user representation, which allows each user to share collaborative signals in the high-order connectivity captured by the user embedding~\cite{hu2023privacy}. 
    \item At any layer, the performance of our LP-GCN (LightGCN+) is better than that of P-GCN, which again shows that our LP-GCN (LightGCN+) is lossless, while P-GCN is not. It also demonstrates that successfully completing the backward gradient propagation is essential for further improving the recommendation performance.
\end{itemize}    
\subsubsection{\textbf{RQ4: Faster Convergence}}
\begin{figure}
	\centering
	\subfigure{ \includegraphics[width=0.3\linewidth]{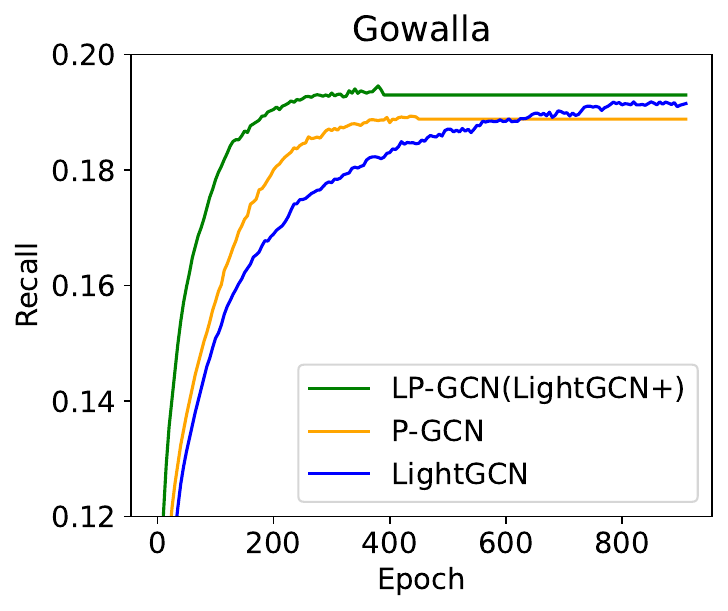} }
	\subfigure{ \includegraphics[width=0.3\linewidth]{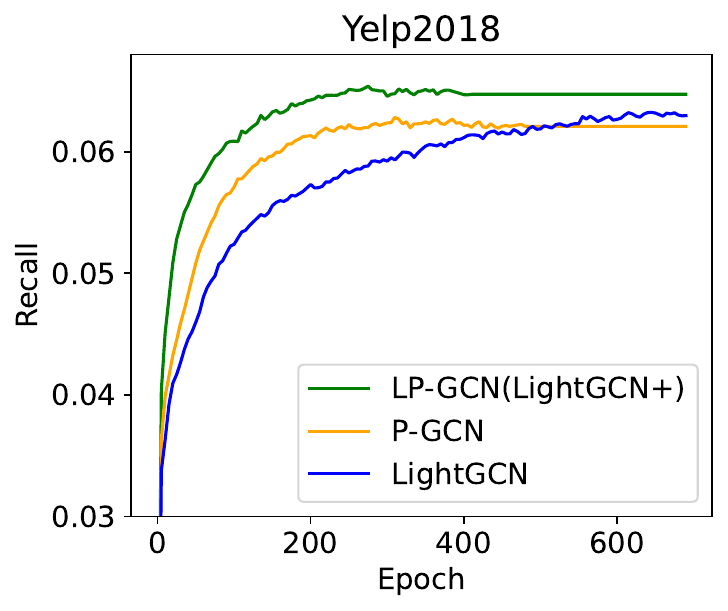} }
	\subfigure{ \includegraphics[width=0.3\linewidth]{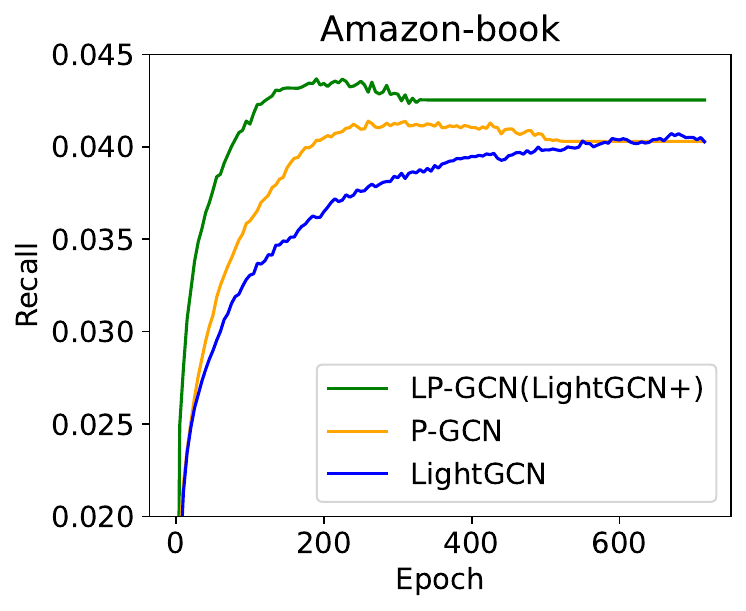} }
	
	\subfigure{ \includegraphics[width=0.3\linewidth]{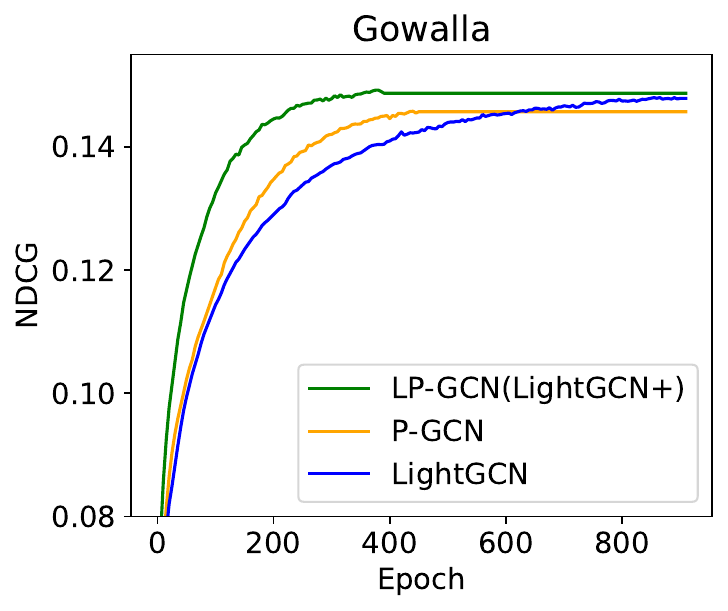} }
	\subfigure{ \includegraphics[width=0.3\linewidth]{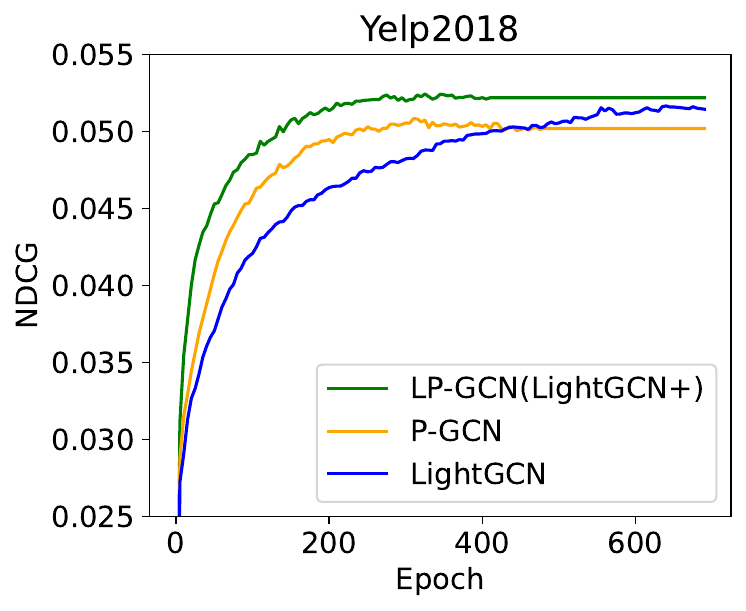} }
	\subfigure{ \includegraphics[width=0.3\linewidth]{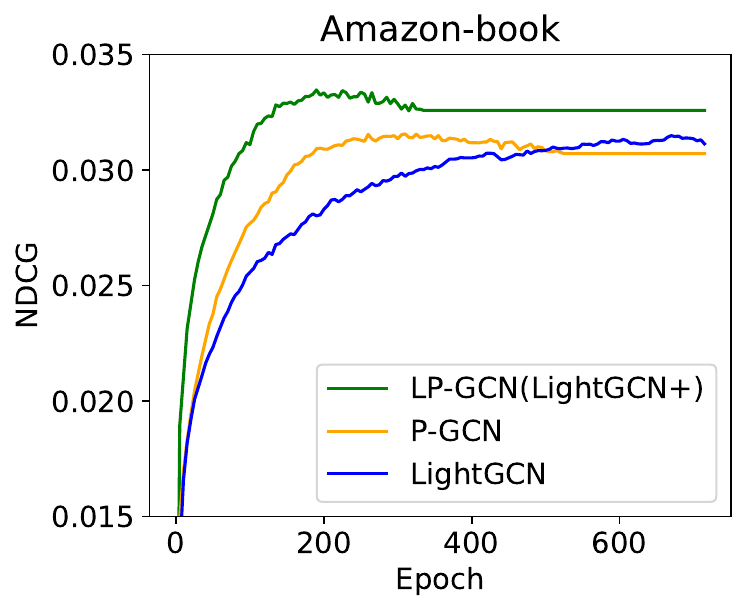} }
	\caption{The convergence curves of P-GCN, LightGCN and our LP-GCN (LightGCN+).}
	\label{fig:convergence}
\end{figure}


We show the convergence curves of LightGCN, P-GCN and our LP-GCN (LightGCN+) in Figure~\ref{fig:convergence}. We have excluded LightGCN+ from this comparison as we have demonstrated that the performance curves of our LP-GCN (LightGCN+) and LightGCN+ are identical in Section~\ref{sec:Losslessness result}. As illustrated in Figure~\ref{fig:convergence}, we have the following observations:

It is evident that the convergence rates for LightGCN, P-GCN and our LP-GCN (LightGCN+) increase in order. P-GCN performs better than LightGCN, primarily because it integrates item-based user representations. Though P-GCN and LP-GCN (LightGCN+) share the same centralized backbone model, P-GCN performs worse because it suffers from gradient loss. This attribute enables our LP-GCN (LightGCN+) to learn more accurately in each update iteration, resulting in faster convergence. This also suggests a more advantageous communication cost for our LP-GCN (LightGCN+).
\subsubsection{\textbf{RQ5: Communication cost}}
\begin{figure*}[ht]
	\centering
	\includegraphics[width=0.99\linewidth]{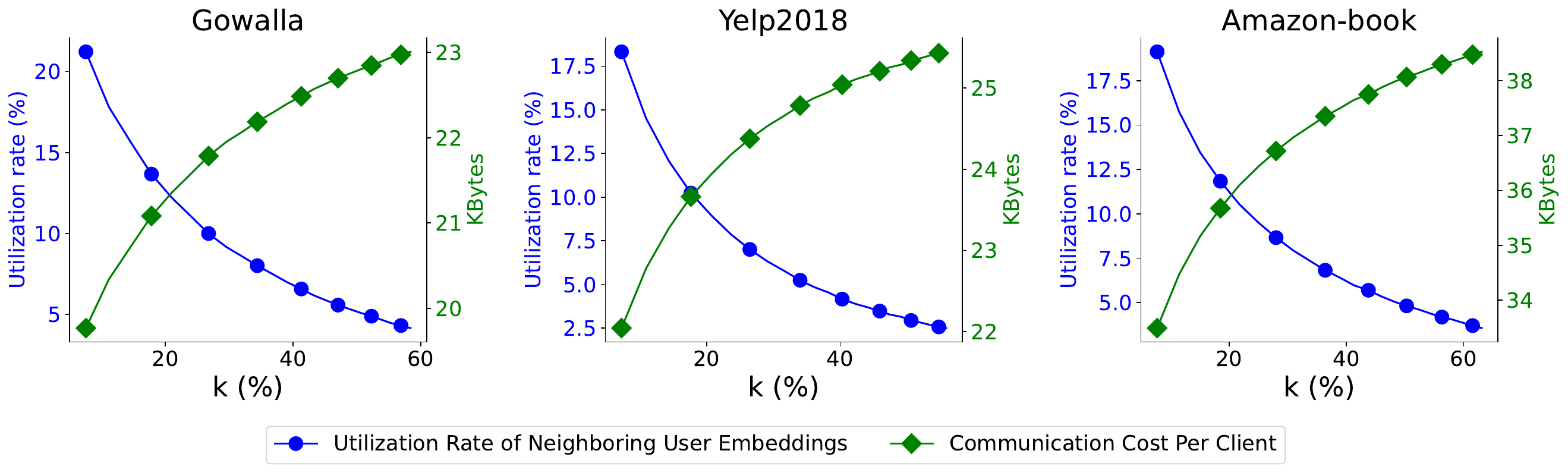}
	\caption{The utilization rate of the convolution-item based neighboring users' embeddings and communication cost per client w.r.t. different proportions of convolution-clients $k$.}
	\label{fig:communication_cost}
	\Description{The utilization rate of the convolution-item based neighboring users' embeddings and communication cost per client w.r.t. different proportion of convolution-clients k.}
\end{figure*}


We show the communication cost in Figure~\ref{fig:communication_cost}, from which we can observe:
\begin{itemize}
    \item The number of convolution-clients, i.e., $k$, varies across different datasets. On Gowalla, Yelp2018, and Amazon-Book, the ranges are [7.57, 59.30], [7.26, 57.05], and [7.72, 64.50], respectively. This variation is closely linked to our method for selecting convolution-clients. Specifically, the selection criteria requires that the union of the item sets interacted with by all convolution-clients contains all the items, which means that there is certainly a minimum number of convolution-clients. 
    \item Since the federated framework is a distributed system, clients are often small electronic devices such as smartphones and tablets, which have limited communication capabilities. The communication cost is therefore a critical factor in determining the practicality of a federated framework. From the perspective of client, we define a new metric to measure the communication cost:
    \begin{equation} \label{eq: communicaition cost}
    C = \frac{ dLB \sum_{\tilde{u} \in \tilde{\mathcal{U}}} |\mathcal{N}_{\tilde{u}}|}{|\mathcal{U}|}
    \end{equation}
    where $C$ represents the average communication cost per client in one iteration. $\sum_{\tilde{u} \in \tilde{\mathcal{U}}} |\mathcal{N}_{\tilde{u}}|$ is the number of the convolution-item based neighboring users' embeddings being transmitted, and $d$ is the embedding size, which is set to 64 in our experiments. $B$ is the size of each element in an embedding, with a 32-bit floating point number taking up 4 bytes. $L$ is the number of convolution layers in the model, where we choose the parameters that give the best performance. For the datasets Gowalla, Yelp2018, and Amazon-Book, the corresponding values are 4, 3, and 4, respectively. As shown in Figure~\ref{fig:communication_cost}, as the value of $k$ increases, the communication cost also rises. Since a larger $k$ reduces the number of convolution-items of each convolution-client on average, which in turn decreases the utilization of the neighboring users' embeddings, leading to higher communication cost. Therefore, we expect $k$ as small as possible.
    The communication cost on Gowalla, Yelp2018, and Amazon-Book are 19.8 KB, 22.0 KB, and 33.5 KB, respectively.
    In our LP-GCN, each node computes its results only once and then synchronizes with other clients. Consequently, all transmitted intermediate parameters are essential and do not incur any additional computational load compared with the centralized counterpart. Therefore, the communication cost is acceptable.
    
    Next, we will provide a more detailed explanation of the metric for the utilization of the convolution-item based neighboring users' embeddings.
    \item We define the utilization of the convolution-item based neighboring users' embeddings:
    \begin{equation} \label{eq: communicaition cost}
    r_{\tilde{u}} = \frac{\left( \sum_{\tilde{i} \in \tilde{\mathcal{I}}_{\tilde{u}}} |\left\{ \mathcal{U}_{\tilde{i}} \backslash \tilde{u} \right\}| \right) - |\mathcal{N}_{\tilde{u}}|}{|\mathcal{N}_{\tilde{u}}|}
    \end{equation}
    \begin{equation} \label{eq: communicaition cost}
    R = \frac{\sum_{\tilde{u} \in \tilde{\mathcal{U}}} r_{\tilde{u}}}{|\tilde{\mathcal{U}}|}
    \end{equation}
    where $\left( \sum_{\tilde{i} \in \tilde{\mathcal{I}}_{\tilde{u}}} |\left\{ \mathcal{U}_{i} \backslash \tilde{u} \right\}| \right) - |\mathcal{N}_{\tilde{u}}|$ represents the number of overlapping convolution-item based neighboring users' embeddings, indicating more frequent reuse of the received convolution-item based neighboring users' embeddings as it increases. $r_{\tilde{u}}$ represents the utilization rate of the convolution-item based neighboring users' embeddings for a single convolution-client, while $R$ represents the average utilization rate of the convolution-item based neighboring users' embeddings across all convolution-clients.
\end{itemize}  
\section{Conclusions and Future Work} \label{sec:conclusions}
In this paper, we are the first to propose a novel lossless GNN-based federated recommendation framework called LP-GCN (lossless and privacy-preserving graph convolution network), which can extend existing GNN-based recommendation models in a federated version equivalently, i.e., building a GNN-based model using decentralized sub-graphs rather than a centralized user-item interaction graph while safeguarding users' privacy well. Concretely, to address the privacy and security concerns when expanding users' sub-graphs, we introduce a method combining hybrid-encrypted item-IDs and virtual item obfuscation, ensuring security even in scenarios where the server colludes with some users. Additionally, we propose an embedding-synchronization mechanism that guarantees a lossless graph convolution process, including both forward propagation and backward propagation, backed by theoretical analysis and experimental validation. Empirical studies on three datasets demonstrate that our LP-GCN outperforms the existing federated methods in all cases.

For future work, we are interested in incorporating more auxiliary information such as interaction sequences, social networks, and behavior types to further enhance the recommendation performance. We are also interested in developing a federated recommendation system that actively learns user preferences~\cite{zhao2017unified}. Moreover, we plan to explore fairness~\cite{li2023fairness} in federated recommendation so as to encourage long-term user participation and contribution to the recommendation ecosystem. 

\bibliographystyle{ACM-Reference-Format}
\bibliography{sample-base}

\appendix 
\section{Proof} \label{sec:Proof}
We will prove that our federated recommendation framework LP-GCN is equivalent to the corresponding centralized counterpart. We will compare the training processes of our LP-GCN and the centralized model, including initialization, forward propagation, loss construction, local gradients computation and backward propagation. If each process is equivalent, then the overall framework is also equivalent.
\subsection{Initialization}
According to lines 14-19 in Algorithm~\ref{Initialization} and lines 3-6 in Algorithm~\ref{LP-GCN}, when the same random seed is used for embedding initialization, the centralized version and the federated framework of our LP-GCN have a same embeddings at layer 0,
\begin{equation} \label{eq_prove:user_embedding}
\boldsymbol{e}_{u}^{(0)} = \boldsymbol{e}_{u(u)}^{(0)}
\end{equation}
\begin{equation} \label{eq_prove:item_embedding}
\boldsymbol{e}_{i}^{(0)} = \boldsymbol{e}_{i(u_k)}^{(0)} , u_k \in \mathcal{U}_i
\end{equation}
where $e_u^{(0)}$ and $e_i^{(0)}$ are user $u$'s embedding and item $i$'s embedding in the centralized version, respectively, and $e_{u(u)}^{(0)}$ and $e_{i(u_k)}^{(0)}$ are user $u$'s embedding in client $u$ and item $i$'s embedding in client $u_k$ in the centralized version, respectively.
\subsection{Forward Propagation}
\textbf{The centralized version:} For each user node $u \in \mathcal{U}$ in the server, the aggregation function is
\begin{equation} \label{eq_prove_cen:user_embedding_convolution}
\boldsymbol{e}_{u}^{(l+1)} = AGG_u\left(\boldsymbol{e}_{u}^{(l)}, \{\boldsymbol{e}_{i'}^{(l)} : i'\in \mathcal{I}_u\}\right)
\end{equation}
, which integrates the representations of both the target node and its neighboring nodes at the $l$-th layer.

For each item node $i \in \mathcal{I}$ in the server, the aggregation function is
\begin{equation} \label{eq_prove_cen:compute_convolution_item_embedding}
\boldsymbol{e}_{i}^{(l+1)} = AGG_i (\boldsymbol{e}_{i}^{(l)},  \{\boldsymbol{e}_{u'}^{(l)} : u'\in \mathcal{U}_i\})
\end{equation}
, which integrates the representations of both the target node and its neighboring nodes at the $l$-th layer.

\textbf{The proposed federated framework LP-GCN:} For each user node $u \in \mathcal{U}$ in the client $u$, the aggregation function is
\begin{equation} \label{eq_prove_fed:user_embedding_convolution}
\boldsymbol{e}_{u(u)}^{(l+1)} = AGG_u\left(\boldsymbol{e}_{u(u)}^{(l)}, \{\boldsymbol{e}_{i'(u)}^{(l)} : i'\in \mathcal{I}_u\}\right)
\end{equation}
where $\boldsymbol{e}_{u(u)}^{(l+1)}$ represents the user embedding stored in client $u$ at layer $(l+1)$. 

According to lines 4-9 in Algorithm~\ref{Forward Propagation}, each client $u_k \in \mathcal{U}_{\tilde{i}} \backslash \{\tilde{u}_{\tilde{i}}\}$ sends its user embedding to client $\tilde{u}$ through the server. Then client $\tilde{u}$ computes the embedding of item $\tilde{i}$,
\begin{equation} \label{eq_prove_fed:compute_convolution_item_embedding}
\boldsymbol{e}_{\tilde{i}(\tilde{u}_{\tilde{i}})}^{(l+1)} = AGG_i (\boldsymbol{e}_{\tilde{i}(\tilde{u}_{\tilde{i}})}^{(l)}, \boldsymbol{e}_{\tilde{u}_{\tilde{i}}(\tilde{u}_{\tilde{i}})}^{(l)}, \{\boldsymbol{e}_{u'(\tilde{u}_{\tilde{i}})}^{(l)} : u' \in \mathcal{U}_{\tilde{i}} \backslash \{\tilde{u}_{\tilde{i}}\} \})=AGG_i (\boldsymbol{e}_{\tilde{i}(\tilde{u}_{\tilde{i}})}^{(l)},  \{\boldsymbol{e}_{u'(\tilde{u}_{\tilde{i}})}^{(l)} : u'\in \mathcal{U}_{\tilde{i}}\})
\end{equation}
where $\boldsymbol{e}_{\tilde{i}(\tilde{u}_{\tilde{i}})}^{(l+1)}$ is the $(l+1)$-th embedding of the convolution-item $\tilde{i}$ in client $\tilde{u}_{\tilde{i}}$, $\boldsymbol{e}_{u'(\tilde{u}_{\tilde{i}})}^{(l)}$ is the $l$-th user embedding from the neighboring user $u'$ stored in the convolution-client $\tilde{u}_{\tilde{i}}$, $\{\boldsymbol{e}_{u'(\tilde{u}_{\tilde{i}})}^{(l)} : u' \in \mathcal{U}_{\tilde{i}} \backslash \{\tilde{u}_{\tilde{i}}\} \}$ represents the set of embeddings of neighboring users when focusing on item $\tilde{i}$, and $\mathcal{U}_{\tilde{i}} \backslash \{\tilde{u}_{\tilde{i}}\}$ represents the users who has interacted with item $\tilde{i}$ exclude $\tilde{u}$ where $\tilde{i}$ locates.\\
By comparing Eq.(\ref{eq_prove_cen:user_embedding_convolution}) with Eq.(\ref{eq_prove_fed:user_embedding_convolution}), and Eq.(\ref{eq_prove_cen:compute_convolution_item_embedding}) with Eq.(\ref{eq_prove_fed:compute_convolution_item_embedding}), we can observe that the results of forward propagation are the same between the centralized version and the federated framework LP-GCN. The only difference is that the centralized computation takes place on the server, while the federated method performs calculations on clients.

According to lines 11-14 in Algorithm~\ref{Forward Propagation}, the embedding of item $\tilde{i}$ is synchronized to the other clients $u \in \mathcal{U}_{\tilde{i}} \backslash \{\tilde{u}_{\tilde{i}}\}$. Therefore, for each item $i$ in different clients, its embedding is the same.

\subsection{Loss Construction}
\textbf{The centralized version:} For each user node $u \in \mathcal{U}$ and each item node $i \in \mathcal{I}$ in the server, their embeddings, prediction score and loss are
\begin{equation} \label{eq_prove_cen:construct_final_user_embedding}
\begin{aligned}
\boldsymbol{e}_{u} = f_u(\boldsymbol{e}_{u}^{(0)}, \ldots, \boldsymbol{e}_{u}^{(L)}),\boldsymbol{e}_{i} = f_i(\boldsymbol{e}_{i}^{(0)}, \ldots, \boldsymbol{e}_{i}^{(L)})
\end{aligned}
\end{equation}

 \begin{equation} \label{eq_prove_cen:predicted_score}
\hat{y}_{ui} = f_{ui}(\boldsymbol{e}_{u}, \boldsymbol{e}_{i}) 
\end{equation}

 \begin{equation} \label{eq_prove_cen:compute one of the local train loss}
\text{loss}(u, i) = h(y_{ui}, \hat{y}_{ui})
\end{equation}

The overall loss is as follows,

 \begin{equation} \label{eq_prove_cen:compute the local train loss}
\text{Loss} = \sum_{(u,i)\in P} \text{loss}(u, i)
\end{equation}
where $P$ represents the set of training pairs.

\textbf{The proposed federated framework LP-GCN:} For each user node $u \in \mathcal{U}$ and each item node $i \in \mathcal{I}$ in each client $u$, we have
\begin{equation} \label{eq_prove_fed:construct_final_user_embedding}
\begin{aligned}
\boldsymbol{e}_{u(u)} = f_u(\boldsymbol{e}_{u(u)}^{(0)}, \ldots, \boldsymbol{e}_{u(u)}^{(L)}),\boldsymbol{e}_{i(u)} = f_i(\boldsymbol{e}_{i(u)}^{(0)}, \ldots, \boldsymbol{e}_{i(u)}^{(L)})
\end{aligned}
\end{equation}

 \begin{equation} \label{eq_prove_fed:predicted_score}
\hat{y}_{ui(u)} = f_{ui}(\boldsymbol{e}_{u(u)}, \boldsymbol{e}_{i(u)}) 
\end{equation}

 \begin{equation} \label{eq_prove_fed:compute one of the local train loss}
\text{loss}(u, i)_{(u)} = h(y_{ui(u)}, \hat{y}_{ui(u)}),\text{Loss}(u)_{(u)} = \sum_{(u,i)\in P_u} \text{loss}(u, i)_{(u)}
\end{equation}

The overall loss is as follows,

 \begin{equation} \label{eq_prove_fed:compute the local train loss}
\text{FedLoss} = \sum_{u\in \mathcal{U}} \text{Loss}(u)_{(u)}
\end{equation}
Compared the formulas between the centralized version and the federated framework LP-GCN, we could make a result that FedLoss (i.e., the loss of our LP-GCN) is equal to Loss (i.e., the loss of centralized version).

\subsection{Local Gradients Computation}
Before back-propagating the gradients, we need to analyse the gradients of last layer $L$. \\
\textbf{The centralized version:} For each user node $u \in \mathcal{U}$ in the server, we have its gradient,
 \begin{equation} \label{eq_prove_cen:computes user gradients of layer L}
\boldsymbol{g}_{u}^{(L)} = \frac{\partial \text{Loss}}{\partial \boldsymbol{e}_u^{(L)}}= \sum_{(u,i')\in P_u} \frac{\partial \text{loss}(u, i')}{\partial \boldsymbol{e}_u^{(L)}}
\end{equation}
where $P_u$ represents the sample pairs w.r.t. user $u$ which are used to construct loss.
 \begin{equation} \label{eq_prove_cen:computes item gradients of layer L}
\boldsymbol{g}_{i}^{(L)} = \frac{\partial \text{Loss}}{\partial \boldsymbol{e}_i^{(L)}}= \sum_{(u',i)\in P_i} \frac{\partial \text{loss}(u', i)}{\partial \boldsymbol{e}_i^{(L)}}
\end{equation}
where $P_i$ represents the sample pairs w.r.t. item $i$ used to construct the loss. Similarly, for each item node $i \in \mathcal{I}$, we have its gradient, 

\textbf{The proposed federated framework LP-GCN:} For each user node $u \in \mathcal{U}$, we have its gradient,
 \begin{equation} \label{eq_prove_fed:computes user gradients of layer L}
\boldsymbol{g}_{u(u)}^{(L)} = \frac{\partial \text{Loss}(u)_{(u)}}{\partial \boldsymbol{e}_{u(u)}^{(L)}} = \sum_{(u,i')\in P_u} \frac{\partial \text{loss}(u, i')_{(u)}}{\partial \boldsymbol{e}_{u(u)}^{(L)}}
\end{equation}
where $P_u$ represents the sample pairs w.r.t. user $u$ used to construct loss.

According to lines 7-11 in Algorithm~\ref{Loss Construction and Gradients Computation}, the client $u' \in \mathcal{U}_{\tilde{i}} \backslash \{\tilde{u}_{\tilde{i}}\}$ will send $\sum_{(u',i)\in P_i} \frac{\partial\text{loss}(u', i)_{(u')}}{\partial \boldsymbol{e}_{i(u')}^{(L)}}$ to the server. Then the server aggregates them to obtain $\sum_{u' \in \mathcal{U}_{\tilde{i}} \backslash \{\tilde{u}_{\tilde{i}}\}} \sum_{(u',i)\in P_i}\frac{\partial \text{loss}(u', i)_{(u')}}{\partial \boldsymbol{e}_{i(u')}^{(L)}}$ and then sends it to client $\tilde{u}_{\tilde{i}}$. In client $\tilde{u}_{\tilde{i}}$, we have the item gradient,
 \begin{equation} \label{eq_prove_fed:computes item gradients of layer L}
\boldsymbol{g}_{\tilde{i}(\tilde{u}_{\tilde{i}})}^{(L)}= \sum_{(\tilde{u}_{\tilde{i}},\tilde{i})\in P_{\tilde{i}}}\frac{\partial \text{loss}(\tilde{u}_{\tilde{i}}, \tilde{i})_{(\tilde{u}_{\tilde{i}})}}{\partial \boldsymbol{e}_{\tilde{i}(\tilde{u}_{\tilde{i}})}^{(L)}}+\sum_{u' \in \mathcal{U}_{\tilde{i}} \backslash \{\tilde{u}_{\tilde{i}}\}} \sum_{(u',i)\in P_i}\frac{\partial \text{loss}(u', i)_{(u')}}{\partial \boldsymbol{e}_{i(u')}^{(L)}}=\sum_{(u',\tilde{i})\in P_{\tilde{i}}} \frac{\partial \text{loss}(u', \tilde{i})}{\partial \boldsymbol{e}_{\tilde{i}}^{(L)}}
\end{equation}
By comparing Eq.(\ref{eq_prove_cen:computes user gradients of layer L}) with Eq.(\ref{eq_prove_fed:computes user gradients of layer L}), and Eq.(\ref{eq_prove_cen:computes item gradients of layer L}) with Eq.(\ref{eq_prove_fed:computes item gradients of layer L}), we can observe that the results of the computation of local gradients are the same between the centralized version and the federated framework LP-GCN, though they are calculated in different places, i.e., on the server and on clients.

\subsection{Backward Propagation}
\textbf{The centralized version:} For each user node $u \in \mathcal{U}$ and each item node $i \in \mathcal{I}$ in the server, we have the gradients,
 \begin{equation} \label{eq_prove_cen:back propagation of user gradients}
\boldsymbol{g}_{u}^{(l-1)} =\frac{\partial \boldsymbol{g}_{u}^{(l)}}{\partial \boldsymbol{e}_u^{(l-1)}} + \sum_{(u,i')\in P_u} \frac{\partial\text{loss}(u, i')}{\partial \boldsymbol{e}_u^{(l-1)}}+ \sum_{i' \in \mathcal{I}_u}\frac{\partial \boldsymbol{g}_{i'}^{(l)}}{\partial \boldsymbol{e}_u^{(l-1)}}
\end{equation}

 \begin{equation} \label{eq_prove_cen:back propagation of item gradients}
\boldsymbol{g}_{i}^{(l-1)} =\frac{\partial \boldsymbol{g}_{i}^{(l)}}{\partial \boldsymbol{e}_i^{(l-1)}} + \sum_{(u',i)\in P_i} \frac{\partial\text{loss}(u', i)}{\partial \boldsymbol{e}_i^{(l-1)}}+ \sum_{u' \in \mathcal{U}_i}\frac{\partial \boldsymbol{g}_{u'}^{(l)}}{\partial \boldsymbol{e}_i^{(l-1)}}
\end{equation}

\textbf{The proposed federated framework LP-GCN:} According to lines 6 and 15 in Algorithm~\ref{Loss Construction and Gradients Computation}, each client $\tilde{u}_{i'}(i' \in \mathcal{I}_u)$ will compute $\frac{\partial \boldsymbol{g}_{i'(\tilde{u}_{i'})}^{(l)}}{\partial \boldsymbol{e}_{u(\tilde{u}_{i'})}^{(l-1)}}$ and send it to the server. Then the server aggregates them, acquiring $\sum_{i' \in \mathcal{I}_u}\frac{\partial \boldsymbol{g}_{i'}^{(l)}}{\partial \boldsymbol{e}_u^{(l-1)}}$, and then sends it to client $u$. In client $u$, we have the gradient of user $u$,
 \begin{equation} \label{eq_prove_fed:back propagation of user gradients}
\boldsymbol{g}_{u(u)}^{(l-1)} =\frac{\partial \boldsymbol{g}_{u(u)}^{(l)}}{\partial \boldsymbol{e}_{u(u)}^{(l-1)}} + \sum_{(u,i')\in P_u} \frac{\partial\text{loss}(u, i')_{(u)}}{\partial \boldsymbol{e}_{u(u)}^{(l-1)}}+ \sum_{i' \in \mathcal{I}_u}\frac{\partial \boldsymbol{g}_{i'}^{(l)}}{\partial \boldsymbol{e}_u^{(l-1)}}
\end{equation}

According to lines 7-11 in Algorithm~\ref{Loss Construction and Gradients Computation}, the client $u' \in \mathcal{U}_{\tilde{i}} \backslash \{\tilde{u}_{\tilde{i}}\}$ will send $\sum_{(u',i)\in P_i}\frac{\partial\text{loss}(u', i)_{(u')}}{\partial \boldsymbol{e}_{i(u')}^{(l-1)}}$ to the server. Then the server aggregates them to obtain $\sum_{u' \in \mathcal{U}_{\tilde{i}} \backslash \{\tilde{u}_{\tilde{i}}\}} \sum_{(u',i)\in P_i}\frac{\partial \text{loss}(u', i)_{(u')}}{\partial \boldsymbol{e}_{i(u')}^{(l-1)}}$, and then sends it to client $\tilde{u}_{\tilde{i}}$.

According to lines 5 and 7 in Algorithm~\ref{Backward Propagation}, the client $u' \in \mathcal{U}_{\tilde{i}} \backslash \{\tilde{u}_{\tilde{i}}\}$ will send $\frac{\partial \boldsymbol{g}_{u'(u')}^{(l)}}{\partial \boldsymbol{e}_{i(u')}^{(l-1)}}$ to the server. Then the server aggregates them to obtain $\sum_{u' \in \mathcal{U}_{\tilde{i}} \backslash \{\tilde{u}_{\tilde{i}}\}}\frac{\partial \boldsymbol{g}_{u'(u')}^{(l)}}{\partial \boldsymbol{e}_{i(u')}^{(l-1)}}$, and then sends it to client $\tilde{u}_{\tilde{i}}$. In client $\tilde{u}_{\tilde{i}}$, we have the item gradient, 
 \begin{equation} \label{eq_prove_fed:back propagation of item gradients}
 \begin{split}
\boldsymbol{g}_{\tilde{i}(\tilde{u}_{\tilde{i}})}^{(l-1)} &=\frac{\partial \boldsymbol{g}_{\tilde{i}(\tilde{u}_{\tilde{i}})}^{(l)}}{\partial \boldsymbol{e}_{\tilde{i}(\tilde{u}_{\tilde{i}})}^{(l-1)}}+\sum_{(\tilde{u}_{\tilde{i}},\tilde{i})\in P_{\tilde{i}}}\frac{\partial\text{loss}(\tilde{u}_{\tilde{i}}, \tilde{i})_{(\tilde{u}_{\tilde{i}})}}{\partial \boldsymbol{e}_{\tilde{i}(\tilde{u}_{\tilde{i}})}^{(l-1)}} + \sum_{u' \in \mathcal{U}_{\tilde{i}} \backslash \{\tilde{u}_{\tilde{i}}\}} \sum_{(u',i)\in P_i}\frac{\partial \text{loss}(u', i)_{(u')}}{\partial \boldsymbol{e}_{i(u')}^{(l-1)}} + \frac{\partial \boldsymbol{g}_{\tilde{u}_{\tilde{i}}(\tilde{u}_{\tilde{i}})}^{(l)}}{\partial \boldsymbol{e}_{\tilde{i}(\tilde{u}_{\tilde{i}})}^{(l-1)}} + \sum_{u' \in \mathcal{U}_{\tilde{i}} \backslash \{\tilde{u}_{\tilde{i}}\}}\frac{\partial \boldsymbol{g}_{u'(u')}^{(l)}}{\partial \boldsymbol{e}_{i(u')}^{(l-1)}}\\
&=\frac{\partial \boldsymbol{g}_{\tilde{i}(\tilde{u}_{\tilde{i}})}^{(l)}}{\partial \boldsymbol{e}_{\tilde{i}(\tilde{u}_{\tilde{i}})}^{(l-1)}}+\sum_{(u',i)\in P_i} \frac{\partial\text{loss}(u', i)}{\partial \boldsymbol{e}_i^{(l-1)}}+\sum_{u' \in \mathcal{U}_i}\frac{\partial \boldsymbol{g}_{u'}^{(l)}}{\partial \boldsymbol{e}_i^{(l-1)}}
\end{split}
\end{equation}
By comparing Eq.(\ref{eq_prove_cen:back propagation of user gradients}) with Eq.(\ref{eq_prove_fed:back propagation of user gradients}), and Eq.(\ref{eq_prove_cen:back propagation of item gradients}) with Eq.(\ref{eq_prove_fed:back propagation of item gradients}), we can observe that the results of backward propagation are the same between the centralized version and the federated framework LP-GCN.

Therefore, we can prove that the graph convolution process of our LP-GCN is lossless, meaning it is equivalent to the non-federated centralized counterpart.
\end{document}